\documentclass[11pt]{article}

\usepackage{a4}
\usepackage[breaklinks,colorlinks,citecolor=blue]{hyperref}

\usepackage{graphicx}

\usepackage{natbib}
\bibpunct{(}{)}{,}{a}{}{;}

\newcommand{\be}{\begin{equation}}
\newcommand{\ee}{\end{equation}}
\newcommand{\bea}{\begin{eqnarray}}
\newcommand{\eea}{\end{eqnarray}}
\newcommand{\Dd}{\mathrm{d}}
\newcommand{\doi}[1]{{\sc doi:} \href{http://dx.doi.org/#1}{\detokenize{#1}}}
\newcommand{\ADS}[1]{{ADS BibCode:} \href{http://adsabs.harvard.edu/abs/#1}{\detokenize{#1}}}

\usepackage{caption}
\begin{document}

\title{\bf Relatively complicated?\\[0.25em] \large Using models to teach general relativity at different levels}
\author{Markus P\"ossel\\ \small Haus der Astronomie and Max Planck Institute for Astronomy}
\date{\normalsize Invited Talk at the session of the section {\em Gravitation and Relativity}\\ at the Spring Meeting 2017 of the German Physical Society (DPG)\\ Bremen, 16 March 2017}
\maketitle

\begin{abstract}
This review presents an overview of various kinds of models -- physical, abstract, mathematical, visual -- that can be used to present the concepts and applications of Einstein's general theory of relativity at the level of undergraduate and even high-school teaching. After a general introduction dealing with various kinds of models and their properties, specific areas of general relativity are addressed: the elastic sheet model and other models for the fundamental geometric properties of gravity, models for black holes including the river model, cosmological models for an expanding universe, and models for gravitational waves as well as for interferometric gravitational wave detectors.
\end{abstract}

\newpage

\tableofcontents

\newpage

\section{Introduction}
The spring meeting 2017 of the German Physical Society (Deutsche Physikalische Gesellschaft, DPG) featured a session on aspects of general relativity education, organised by Domenico Giulini (Leibniz University Hannover). The present text is an extended version of my opening talk. My aim was to provide a systematic survey of the manifold applications and pitfalls of pedagogical models in teaching general relativity. 

Such models are a central tool in teaching about Einstein's theory of gravity, in particular in those settings where learners will not have access to the theory's full formalism, such as an undergraduate or high school course. On the other hand, general relativity is sufficiently complex, as well as sufficiently subtle, to create problems for those who adhere too closely to simple models, and whoever employs models in their teaching will need to exercise proper caution. 

The following review is primarily meant for teachers and science communicators, but may also be useful for learners who are using simple models in an attempt to gain an understanding of the basics of Einstein's theory. The text is structured as follows: Section \ref{ModelSection} reviews models, specifically teaching models for general relativity, and their basic properties. Section \ref{GoodBad} asks what makes a model good or bad, and lists appropriate quality criteria. Section \ref{ElasticSheet} discusses what is surely the most famous model for teaching general relativity: the warped elastic sheet. The following sections discuss models for specific aspects of general-relativistic physics: black holes in section \ref{BlackHoles}, cosmology (with a focus on cosmic expansion) in section \ref{Cosmology}, gravitational waves and interferometric gravitational wave detectors in section \ref{GWSection}. A brief wrap-up is found in section \ref{Conclusion}.

\section{Models}
\label{ModelSection}

General relativity, from curved space to black holes and Big Bang cosmology, holds great fascination for a large portion of the general public. On the other hand, a proper understanding of how the theory works requires fairly advanced mathematics -- so advanced that even at the level of graduate, teaching the relevant mathematical concepts (notably differential geometry) is commonly an integral part of general relativity courses. 

Given the theory's broad appeal, efforts to make general relativity accessible in a simplified way are nearly as old as the theory itself \citep{EinsteinAllgemein,Weyl1920,Eddington1921,BornRelativitaet}, and range from curricula tailored to undergraduate students in physics \citep{Hartle2006} or part of a general education syllabus \citep{Hobson2008} to highly popular expositions of the theory and/or closely related subjects such as cosmology, black holes, or quantum gravity \cite[and others]{Gardner1976,Weinberg1977,Sagan1980,GreensteinFrozen,HawkingBrief,Wheeler1990,Thorne1994,Bartusiak2000,Levin2002,GreeneElegant,Randall2005,Levin2016}. Regarding the variety of levels at which general relativity can be taught, I find the ``in-between'' particularly interesting: the books somewhere in between the maths-free style of popular science and the bring-on-the-formalism style of fully fledged general relativity text books, which require more limited knowledge of mathematics (such as the typical mathematical toolkit of high school or undergraduate students) to explore general relativity and its applications \citep{Synge1970,Sexl1975,Berry1976,Geroch1981,Liddle1988,Taylor2000,Hartle2003,Schutz2004,Foster2006,BeyversKrusch2009}. Notably, there have been several proposals to revert the usual order of presentation in physics courses, introducing a simplified version of general relativity first, and then move on to physics topics commonly considered more elementary. Examples are the ``general before special relativity'' approach of \citet{Rindler1994} and, regarding high school physics teaching, the successful ``Einstein First'' approach of introducing concepts from relativity (and quantum mechanics) right at the start \citep{Pitts2013,Kaur2017a,Kaur2017b,Choudhary2018,Foppoli2018}.\footnote{More information about the project can be found at \href{https://www.einsteinianphysics.com/}{https://www.einsteinianphysics.com/}}

To replace or support the teaching of the full mathematical formalism, such expositions, whether in the form of a book, article, lecture, or video, make use of {\em models}. The present text is an attempt to give an overview of such models, as well as a discussion of their physical properties. While the review will list the advantages and disadvantages of specific models from the point of view of physics, it will not address the usefulness of specific models in specific contexts; indeed, education research addressing that question, involving tests of the suitability of specific models and forms of presentations of information about general relativity for a given audience, appears to be largely uncharted territory, although there have been relevant projects both on a secondary school level \citep{Baldy2007,Henriksen2014} and involving undergraduate students \citep{Bandyopadhyay2010,Watkins2014}.

\subsection{Models defined}

In the context of this review, ``model'' is taken to be synonymous with a generalised version of ``teaching model'' as follows: 
\begin{quotation}
\noindent A {\bf (teaching) model of general relativity, or of a part of general relativity}, is a concrete or abstract entity which represents some of the structure of general relativity, or structures related to general relativity's concepts and applications, in a simplified way, for the purpose of teaching about general relativity and its applications.
\end{quotation}
This includes teaching models in the narrow sense, that is, models specifically created for teaching, as well as expressed models, consensus models and scientific models that were created as part of the research process, but can be used for teaching \citep{GilbertEtAl2000,Gilbert2004}.\footnote{Teaching models fit into various theories of teaching and education in different ways. Their simplifications are an important part of ``pedagogical reduction'': the effort to transform the complexities of subjects including physics into content suitable for learners \citep{Gruener1967}. In parts of physics education research, this kind of pedagogical simplification is known as ``elementarization'' \citep{Bleichroth1991, Reinhold2010, Kircher2015}, although one should be careful since the word has at least one other meaning --- the selection of elementary topics deemed to be of fundamental importance for a given subject. As such, models have long been a key ingredient of astronomy and physics teaching \citep{Lindner1997}. }
  
Models can be physical objects, such as the well-known rubber balloon used to demonstrate cosmic expansion \citep{Lotze1995a,Lotze2002c,Hawking1996}, or the glass or plastic lenses that model the effects of gravitational lensing \citep{Liebes1969,Icke1980,Higbie1981,Adler1995,Falbo-Kenkel1996,Lotze2004,Brockmann2007,Brockmann2009,Huwe2015}. They can also be actions, such as roleplaying \citep[in our special case relating to astrophysical objects and/or processes]{Aubosson2006a}. Visualisations that showcase specific relativistic effects, frequently introducing auxiliary structures to do so and/or using a first-person perspective, are another type of model \citep{Kraus2005,Kraus2005b,Kraus2007,Nollert2005,Weiskopf2006,Kraus2008,Ruder2008,Mueller2011,Kortemeyer2013,Boblest2015,Mueller2015}. Near the opposite end of the photorealism scale, infographics make use of simple graphical structures and arrangements to convey key information \citep{Lowe2015,Watzke2017}. 3D printing, on the other hand, takes visualizations into the third dimension \citep{Clements2017,Arcand2017}.

\begin{figure}[htbp]
\begin{center}
\includegraphics[width=0.45\textwidth]{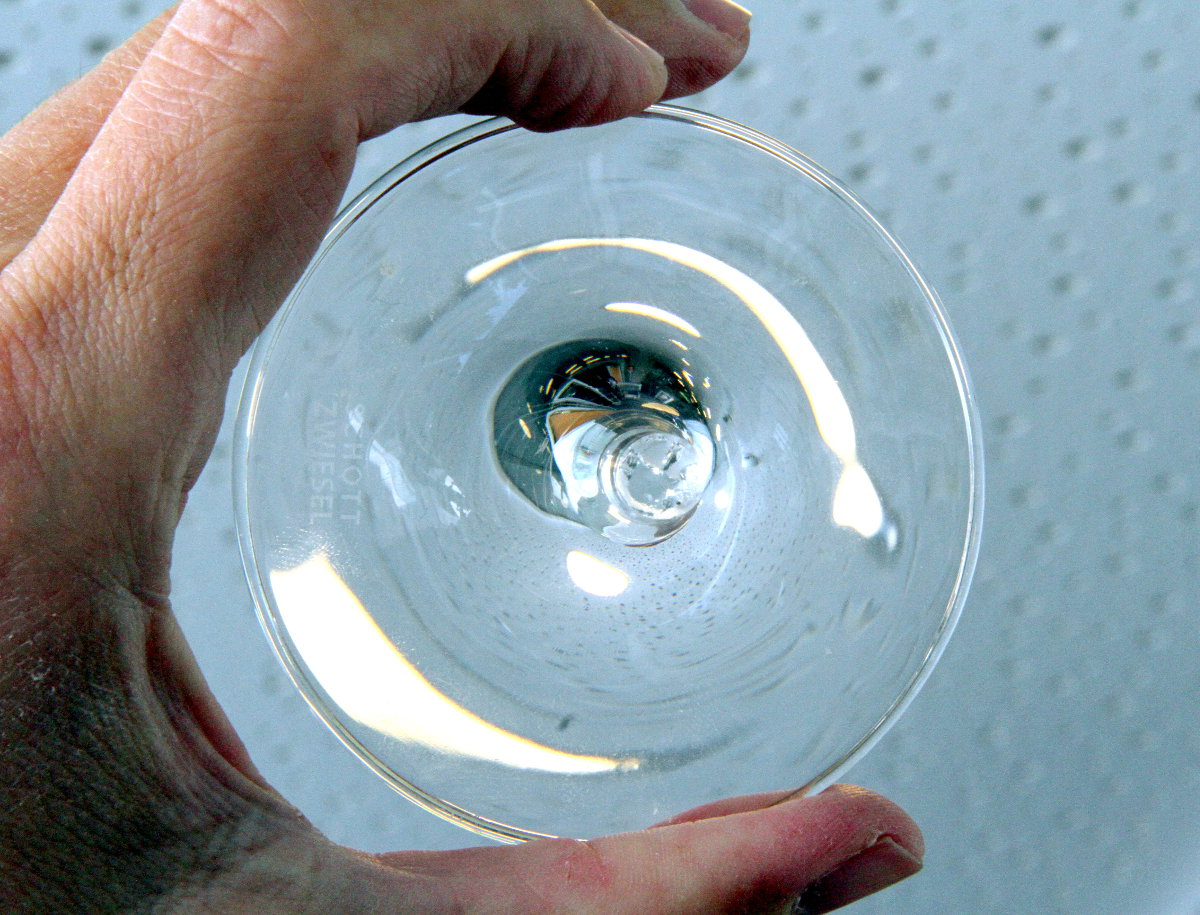}\hspace*{1.5em}\includegraphics[width=0.45\textwidth]{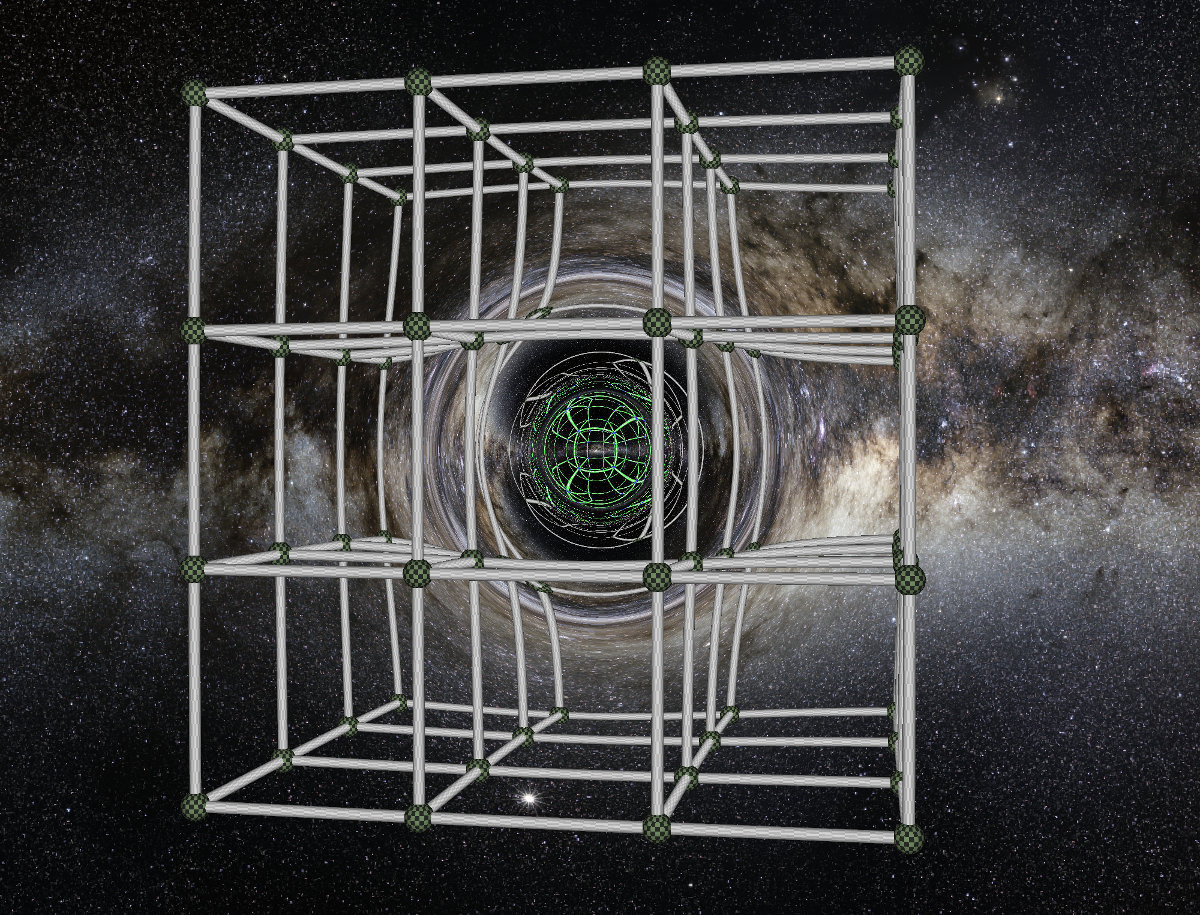}\\[1em]
\includegraphics[width=0.45\textwidth]{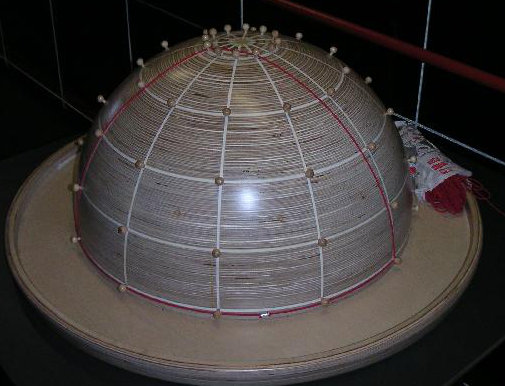}\hspace*{1.5em}\includegraphics[width=0.45\textwidth]{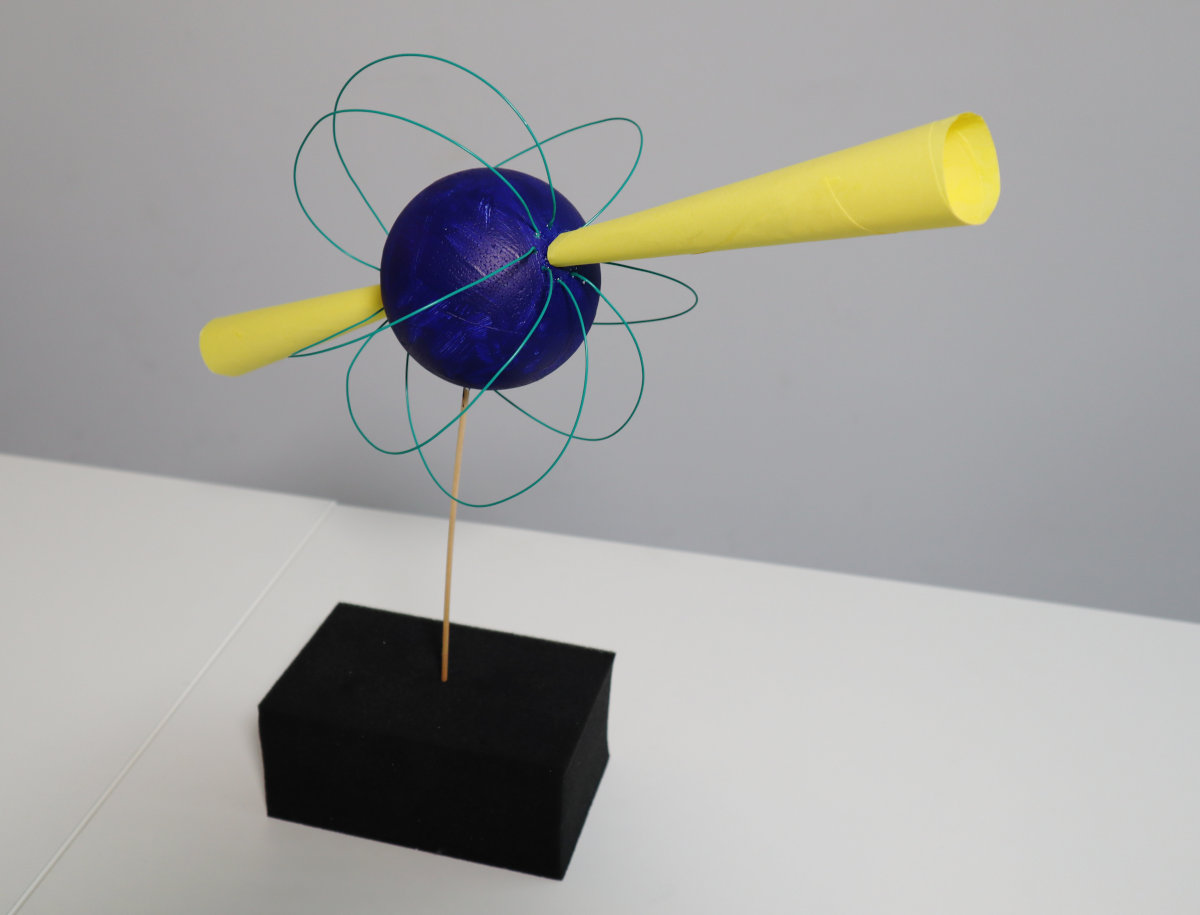}\\[1em]
\caption{Examples for relativity-related models. Top left: physical model of a gravitational lens (own image), Top right: visualisation with auxiliary structure (image: Thomas M\"uller). Bottom left: Wooden half-sphere to illustrate the curvature of a surface; pegs and elastic string allows for the easy construction of geodetic triangles (own image). Bottom right: Model of a pulsar which, when rotated, can be used to illustrate the lighthouse effect of the pulsar's radiation (own image)}
\label{TwoModels}
\end{center}
\end{figure}
The broad concept of a teaching model includes uses of analogy, metaphor, and simile, all of which employ structural similarities to indirectly convey information about an entity, exploiting prior knowledge on the part of the target audience \citep{Aubosson2006b,McCool2008,McCool2009a,Niebert2012}.

Use of models and visual analogies can even cross the boundaries between science and art --- an impressive example is the design-award-winning book by \citet{Leitner2013}, which uses carefully arranged everyday objects to illustrate relativity, quantum theory and cosmology.
 
Various kinds of models have been used to explain the mathematics underlying general relativity. Rectangular spatial coordinates can be visualized by measurements inside a room whose walls and floor provide the three orthogonal coordinate planes \citep{Sullivan1926}, or by using the rectangular layout of some cities in the United States, plus the information of which floor of a building happened on \citep{Slosson1920,Harrow1920}. General coordinate transformations have been visualised by alternative city layouts \citep[chapter 3]{Greene2004}, or by marking the usual Cartesian coordinates on a rubber sheet, and then continuously deforming that sheet (\citealt{Ames1920}, chapter 8 in \citealt{Russell1925}, \citealt{Poessel2016b}), and even by replacing the usual straight measuring rods by wriggling live eels \citep[chapter 8]{Russell1925}. Distorted images in a fun-house mirror have been used as a simple image for introducing the necessity of measurement (in other words, a metric) in a situation where arbitrary coordinates are admissible --- even if you are looking into a distorting mirror, you will be able to accurately measure your height using a yardstick you hold next to yourself \citep{Slosson1920}. The same author introduces the notion of spacetime by cutting a film strip into single frames, and stacking those on top of each other, an idea that can also be found in \cite{Royds1921}. Tensors have been modelled as ``black boxes'' with rigid rods pushed in, or pulled out as input, and moving in response as output \citep{Synge1970}, and singularities via the focussing properties of an optical system \citep[sec.\ 12.1]{Filk2004}.

Last but not least, models can themselves be mathematical in nature --- similar in nature to the astrophysical models used not for teaching, but as an integral part of the research process (\citealt{Hacking1989}, \citealt{Anderl2016},
 ch.\ 8 in \citealt{Anderl2017}). Examples include a toy model of non-instantaneous (scalar) gravity used to demonstrate properties of gravitational waves, for instance \citep{Schutz1984}, or a (faithful!) reformulation of Einstein's equations in terms of the changing volume of a spherical formation of test particles \citep{Baez2005}. In a very simple type of mathematical model, more complex functional relationships are replaced by proportionalities --- as is the case in dimensional analysis, a surprisingly powerful way of deriving physical laws, in our particular context e.g. for the properties of gravitational wave sources \citep{Mathur2017}.

In explaining physical experiments, a common simplification is to leave out various sources of noise, as well as the features built into the experimental setup to counter specific noise. The result is a simplified model of the experiment. We will encounter such models in section \ref{InterferometricDetector} and following when we talk about interferometric gravitational wave detectors. We will meet additional examples for the different kinds of models throughout these notes.

There is a biological, evolutionary context to the use of many of these models. Our brains did not evolve to (explicitly) solve differential equations, or to manipulate algebraic expressions. These are skills that need to be learned, over time. In contrast, models play to our brains' natural strengths. Notably, our brains are good at recognising visual patterns, and configurations in three-dimensional space (even though we are handicapped by the fact that what our eyes see is, in fact, not truly three-dimensional, but the combination of two two-dimensional projections). Whenever we manage to encode part of the relative structure into patterns or relations, whether in an object that is meant to be manipulated hands-on or a visualisation, we exploit this particular type of brainpower.

In one respect, general relativity is at a disadvantage relative to other areas of physics, when it comes to models meant to show geometrical structure. Precisely because the geometry of space and time is curved and distorted in general relativity, it does not conform to our everyday experiences --- and sometimes, simple visual models of spacetime geometry can be misleading for this very reason. A poignant example are representations of the interior of a black hole, as shown in section \ref{BHGeometry}, below.

\subsection{Narratives}

Our brains also appear to be good at listening to stories, at following them and understanding them. Narrative structure is commonly used in teaching, whenever we give our lecture a narrative: introducing some context that motivates the introduction of new concepts, which leads to consequences/applications and/or the introduction of additional concepts, and so on, to make a particular kind of model: an {\em explanatory story} \citep{Beyond2000}.

As an example, consider the narrative of the ``procrastination principle,'' told in terms of spacetime diagrams and the concepts of special relativity (\citealt{Wheeler1990,Poessel2005}, with the mathematical description worked out in \citealt{Gould2016}, cf. \citealt{Stannard2017}):
\begin{enumerate}
\item Classical free particles move along straight lines at constant speeds
\item In a spacetime picture, this corresponds to straight worldlines
\item Special-relativistic time dilation shows that the straight worldline joining two given events is the worldline of maximal proper time 
\item Thus, free particles move so that maximal proper time elapses between any two events on their worldlines. This is the procrastination principle: ``free particles procrastinate''
\item Introduce regions where time passes more quickly and regions where time passes more slowly, as measured by the proper time of observers whose worldlines
pass through these regions 
\item Observe how location-dependent time plus the procrastination principle change the orbits of free particles. The result is the Newtonian limit of general relativity: classical gravity encoded in the time-time coefficient of the metric.
\end{enumerate}
While this narrative stays close to the physical concepts, narratives for a more general audience often employ embellishments, and more elaborate scenarios. The first step is often to introduce roles your audience is familiar with: those of a person who observes and/or experiences.

An idealised observer, proper time and radar ranging are abstract. But consider an everyday kitchen scene, with a man cooking a breakfast egg. Now, use a rope to lower the kitchen further and further towards a black hole, and observe events from afar \citep{GreensteinFrozen}. This is a narrative scenario that allows the reader to identify with the observer, and experience relativistic effects in the same way that we share the lives of our favourite protagonists in films or novels. For many scenarios of relevance to general relativity, it helps when readers are familiar both with media coverage of space travel, and with science fiction; in fact, the fascination of science fiction stories can be used deliberately as a teaching tool \citep{Morris1988,James2015}.

\begin{figure}[htbp]
\begin{center}
\includegraphics[width=0.4\textwidth]{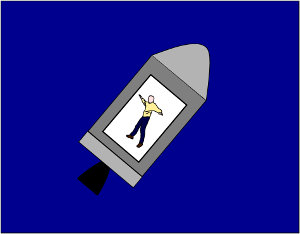} \hspace*{8em} \includegraphics[width=0.2\textwidth]{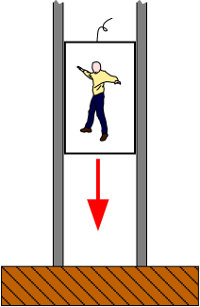}
\caption{This is your body in free fall: personalising the physics of the equivalence principle. Images from \citeauthor{Poessel2005} \citeyear{Poessel2005}}
\label{FreeFallFigure}
\end{center}
\end{figure}

The equivalence principle is another good candidate for personalisation (cf. figure \ref{FreeFallFigure}). The statement that, in a suitably small cabin over a suitably short period of time, free fall is equivalent to a situation with no gravitational influence whatsoever, is so general that readers can immediately identify with a hypothetical (although, possibly, doomed) person in such an elevator, or otherwise in free fall (\citealt{BornRelativitaet}, \citealt{Slosson1920}, \citealt{Russell1921}, ch.\ XI in \citealt{Barnett1948}, ch. 5 in \citealt{Gardner1976}, \citealt{Davies1977}, \citealt{Zeilik1991}, \citealt{Mielke1997}), or even imagine a fictitious situation in which they themselves find themselves in a windowless cabin. Alternatively, accelerating such a closed cabin can produce a gravitation-like downward acceleration felt by everyone who is in the cabin.In modern texts, elevator cabins are the most common example. Other possibilities are spaceships, or the giant habitable bullet employed in Jules Verne's {\em From the Earth to the Moon} and {\em Around the Moon} \citep{Langevin1922,Nordmann1922}.\footnote{Verne himself did not describe the situation correctly --- in his account, the travellers within the projectile feel no weightlessness but in the brief phase where they are at the ``point of equal [gravitational] attraction'' between the Earth and the Moon, described in chapter 8 of {\em Autour de la Lune} (1870).}
 (Alternatively, beyond a merely narrative approach, microgravity in free fall can of course be simulated in the classroom, for instance with the help of a smart phone camera \citep{Kapotis2016}.)

Extreme examples of elaborate stories are exploration narratives for whole counter-factual worlds -- Edwin Abbott's flatland, with its elaborate geometry-based caste system \citep{AbbottFlatland}, its successor sphereland with an ambitious triangulation programme \citep{Caplan2015} or following the fate of rulers moved along geodesics \citep{Will1986}, or the relativistic world of George Gamow's Mr. Tompkins \citep{GamowTompkins}. More common are historical narratives: expositions that introduce scientific concepts, phenomena, and theories, by following the historical path of discovery; examples are
\citet{Ferris1988,Overbye1991,Bartusiak2000,Bartusiak2009,Ferreira2014,Bartusiak2015,Bartusiak2017}. Such historically-oriented narratives are quite powerful: they show the scientists themselves and make the process of scientific discovery visible; they automatically have more authenticity than artificial worlds, and show us science as a deeply human endeavour. 

One possible downside particular to historical narratives is that once history is used as a tool for explaining physics there is a temptation to oversimplify in a specific way: Real science history is messy, progress in research anything but straight or straightforward; byways and distractions abound; scientists may find the right answer for the wrong reasons, or wrong answers for the right reasons. In moderation, such elements serve to add verisimilitude to the narrative, but more often than not, historical accuracy conflicts with the goal of simplified exposition. If you are trying to get your (or your reader's) head around the fundamentals of general relativity, too early an encounter with Einstein's ``Entwurf theory,'' or the various culs-de-sac of Einstein's long struggle towards his field equations, are likely to add inacceptable amounts of complication and confusion. 

A common solution to this dilemma are pseudo-historical accounts of physics (or other sciences), deplored by historians of science, but useful as, yes, simplified models of physics history. Useful, that is, in the context of developing a simplified account of a physical theory, or experimental results -- certainly less than useful, and in many cases positively harmful, for understanding real-life scientific progress. Which brings us to the more general question of when models are good, and when they are bad, to be addressed later on in section \ref{GoodBad}.

\subsection{General and specialised models}

The common ground provided by images, or narratives, in particular with historical narratives and the conventions and basic structure of a life story, make these various model techniques suitable for a general audience. Most of the audience will share similar experiences when it comes to three-dimensional space, perspective, temporal and narrative structure.

But experiences are not universal --- and that certainly includes some of the experiences utilized to create simple models. Various types of disabilities mean that models that rely on visual or aural perception will not be accessible for a subset of the intended audience. This is where a healthy diversity of models is beneficial, providing alternatives to understand a specific concept using, say, a visual model or a suitable narrative. The natural diversity of models is complemented by specific efforts to develop models suitable for people with disabilities \citep{Grady2003,Grice2006,Arcand2010,Ortiz-Gil2011,Kraus2016}.

Restricted, more specialised audiences are likely to share even more common ground than a general audience. High school students of various ages, or undergraduate students (physics or not) have additional knowledge, notably about mathematics and physics, that one can exploit for model building. A prominent example are the various derivations of relativistic effects that assume no more than previous knowledge of special relativity and a simple version of the equivalence principle. Such derivations yield exact results for gravitational time dilation \citep{Schild1960,Sexl1975,Schutz1985,Schroeter2001,Schroeter2002} and, making ample use of the symmetries of a homogeneous and isotropic universe, in cosmology \citep{CallanEtAl1965}. They yield qualitative solutions, typically wrong by a factor of 2 for neglecting the curvature of space, for light deflection, with the same results as those of ballistic theories of light based on Newtonian mechanics (\citealt{Langevin1922}, \citealt{Melcher1978} [section 10.6.2], \citealt{Koltun1982}, \citealt{Lindner2004}, \citealt{Lotze2005}, \citealt{Lovatt2009}).\footnote{There is some controversy about which assumptions, beyond the equivalence principle, are needed to derive certain general-relativistic results \citep[and references cited therein]{Kassner2015}.}

\section{Models good and bad}
\label{GoodBad}

Borrowing some terminology from mathematics, the key part of the definition of any model is a {\em map}: various features and properties of general relativity, its concepts and applications, are mapped to corresponding features and properties of the model. In model descriptions, this is where the relevant verbs are ``to play the role of'',  ``to represent,'' or ``to correspond with.'' In a constructivist framework, that typically involves linking structures from one subdomain of knowledge with those of another subdomain \citep{Zubrowski2009}.

In the expanding substrate models of section \ref{ExpandingSubstrate}, the rubber balloon, or the raisin cake, play the role of expanding space in cosmology; stickers affixed to the balloon in the one case, the raisins in the other, represent galaxies in the Hubble flow. With the concept of a map in mind, we can judge the quality of models by the following criteria:

\begin{enumerate}
\item {\em Faithfulness}: the model represents all important features of the original
\item {\em Parsimony:} non-essential, potentially distracting features are kept to a minimum
\item {\em Confusion avoidance:} the model has no (obtrusive) misleading features
\item {\em Accessibility:} the target audience is capable of understanding the structure of the model
\end{enumerate}

\begin{figure}[htbp]
\begin{center}
\includegraphics[width=0.9\textwidth]{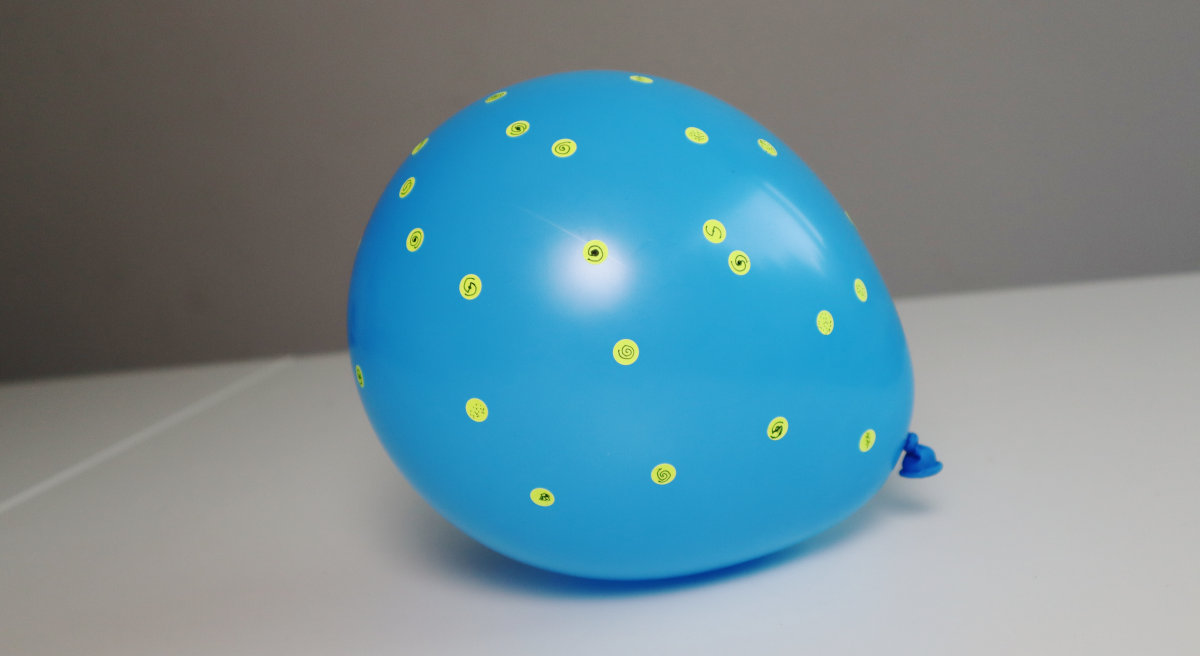}
\caption{Rubber balloon model of the (expanding) universe. The stickers on the balloon's surface represent galaxies in the Hubble flow}
\label{BalloonUniverseOne}
\end{center}
\end{figure}

As for faithfulness, the ideal model represents all important features of general relativity. What is or isn't important is determines by the goal of that particular model: which aspect or part of relativity is to be made accessible to the intended audience?  In the case of the expanding rubber balloon universe (cf. figure \ref{BalloonUniverseOne}), one aspect of faithfulness would be that distances between the sticker-galaxies on the balloon surface indeed grow in proportion to a universal scale factor as the balloon expands. This shows how good models need to have structure: By exploring the relations of various features of the model, one can learn about relations of the corresponding concepts and/or phenomena in general relativity. The galaxy-stickers on the balloon are not just abstract symbols. Once I have understood what they are, I can ask questions such as: how do their distances along the balloon surface change? The answer to this question tells me something about the properties of intergalactic distances in the Hubble flow. Structural correspondences are the key to how models help us understand the original.

If a model represents only specific features of a physical concept or phenomenon, the target audience is likely to judge those aspects to be more important than the unrepresented aspects. If those are indeed the important aspects, the model helps the audience to understand; if they are not, that can hamper understanding \citep{Kampourakis2016}.

Following through with a model, taking the model seriously, is crucial --- up to a point. There will always be model features that do not correspond to any features of the original. The balloon universe has a neck and mouth, necessary to fill the balloon; there is no corresponding property of the mathematical models of general relativity, nor, one hopes, of the real universe. This introduces the criterion of parsimony in the following sense: A model's extraneous features, that is, those features that do {\em not} represent aspects of the physics-to-be-taught, should be kept to a minimum. There are some subtle aspects involved at this point. For instance, as it turns out, color choice for representing dark matter in a cosmological visualization can have a significant effect on the audience's interpretation of the visualization \citep{Buck2013}.

Extraneous properties become problematic whenever they are potentially misleading. In the case of the rubber balloon, mouth and neck will hopefully not confuse anybody; those additions are too obviously artificial. Much more problematic is the embedding of the two-dimensional balloon surface in three-dimensional space. If you have ever explained cosmology using the rubber balloon model, and asked your audience where the center of the universe is, chances are that the majority of respondents identify the center of the universe with the center of the three-dimensional balloon. 

A particularly unfortunate effect of misleading model features is that those in your audience who aim to understand the model and to build on it tend to be hardest-hit: those who try most actively to understand what you are teaching, by trying to think through the consequences of your model, to reconcile the model with other knowledge about general relativity they have acquired, and to extrapolate from the model to learn more about relativistic physics. Misleading model features can make reconciliation with other aspects of general relativity difficult, are not a solid foundation for extrapolation, and in this way can have a discouraging influence on learners.

In contrast, those who listen to your explanation passively, and do not try to make the model their own by thinking it through, are unlikely to stumble upon misleading features unless you point them that way explicitly. 

The example of the center of the universe suggests an operational criterion for how to test model quality. Make an inventory of the structure of whatever part of general relativity you are trying to teach. What leads to what else? What can be deduced? Then, ask your audience specific questions, designed to make your audience draw the corresponding conclusions from the model. The correct and incorrect conclusions should give you a fairly complete basis on which to judge the quality of your model.\footnote{As far as I am aware, there is at this point in time not yet any general relativity concept inventory, corresponding to the Relativity Concept Inventory for special relativity \citep{Aslanides2013}, which would provide a more general framework, and facilitate comparisons between different model-based teaching situations.}

If you are lucky, the feedback will point towards possible improvements. Where no improvements are possible, you will want to think about ways of mitigating potential confusion -- often by naming and explaining the extraneous, potentially misleading features. In the case of the rubber balloon, there is no direct way around the embedding. But if you tell your audience about this specific problem, pointing out that the universe, in this case, is the two-dimensional surface of the balloon, you can try to minimize the impact this extraneous feature will have. Timing is important, of course. Caveats presented too early, too numerously, may well spoil the effect of your model and serve to create the confusion they are meant to help avoid.

Models should be accessible. Models exploit that your audience is more familiar, or can more readily be familiarized with, your model structure than with the real thing. Even if you should uncover deep analogies between general relativity and, say, dramatic performances in 3rd century Iceland, then at least to non-Icelandic audiences, and quite probably also to modern-day Icelanders, the resulting simplified model would likely be not much less alien than the original.  

Faithfulness and parsimony can be analysed within the realm of physics, without reference to a specific target audience. Confusion avoidance and accessibility are audience-specific; some of their aspects can be analysed in abstract terms, for instance by noting where the model and the physics it is meant to represent diverge. But ultimately, degree and thus significance of confusion and accessibility can only be tested in real-life teaching situations --- and experience shows that, for instance, students in real-life situations often enough find aspects of a teaching model confusing the teacher had not even thought of \citep{Harrison2006}. Testing can occur unsystematically, with teachers reflecting on and documenting their own experiences, or systematically within controlled studies.

With all this in mind, let's look at specific examples for teaching models in the context of general relativity, their strengths and weaknesses!

\section{The elastic sheet}
\label{ElasticSheet}

The elastic sheet or rubber sheet model, sometimes identified as a trampoline or realised in the shape of a pillow \citep{Baldy2007}, is one of the most common models for gravity in general relativity. It is used in a variety of contexts: in text books \citep{Berry1976,Melcher1978,Bennett2017}, in popular books \citep{Harrow1920,Sampson1920,Gamow1947,Gardner1976,Nicolson1981,Nicolson1985,Will1986,GreeneElegant,Greene2004,Hawking2001,Stannard2008,Bennett2014,Egdall2014,Vaas2018} and videos, in the context of astronomy education \citep{White1993,Chandler1994,Brockmann2007,Brockmann2009,Turner2013a,Turner2013b,Ford2015,Kaur2017a,TranRussell2018a,TranRussell2018b} and astronomy outreach \citep{McCool2008} both for Newtonian physics and for general relativity, and famously also in Carl Sagan's {\em Cosmos} television series \citep{CosmosEp9}. Swimwear fabric is a suitable material for demonstration experiments. 

\begin{figure}[htbp]
\begin{center}
\includegraphics[width=0.85\textwidth]{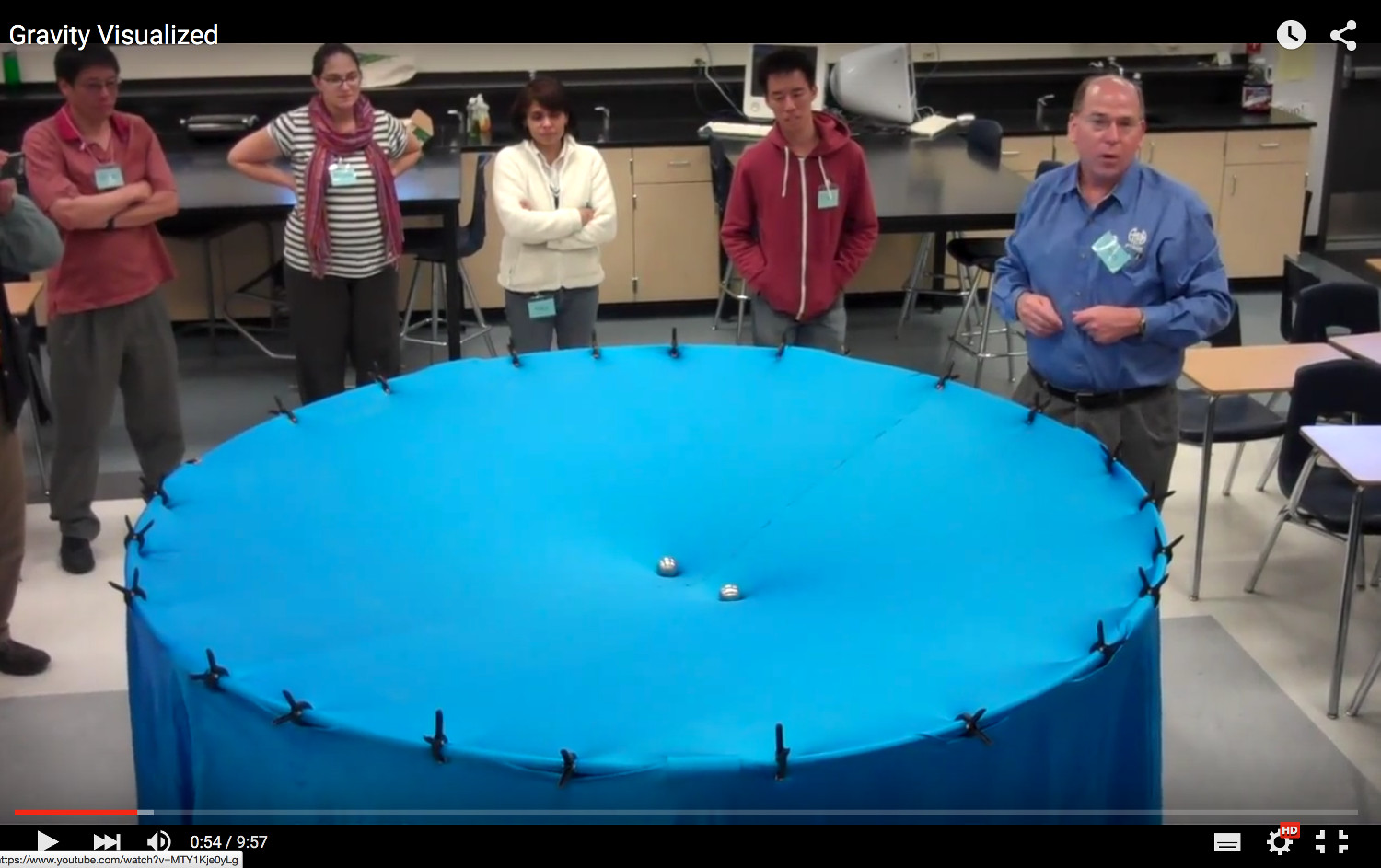}
\caption{Elastic sheet demonstration of relativistic gravity by Dan Burns, Los Gatos High School, Los Gatos, CA.
Screen shot from the video  \href{https://youtu.be/MTY1Kje0yLg}{https://youtu.be/MTY1Kje0yLg} }
\label{RubberSheetBurns}
\end{center}
\end{figure}

Fig.\ \ref{RubberSheetBurns} shows an example of the elastic sheet demonstration, presumably not with rubber but with a fairly elastic piece of fabric, from a highly successful YouTube video by Dan Burns at Los Gatos High School in Los Gatos, California.

As of late December 2018, this video, which was uploaded on March 10, 2012, has more than 53 million views on YouTube. Evidently, this is an appealing demonstration. The main elements are readily accessible to a general audience. In fact, they conform so much to our expectations from everyday experience that the model can be presented as a narrative. Even when we just hear about the set-up, we are likely to be able to picture how the fabric is distorted by the heavy metal sphere that is placed upon it, and how the trajectories of the smaller spheres rolled unto the fabric are bent around the larger sphere, corresponding to (partial) orbits.\footnote{An example can be heard in \citet{Freistetter2013b} from 3:55 to 4:42.}

Let's now talk about the strength and weaknesses of this model, about what works and what doesn't. There are two types of arguments to be made here. One involves questions of experimental education research: given specific criteria about what students should understand about general relativity, do specific elastic sheet demonstrations and activities contribute significantly to understanding, given a specific target groups, as judged by those criteria? I will cite some systematic studies concerning these questions; in addition, articles by education practitioners on the elastic sheet frequently contain judgements based on the authors' practical experience, without a systematic quantification.

More in line with the focus of this text, I will go into more detail concerning another set of questions: which aspects of general relativity are captured well by the elastic sheet model, and which aspects of the model are misleading as they diverge from the underlying physics. The practical relevance of this second set of questions for teaching will depend on the context. If the goal is to introduce students to a limited set of concepts related to general relativity, misleading features that would only become important once students tackle advanced concepts will not matter as much as when the models are merely the beginning of a more thorough course of teaching.

\subsection{What the elastic sheet model delivers}

The map from general relativity to the elastic sheet model is straightforward. The elastic sheet is meant to represent spacetime. The spheres represent masses in spacetime, with the smaller spheres corresponding to smaller masses, possibly even test particles. It is straightforward to see which relationships are preserved. Masses distort\footnote{There is an issue of wording here that different authors handle differently. In the formalism of general relativity, curvature has a specific meaning: spacetime is curved unless the Riemann tensor vanishes everywhere. But even spacetime that is flat, in other words: not curved, can be distorted, and will in general be distorted unless very special coordinates are chosen, namely the usual coordinates of special relativity. In a general spacetime, distortions can be made to vanish at least locally by choosing a reference frame that is in free fall. The only distortions that remain are those directly due to curvature; physically speaking, those correspond to tidal forces, that is, to gravitational influence varying with location or over time. Gravity as we know it and need it to describe, say, planetary orbits is made up of both kinds of effect: tidal effects as well as those effects that can be made to vanish in a free-falling reference frame. I use ``distorted'' for this combination of effects that can be locally transformed away and tidal effects both; other authors use ``curved'' in the same sense.} the surrounding spacetime; the spheres distort the elastic sheet. Spacetime distortion decreases as you move away from a mass; so does sheet distortion as you move away from a heavy sphere. 
 
Spacetime distortion, in turn, determines how objects move. This carries over to the elastic sheet model, where movement depends on sheet distortion. The model goes as far as to reproduce the straight motion of test particles in the absence of a gravitating mass, and, at least qualitatively, the fact that in the presence of a mass, orbits are curved around the mass, leading to smaller masses circling the central mass in the manner of planets orbiting the Sun. In this way, the model reproduces John Wheeler's famous two-sentence version of general relativity: spacetime tells matter how to move; matter tells spacetime how to curve \citep{Wheeler1990}.

As far as orbits around the central heavy mass are concerned, angular momentum is approximately conserved, and particles in somewhat elliptical orbits reproduce a qualitative version of Kepler's second law, moving faster when they are closer to the mass, and slower when they are further away.

The consequences of placing a heavy sphere on a stretched-out piece of fabric, and predictions of what happens when smaller spheres are made to roll on the surface of the distorted sheet, are well within what students can predict beforehand, given their pre-existing knowledge about basic everyday physics \citep[sec. III.B.]{Farr2012}.

More generally, the elastic sheet shows that gravity doesn't just stop at a certain distance from the gravitating mass. Sheet distortion will become less and less as we move away from the central sphere, but there is no definite cut-off point. The sheet is a continuum, smooth, with no edges or steps or other discontinuities. As such, this model is a welcome antidote to people whose mental picture of gravity includes a boundary for, say, the Earth's gravity (marshalled e.g. as an erroneous explanation for why astronauts on the International Space Station ISS, or the Apollo astronauts in propulsion-free phases of their journey, are weightless), including the misconceptions that gravity cannot exist in empty space, or on the moon, as well as contrary misconceptions such as that all gravity comes from the Sun \citep{BarZinnRubin1997,Watts1982,Borun1993, CominsErrors}.\footnote{In cases where the elastic sheet is not set up as a physical model, but visualised e.g. in an animation, this advantageous aspect can be lost. In visualisations not based on simulations, but on graphical manipulation of, say, spline curves, graphics designers might find it easier to draw a dip of the sheet in the vicinity of the mass, but not to continue the dip to the sheet boundary. This is particularly tempting when the combined effects of two neighbouring masses are to be drawn; simulating this properly requires some calculation. An example for finite-sized dips can be found in this video commissioned by the Max Planck Society on the occasion of the first direct detection of gravitational waves: 
\href{https://youtu.be/mtCAmb_Mg1k?t=1m3s}{https://youtu.be/mtCAmb\_Mg1k?t=1m3s}
} 

The elastic sheet also demonstrates that gravity is mediated by changes in the gravitating object's environment \citep{Baldy2007}, thus avoiding the idea of an action-at-a-distance, a concept that has proved difficult to grasp for middle-school students \citep{BarZinnRubin1997}. This provides for a simple way of introducing key concepts that apply both to classical field theory and to general relativity.

The elastic sheet can prepare an audience for the idea of a finite speed of propagation for gravity, although whether or not, say, suddenly removing the central heavy sphere, or, more realistically, rapping on the sheet with a finger, or having binary spheres orbit each other will indeed result in a somewhat retarded change in distortion is likely to depend on the specific setup used. These are the model's simplified versions of gravitational waves in Einstein's theory \citep{SmarrPress1978,Begelman2010}. Orbiting binary spheres in the center of the sheet, and asking students to watch out for changes in distortion near the edge of the sheet, is one possible activity preparing students for the concept of gravitational waves \citep{Farr2012}. We will revisit this connection in section \ref{GWProperties} about gravitational waves, in particular with Fig.~\ref{MouldFigure} and the associated video. 

Given these properties, an elastic sheet approach, in this particular case realised by a pillow with compact spheres placed upon it, proved superior to a particular variant of teaching the Newtonian gravitational force law in French 9th-graders. Specifically, regarding diagnostic questions designed to determine students' adherence to models of attraction of different degrees of realism --- universal attraction, attraction only in the vicinity of celestial bodies, gravity an exclusive property of Earth, and others --- the ``Einstein before Newton'' approach led to better outcomes \citep{Baldy2007}.  A similar approach was successfully evaluated for teaching general undergraduate students, albeit without a control group \citep{Watkins2014}. In both cases, the teaching approach did not address issues beyond Newtonian gravity --- the elastic sheet analogy was not used to teach about the differences between Einstein's and Newton's theoretical frameworks.

\subsection{Problems with the elastic sheet}
\label{ElasticSheetProblems}
\begin{figure}[htbp]
\begin{center}
\includegraphics[width=0.4\textwidth]{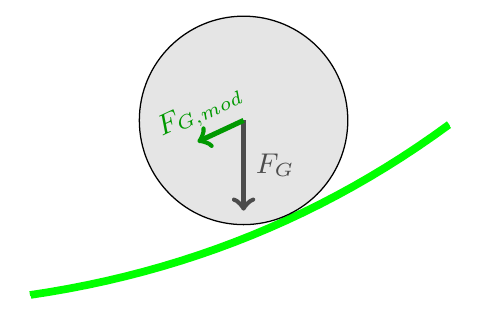}
\caption{Sphere on an elastic sheet: cross section seen from the side. The heavy sphere distorting the sheet is off to the left.}
\label{DoubleGravity}
\end{center}
\end{figure}

What, then, about the elastic sheet's critics? They have a number of valid points, as well. Probably the most common objection to the model is the confusing double role played by gravity \citep{GreeneElegant,Price2016,Janis2018}. The heavy sphere distorts the elastic sheet because of its weight, in other words: because of the Earth's gravity. We are presupposing the existence of gravity in order to explain gravity. In the model, there are two kinds of gravity, as pictured in figure \ref{DoubleGravity}.
The figure shows a sideways sectional view of one of the smaller spheres -- the central, heavy sphere is somewhere to the left -- and the surrounding region of elastic sheet, which in this sectional view appears as a green line. Also shown are the directions (if not the magnitudes) of the two kinds of gravity that play a role in our model: real, external gravity denoted as  $F_G$, and pointing straight down, and model gravity, $F_{G,mod}$, the result of real gravity plus the constraint caused by the sheet, which points towards the central, heavy, sphere, as the model source of gravity. 

\begin{figure}[htbp]
\begin{center}
\includegraphics[width=0.7\textwidth]{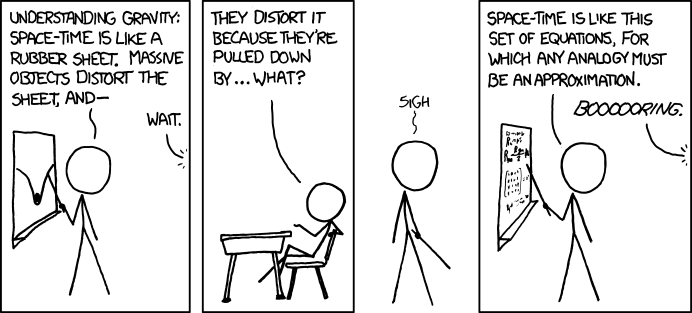}
\caption{Comic strip depiction of the problematic double role of gravity. Image: ``Teaching Physics'' by Randall Munroe at \href{https://xkcd.com/895/}{https://xkcd.com/895/}, licensed under \href{https://creativecommons.org/licenses/by-nc/2.5/legalcode}{CC BY-NC 2.5} }
\label{XKCD}
\end{center}
\end{figure}

Again, a large portion of the audience might not even notice this double role, and focus exclusively on the behaviour of the spheres. But the double role is likely to present at least a stumbling block to those who are trying to think the model through. As shown in figure \ref{XKCD}, the double role has even made it into a subsector of popular culture.

The second objection is that the model, by its nature, emphasises the role of distorted space over the role of time, and students might be confused whether it is space or spacetime that is curved to produce gravity \citep{Janis2018}. After all, a two-dimensional sheet is a natural stand-in for space. Its geometry is Euclidean, with none of the unfamiliar properties of spacetime. In some presentations of the model, this weakness has carried over to the description, and we are told that, in Einstein's theory, ``gravity is only a pucker in the fabric of space'' \citep{CosmosEp9}, implying that time is not affected by the influence of gravitating bodies.

Unless you are talking to an audience of mathematicians, whose first association upon hearing the word ``space''  is an abstract, generalised space, this is misleading. In the usual Newtonian limit of general relativity, after all (slowly moving particles, weak gravitational field, coordinates adapted to those of classical mechanics), the Newtonian part of gravity is purely an effect of time distortion, encoded in the metric coefficient $g_{tt}$. In other words: in the coordinate description arguably closest to classical physics, the familiar part of gravity, which makes planets orbit the Sun, and objects fall to the ground here on Earth, is not a distortion of space, but of time. This time distortion, and thus the effect that amounts to Newtonian gravity producing the familiar orbits, is not shown directly in the model \citep{Bennett2014,Gould2016,Kaur2017a,Stannard2017}.

The model shares one fundamental drawback with other material models of spacetime, including the expanding substrate models of section \ref{ExpandingSubstrate}: one crucial ingredient of special relativity, which carries over to general relativity, is that there is no absolute standard relative to which motion is defined. Any material medium, such as the elastic sheet, introduces a privileged standard of rest.

There are other objections, although not as weighty (no pun intended). The heavy spheres are meant to be located in space, and space is represented by the two-dimensional elastic sheet. If you were to adhere strictly to this representation of space, then whatever matter objects are in that model spacetime should be represented as two-dimensional, as well -- two-dimensional objects that are fully contained in the two-dimensional representation of space. But in the model, the spheres are on top of the two-dimensional sheet, and they are themselves three-dimensional. (Compare and contrast with the two-dimensional balloon model of cosmic expansion, where the galaxies are themselves approximately two-dimensional -- either stickers or markings on the balloon surface.) 

The combination of three-dimensional objects (spheres) together with the special plane in space distinguished by the orientation of the elastic sheet also serves to obscure the underlying symmetry of the situation --- after all, the gravitational action of a spherical mass is spherically symmetric \citep{Gould2016}.
 
Generally, the use of an embedding space is an extra complication likely to require an explanation \citep{Bennett2014}. After all, one of the spatial dimensions in the elastic sheet model does not correspond to any physical space dimension, but is merely an auxiliary dimension, which allows for a geometrically faithful embedding of the two-dimensional curved surface into three-dimensional space.

The representation of four-dimensional space-time by a two-dimensional surface requires a certain capacity for abstraction. But even after this mental leap is taken, there is a key difference between the physics of general relativity and that of the elastic sheet. In general relativity, what counts is the {\em intrinsic curvature} of space-time. In fact, realising the difference between intrinsic curvature as a property of a curved two-dimensional surface (in three-dimensional space), and extrinsic curvature as a property of the embedding of that two-dimensional surface in space, was the key idea that allowed Gauss to develop differential geometry. But in the elastic sheet model, the embedding, and thus extrinsic curvature, is an integral part of the model, as is shown in figure \ref{DoubleGravity}. Movement is not determined solely by the geometry of the deformed sheet; instead, the angle between the surface normal and the vertical direction defined by Earth's gravity determines the acceleration. Gravity, in this model, is not just geometry. Contrary to what happens in general relativity, gravity in the elastic sheet model arises from the interplay of geometry and an external influence, which can be expressed as a field defined on the elastic sheet. 

Last and probably least, there is unrealistic friction in the motion of spheres on the spandex sheet, unlike the motion of objects in empty space. And even without friction, motion on a distorted elastic sheet differs significantly from motion in a gravitational field --- and not only because the spheres are rolling, but in more important respects, notably concerning the shape of the orbit and the orbital speed at various points of the trajectory \citep{Haehnel2007,Middleton2014,Middleton2016}.\footnote{If you think that complete faithfulness in reproducing orbits is too much too ask of a model: the river model in section \ref{RiverModelBlackHolesSection} can do exactly that.}

There are two additional connections; whether those are seen as sources of confusion, and thus as disadvantages, or as helpful associations, will depend on the teaching context. The first is with the classic concept of the {\em gravity well}, that is, of a three-dimensional representation of the gravitational potential around a spherical object. That plot, with the x and y directions representing space and the z direction showing the numerical value of the gravitational potential, looks similar to the elastic sheet deformations around a massive object. The interpretation of the potential is, of course, much more specific. Both share the property of gravity pulling an object in what in the visualisation is the downward direction. The second is with (half of) the Flamm paraboloid: the two-dimensional surface of revolution embedded into three-dimensional space in just the right way to faithfully reproduce spatial geometry in the equatorial plane of a Schwarzschild black hole \citep[section 11.3]{Rindler2001}. In that case, the interpretations are even less related. The spatial geometry shown does contribute a little bit to the black hole's gravitational attraction, but, as stated above, the main contribution in this case that includes the well-known Newtonian effects amounts to time distortion.

While the arguments given so far are physics-based, some of the misunderstandings have been reproduced ``in the field,'' in tests on upper secondary school students who had been taught about gravity using the elastic sheet model \citep{Kersting2018a,Kersting2018b}.

\subsection{Non-elastic two-dimensional geometry}
\label{Inelastic2D}

At least half of Wheelers dictum, namely that curved spacetime tells matter how to move, can be realised not with an elastic surface, but a fixed surface. In an early example, Bertrand Russell likened the influence of geometry on motion to people holding lanterns, moving in a landscape that consists of a plane with a single hill, on which there is a bright beacon. As the people move from village to village, they avoid going up the hill, which results in curved paths. An observer in a balloon, at night, can deduce from the movement of the lanterns that there is something surrounding the beacon that changes the motion of the people \citep{Russell1925}. Another image is that of a golf course, where curvature influences the trajectory of a rolling golf ball (F. Lindemann in the introduction to \citealt{Schlick1920}; \citealt{Luminet1987,Zeilik1991,Egdall2014}).

In some illustrations, the curvature of a two-dimensional surface is merely part of an illustration showing celestial bodies and deflected light rays passing from one to the other \citep{Gamow1947,Braginski1989}.

Distorted two-dimensional geometries can, more generally, be used to demonstrate the concepts of curvature and of geodesic motion. A common example are geodesics on the Earth's surface, which have a practical application in determining the flight routes between distant cities \citep{Gould2016,Stannard2017}. Steven Weinberg has posed for readers of his 1972 book on gravitation and cosmology the problem of determining, from the given mutual distances between four locations, whether or not J.~R.~R. Tolkien's Middle Earth is flat \citep[Fig.\ 1.1]{Weinberg1972}. Alternatively, Eddington tells the story of charting the surface of the Earth on a flat map; in order to account for distant travellers' different statements of distances (caused by the Earth's curvature), such flat-mappers must introduce a demon that influences travellers' propagation in those distant regions. In this simile, the demon is Newton's gravitational force, while Einstein's geometric theory amounts to realizing that the Earth is, really, a sphere \citep{Eddington1922}.

An unusual model is the one that posits a physical mechanism for the distorted geometry: The mass responsible for the non-Euclidean geometry is represented by a very hot body; geometric measurements in the surrounding space are performed using metal measuring rods, which become longer when they are closer to the central body, and heat up, and which shrink at greater distance from the central body \citep{Andrade1921}.

\begin{figure}[htbp]
\begin{center}
\includegraphics[width=0.48\textwidth]{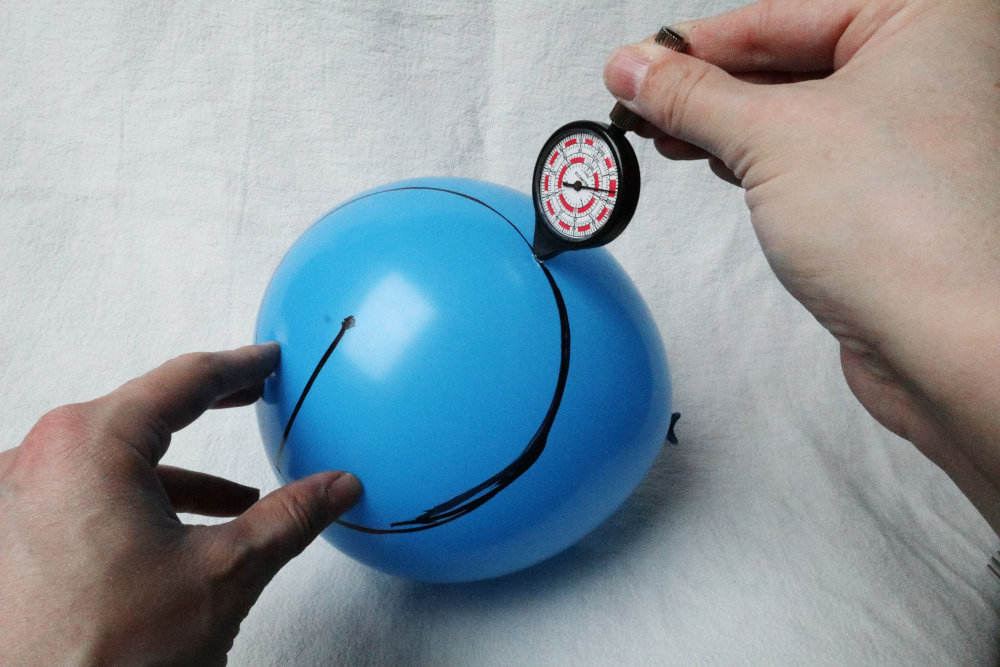} \hspace*{0.5em}
\includegraphics[width=0.48\textwidth]{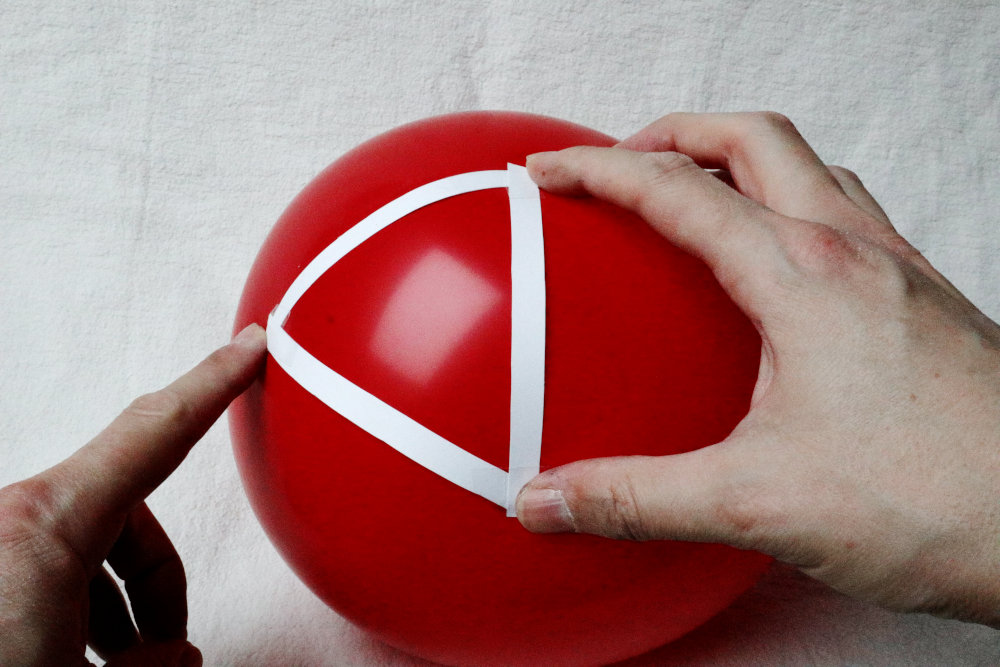}
\caption{Tracing the curved geometry of balloons: Measuring radius and circumference of a drawn-on circle with a map-measuring device (left), and constructing a paper-strip triangle on a balloon (right) whose three angles are then measured, separately, with a protractor, by flattening each of the three triangle edges onto a plane. The activities themselves can be seen in the YouTube video \href{https://youtu.be/-czyukZm94I}{https://youtu.be/-czyukZm94I}}
\label{BalloonGeometry}
\end{center}
\end{figure}
For hands-on activities, objects with curved surfaces and more accessible scales are suitable.  Some balloon-based measurements are illustrated in Fig.\ \ref{BalloonGeometry}. An example are circles drawn on the surface of a squash, their radii measured using a paper strip and their circumferences with a map-measuring device (opisometer), and the surface's curvature derived via the ratio of circumference to radius \citep{Effing1977}. Alternatively, paper-strip triangles taped on spherical or saddle-like surfaces \citep{Wood2016}, or drawn onto rubber strips or the surface of a balloon \citep{Sellentin2013}, allow for the exploration of surface curvature via the deviation of the sums of the triangles' angles from 180 degrees.

A particularly ingenious way of demonstrating the concepts of parallel transport and of curvature employs a model of the ``Chinese South-Seeking Chariot,'' a chariot whose wheels are connected to a configuration of gears in just the right way for a direction-pointer to keep pointing in one and the same direction, in effect parallel-transporting the direction vector in question \citep{Santander1992,Liebscher1999}.

\begin{figure}[htbp]
\begin{center}
\includegraphics[width=0.6\textwidth]{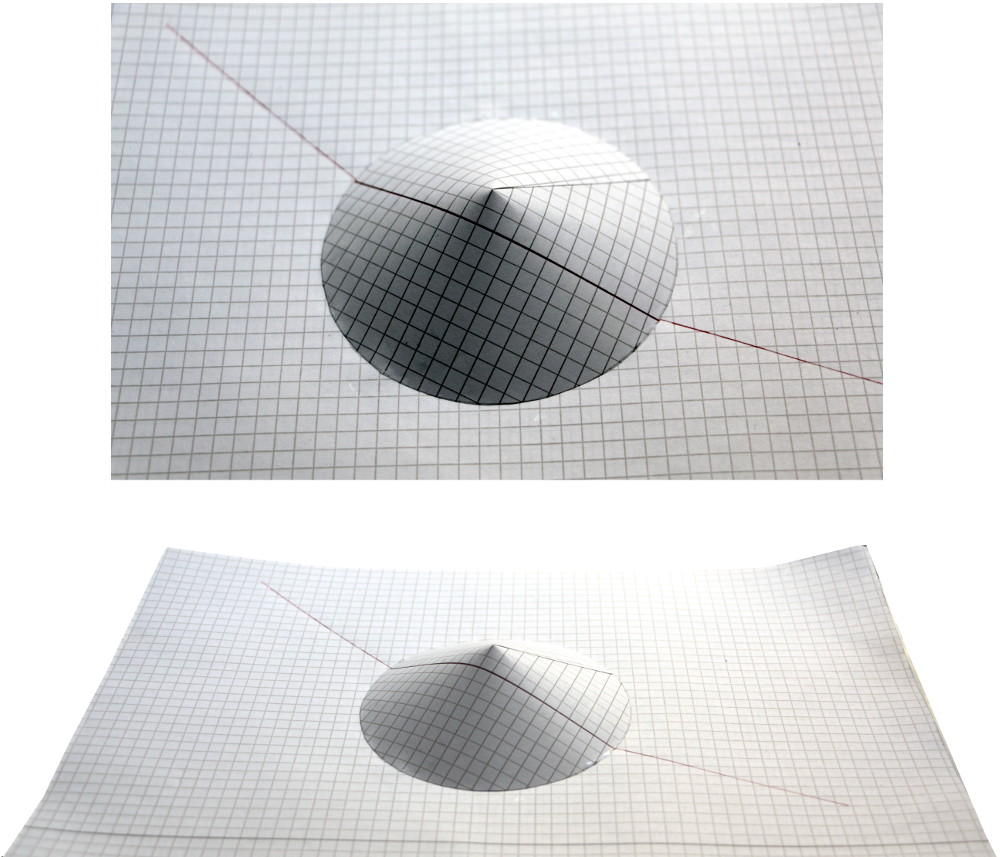}
\caption{Flat paper with attached paper cone. At the apex of the paper cone is the point mass, and the cone represents the distortion of space near the mass}
\label{PointMassCone}
\end{center}
\end{figure}
A simple version of a geometric model is the following \citep{Kornelius2005}: Imagine two-dimensional space as a flat sheet of paper, and represent the influence of a point mass on the curvature of space by a cone, which has the point mass at its apex, as in figure \ref{PointMassCone}. Locally, the surface of the cone is flat, so it is easy to draw straight lines. By intermittently flattening the local cone region to the paper, lines can be extended from the flat paper onto the cone, and vice versa. Light passing through the ``zone of influence'' of the mass is deflected. This model shares a number of properties with proper general relativity in 2+1 dimensions: there, too, point masses correspond to the apices of cones, and space is locally flat everywhere \citep{Deser1984,Carlip1998}.

A simple alternative setup is to use an ordinary curved plate, put upside down, as a stand-in for the cone, and a sticky tape to map out the path of the light, resulting in a deflection where the tape is taped across the curved underside of the plate \citep{TranRussell2018b}. Similar two- and three-dimensional models, but adapted specifically to space geometry around a black hole, will be described in section \ref{BHGeometry}.

\section{Black holes}
\label{BlackHoles}

\begin{figure}[htbp]
\begin{center}
\includegraphics[width=0.9\textwidth]{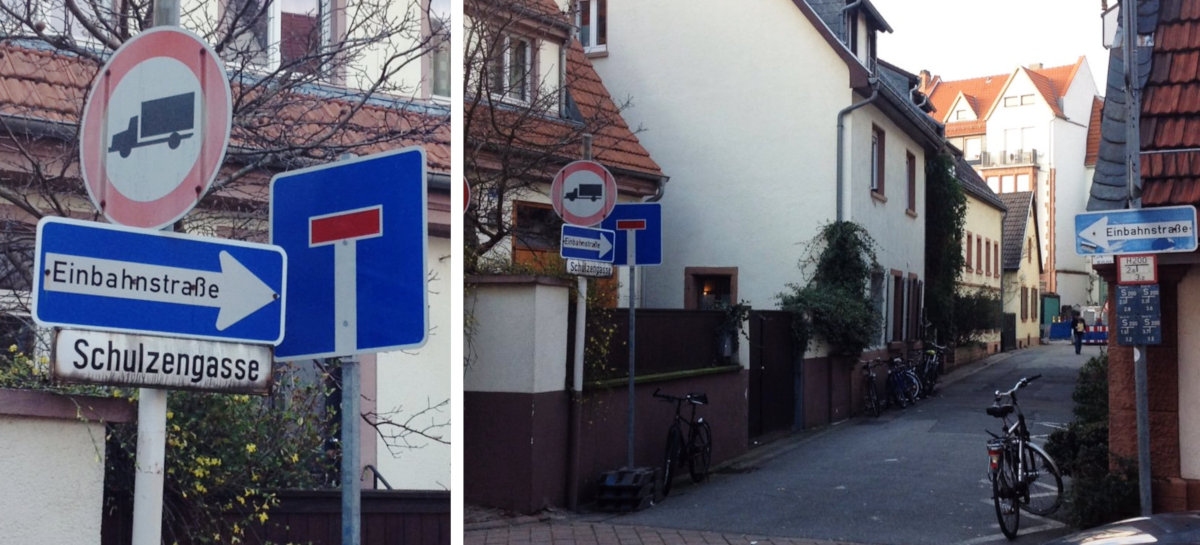}
\caption{Traffic black hole in Heidelberg, Germany, in January 2016: A street that is both a one-way street and a dead end. Left: the black-hole combination of traffic signs. To the right is the dead-end sign, while the middle sign to the left (with the white arrow and ``Einbahnstra\ss e,'' the German word for one-way street) is the one-way street sign. Right: street that, by the combination of signs, has become a traffic black hole. The one-way street sign is temporary, put there (somewhat thoughtlessly) because of a construction site sealing off the street some way back}
\label{TrafficBH}
\end{center}
\end{figure}

Perhaps the simplest model for a black hole is based on a combination of traffic signs: a black hole is a one-way street that is also a dead end. This combination of street signs is rather uncommon in real life, for obvious reasons (figure \ref{TrafficBH} shows a rare sighting of such a combination of German traffic signs, effectively a ``car trap,'' in the wild). But it does encapsulate the fundamental properties of a black hole rather concisely: a black hole is a region of space set apart from the rest. Its boundary, the event horizon, is the entry to a cosmic one-way street: matter and light can cross the boundary surface in one direction, namely from the outside to the inside, but not in the opposite direction --- nothing from the inside can get out. And as far as we know, black holes have no hidden exits inside. They are dead ends; what falls inside, stays inside.

\subsection{Newtonian calculations}

For a non-rotating, spherically symmetric black hole, the event horizon is a spherical surface whose area corresponds to the surface area of a sphere with radius
\be
{\cal R}\equiv \frac{2GM}{c^2},
\ee
known as the Schwarzschild radius. The value of this critical radius can be derived in the framework of a simple classical model that employs the Newtonian concept of escape velocity \citep{Hall1985,Froehlich1987,Braginski1989,Lotze2000b,Ellwanger2008,Kraus2009}. In fact, for a massive sphere with mass $M$ and with the radius ${\cal R}$ given by that formula, the Newtonian escape speed is equal to the speed of light (which can most quickly be calculated from energy conservation). 

The interpretation that this Newtonian model allows for a {\em derivation} of the Schwarzschild model is problematic, though. The light particles in this simple model are slowed down, and do not keep moving with the speed $c$, which is in contravention of the constancy of the speed of light. The escape speed scenario does not mean that no light can cross the horizon sphere of radius ${\cal R}$; it merely means that light cannot escape to infinity from inside that radius. This is markedly different from the role the horizon plays for general relativity's black holes. Given these differences, the fact that the Newtonian calculation gives the exact value of the Schwarzschild radius must be considered a coincidence. While the combination of quantities that make up the Schwarzschild radius is fixed on purely dimensional grounds, there is no reason to expect a simplified derivation to yield the correct numerical factor, namely 2. Nonetheless, the tools of Newtonian mechanics prove useful when, say, calculating orbits around the Schwarzschild black hole, which can be achieved using a modified effective potential within the framework of classical mechanics \citep{Lotze2000b}.

\subsection{Astrophysics}

The astrophysical phenomena associated with black holes, as well as the observational methods employed, make for a compelling narrative of research and discovery \citep{GreensteinFrozen,Thorne1994,Mueller2009,Bartusiak2015}. Given the important role played by black holes in modern astrophysics --- notably as an end state for massive stars, and as the engine behind the considerable luminosity of quasars --- there are numerous teaching models for astronomical phenomena that are directly or indirectly connected to Einstein's theory. 

\begin{figure}[htbp]
\begin{center}
\includegraphics[width=0.465\textwidth]{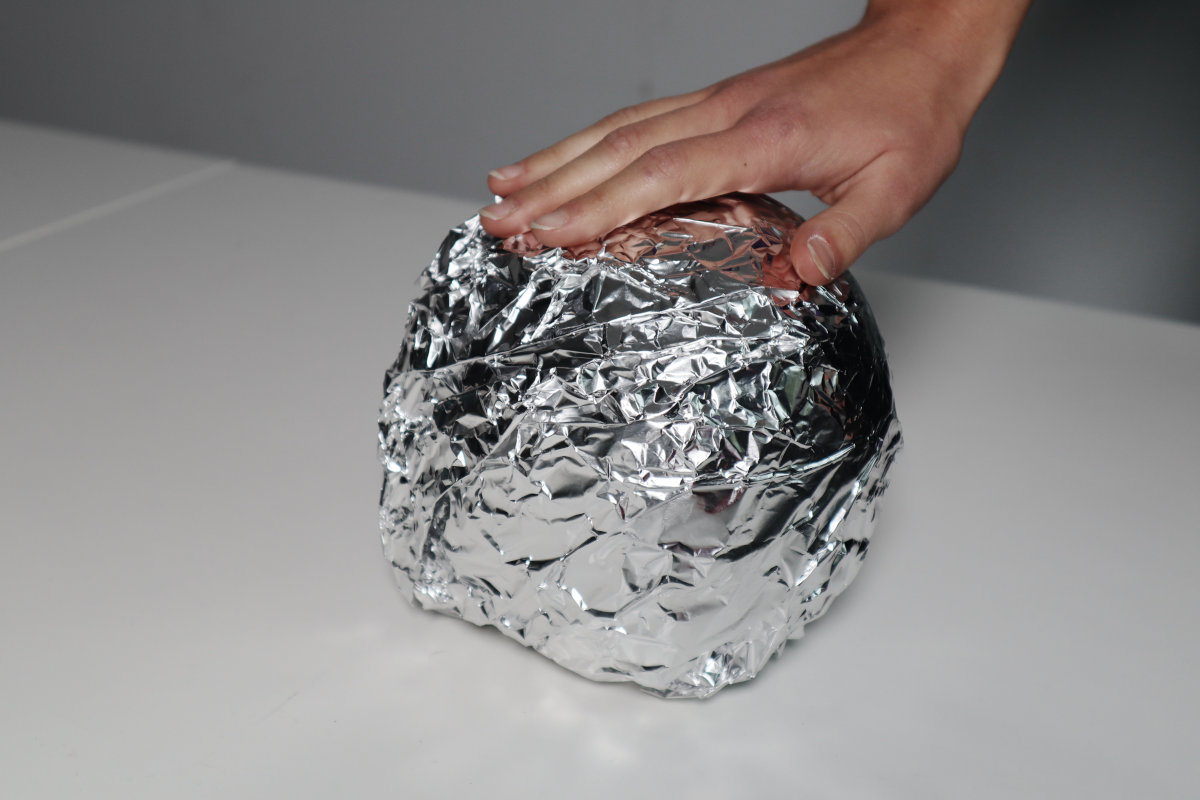} \hspace*{1.5em}\includegraphics[width=0.465\textwidth]{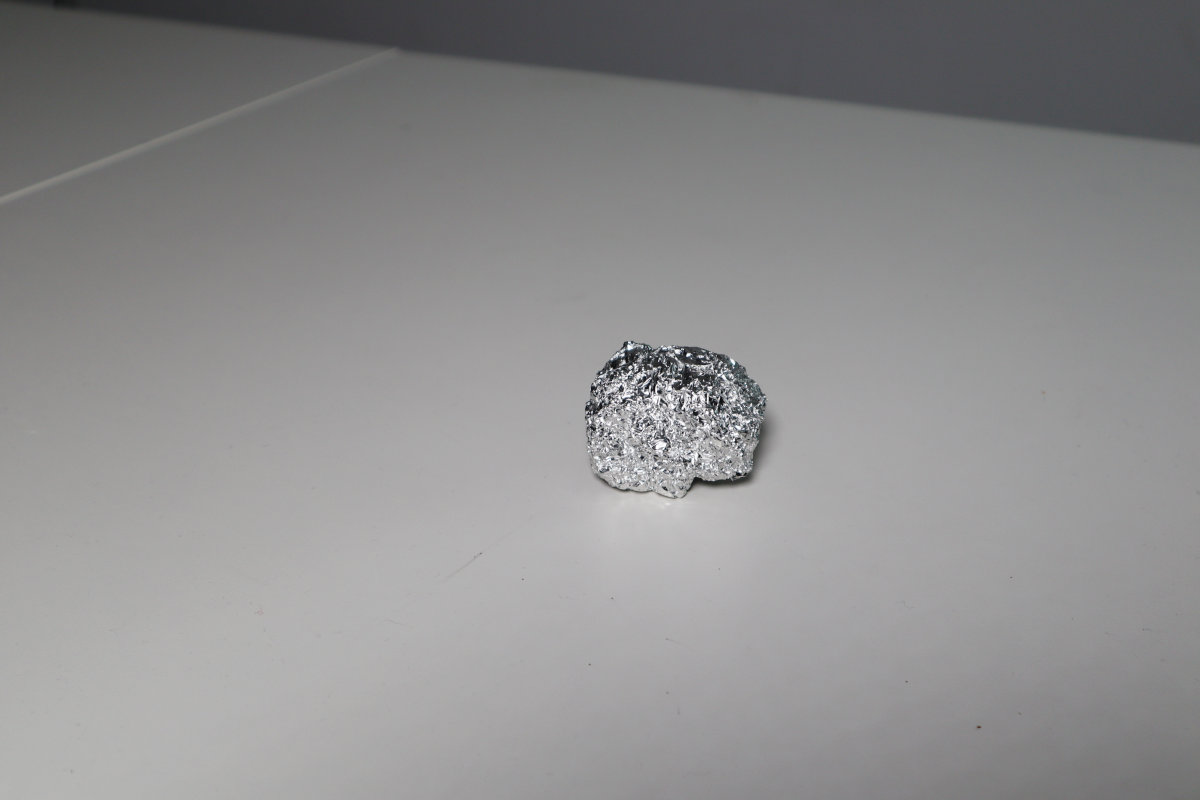}
\caption{Stellar collapse model: an aluminium foil star supported by an inflated rubber balloon core (left) collapses to a compact crumpled aluminium foil remnant (right). Own image after the descriptions in \citet{Whitlock2001} and \citet{Turner2013b} }
\label{StellarCollapse}
\end{center}
\end{figure}

For younger children in particular, such models can focus on very basic processes and changes. An example is a model of stellar collapse, figure \ref{StellarCollapse}, which begins with an ordinary star. 
The star is represented by an inflated rubber balloon (the pressure-producing stellar core in which fusion takes place), covered in several layers of aluminium foil (the star's outer atmospheric layer). Participants can squeeze the model star, their hands standing in for the gravitational force, feeling for themselves the counteracting inner pressure. As the star's hydrogen fuel is exhausted, the counter-pressure cannot be kept up; participants simulate this by poking a hole into the rubber balloon with a pin or other sharp object. The aluminium foil (with loose rubber skin inside) is then crumpled to simulate stellar collapse and the compact stellar remnant \citep{Whitlock2001,Turner2013b}.

Stellar collapse can also be demonstrated using a ``dynamic human model,'' a kind of roleplaying: In the initial scene, an inner circle of children pushing outwards represents the star's stabilising pressure, while an outer circle of children plays the role of gravity, pushing inwards. On command, the inner pressure ceases, and the star collapses, with children scattering into space as supernova ejecta and one remaining child, suitably labelled and spinning quickly, representing the stellar remnant, the neutron star or black hole \citep{Bishop1990}. While such models reinforce the popular perception of black holes as very dense, it is worth noting that black hole density is mass dependent. A simple estimate for this computes a mean density for a black hole, or an object about to become a black hole, using the usual Euclidean geometry. Such an estimate shows that, while solar-mass black holes are indeed exceedingly dense, supermassive black holes with millions or billions of solar masses can have average densities comparable to the density of water, or of air \citep{Hall1985,LoPresto2001}.

\begin{figure}[htbp]
\begin{center}
\includegraphics[width=0.9\textwidth]{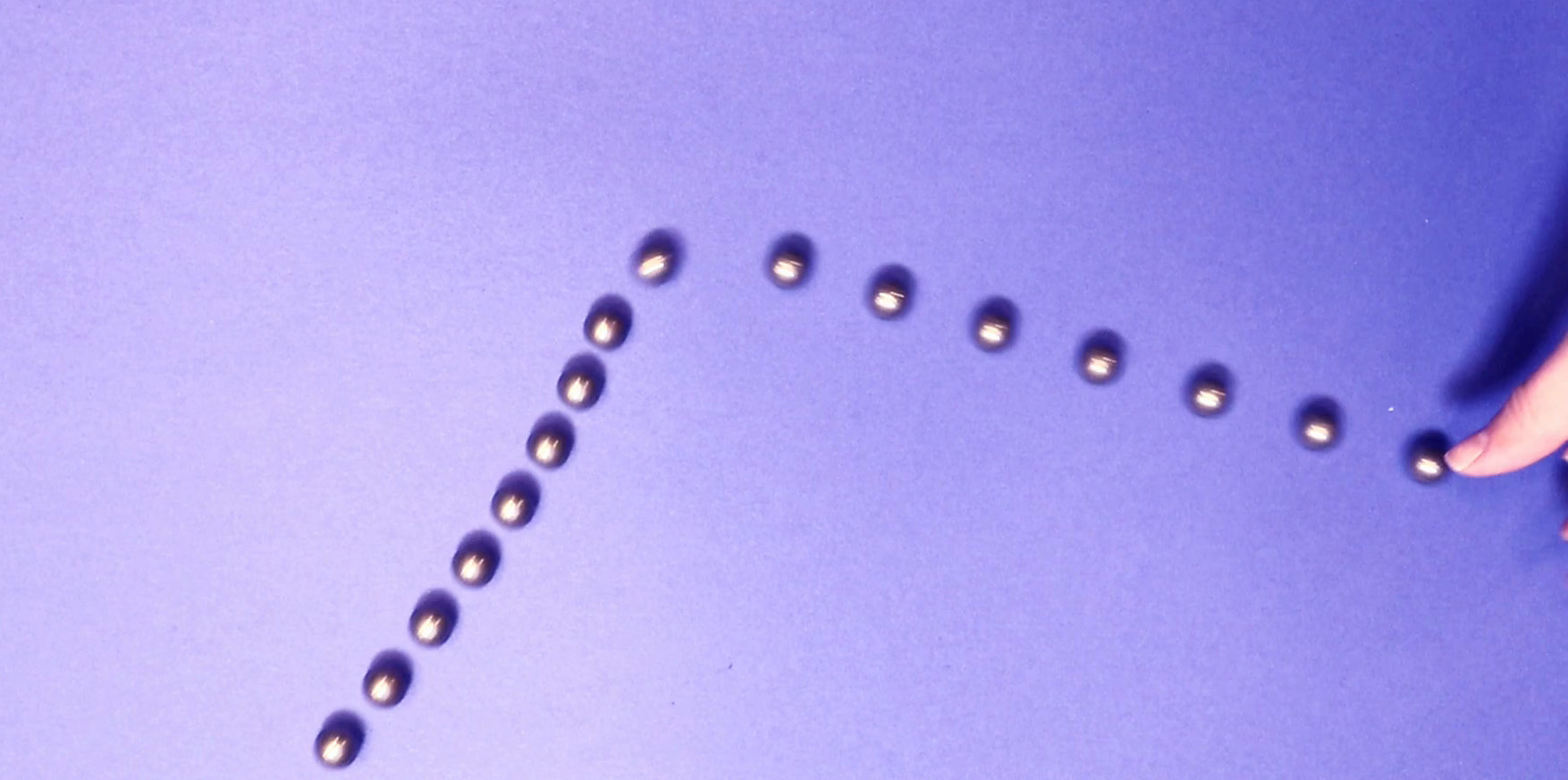}
\caption{Trajectory of a steel ball rolling on cardboard, with a magnet hidden underneath. The image shows a montage of positions for the steel ball, separated by equal time intervals. The set-up can be seen in action in the YouTube video \href{https://youtu.be/IO4QgudwW4s}{https://youtu.be/IO4QgudwW4s} }
\label{magnet-bh}
\end{center}
\end{figure}

Black holes can by definition not be seen directly, but they can be detected by their gravitational effects on nearby objects. A simple teaching model for this involves placing magnets underneath a cardboard, signboard, or thin styrofoam sheet, each magnet representing a black hole. While these ``black holes'' are invisible, their presence is revealed when magnetic marbles, or loose ball bearings, are rolled across the cardboard. Some of the marbles get deflected, and some might even go into orbit around one of the black hole magnets \citep{ASP2005}. An example for this kind of set-up is shown in Fig.\ \ref{magnet-bh}.

While the simple model is a vivid demonstration of how we can use the influence of a hidden object to detect that object's presence, there are several ways in which the model diverges from reality. For one, the steel balls can slow down when passing the magnet; that is not an effect expected for ordinary gravitational interactions. The case when one of the balls ``gets stuck'' on the magnet can charitably be interpreted as the ball having plunged into the black hole, though!

\begin{figure}[htbp]
\begin{center}
\includegraphics[width=0.6\textwidth]{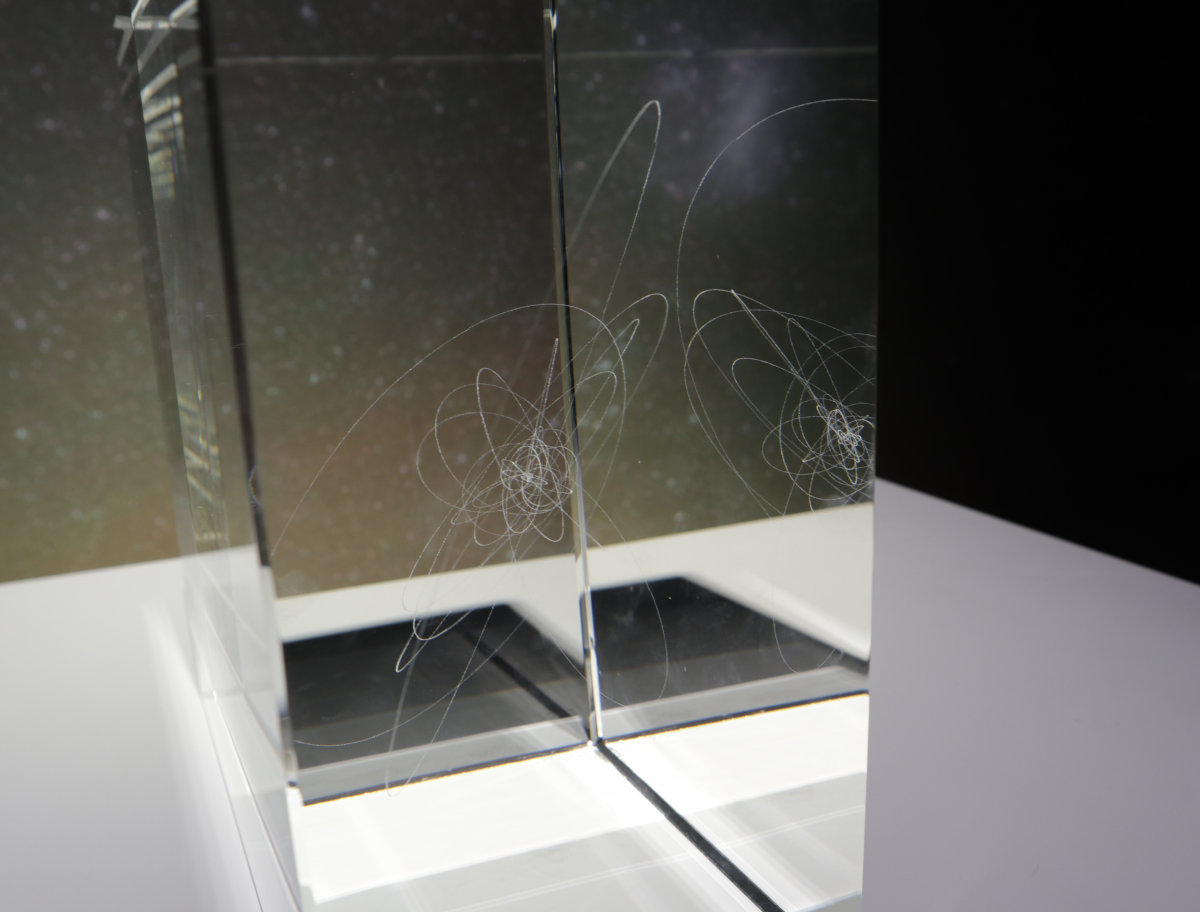}
\caption{Three-dimensional model of the orbits of the stars closest to our home galaxy's central black hole, created for the ESO supernova. Visible are both the etched-in orbits and, on the right, their internal reflection at the wall of the transparent block. The model can be seen from slightly different perspectives in the YouTube video \href{https://youtu.be/GVvcJAtCPF4?t=90}{https://youtu.be/GVvcJAtCPF4} from time mark 1:30 onwards}
\label{SgrAOrbits}
\end{center}
\end{figure}

The most important application for this kind of detection-by-effect is the supermassive black hole in the center of our home galaxy, the Milky Way, which has been studied by tracking the motion of nearby stars over the past decades, beginning with (independent) pioneering work by the groups of Reinhard Genzel \citep{Schoedel2002} and Andrea Ghez \citep{Ghez2003} in the early 2000s. A 3D model of the reconstructed orbits of the closest stars\footnote{An impressive animation based on real observations with the NACO instrument at the European Southern Observatory's Very Large Telescope in Chile over the past 20 years can be found on \href{https://www.eso.org/public/videos/eso1825e/}{https://www.eso.org/public/videos/eso1825e/}} can be seen in Fig.\ \ref{SgrAOrbits}. Given the orbital data of these stars, it is a straightforward exercise to use Kepler's third law to determine the central black hole mass \citep{Ruiz2008}; given an orbital plot, the same calculations are possible, but projection effects need to be taken into account \citep{Fischer2006}.

Interestingly, in the most recent and most precise observations to date, the central source is not invisible any more: Using the GRAVITY instrument which combines the light of all four of the 8-meter main telescopes of ESO's Very Large Telescope interferometrically, the distance between the optical counterpart of the central black hole and the nearby star S2 was tracked with unprecedented precision. The dim near-infrared glow at the position of the black hole, and the more easily detectable radio glow that has earned the object the name of ``Sagittarius A*'' as a radio source, are due to matter in the black hole's immediate vicinity. Tracking both the infrared glow indicating the position of the black hole and the star S2, the astronomers in Reinhard Genzel's group were even able to determine the relativistic redshift affecting S2, a combination of the gravitational redshift and the (transverse) relativistic Doppler shift \citep{GRAVITY2018}.

When matter is falling onto a supermassive black hole at a markedly higher rate than in our own galaxy, the result is an active galactic nucleus (AGN). The unified model of active galactic nuclei lends itself to modelling since the different faces an AGN presents to astronomical observers are a matter of perspective: Recreate the basic geometry of the different components of an AGN, of the central hole and accretion disk, the dust torus and the two jets, and you can recreate the perspectives that make an AGN appear as a blazar (looking directly into the jets), as a quasar or Seyfert galaxy of type I (with the broad-line region near the accretion disk visible) or type II (with the broad-line region obscured). The recreation can be in the form of a cardboard-and-styrofoam-ball model, a pop-up book, and even an edible model complete with bagel accretion disk and ice-cream cone jets \citep{Cominsky2013}. Accretion itself can readily described using simple Newtonian models \citep{Fischer2006,Vieser2015}.

Focussed on more general properties of black holes, and their detection, is the ``Black Hole Explorer Board Game,'' a board-game model for teaching. The board represents space; the closer the spaceship comes to the black hole, the more energy points need to be expended to regain distance from the black hole \citep{ASP2005}. A number of such models are described in the recent {\em astroEDU} collection of black-hole related activities for children 6 years and older \citep{TranEtAl2018}. 

\subsection{Geometry}
\label{BHGeometry}

There are different ways of modelling the distorted geometry around a black hole. Arguably the most common is based on the elastic sheet model analysed in section \ref{ElasticSheet}. Black holes are the closest general relativity can get to a point particle --- the simplest model of a gravitational source in Newtonian gravity. Indeed, when Karl Schwarzschild, in 1916, published the first exact solution for describing spacetime around what we today understand as a black hole, he presented it as a point particle solution for general relativity. The Birkhoff theorem tells us that, from the outside, the gravitational attraction of a spherically symmetric black hole is the same as that of any other spherically symmetric object of the same mass.

Given this context, it makes sense to model the neighbourhood of a black hole in the same way as that of any other spherical mass. Elastic sheet models have been used to illustrate the concept of a black hole \citep[Fig.\ 1.15]{Hawking2001}, sometimes as a bottomless well or pit \citep{CosmosEp9,Whitlock2001,Begelman2010,Mueller2013,Bahr2016,Bennett2017}, sometimes with the funnel either narrowed to a point, or a small sphere added to the tip, to denote the black hole's central singularity \citep[chapter 6]{Hawking1996}, or even with the great compactness of the object producing an actual hole at the bottom of the funnel \citep{Bennett2014}.  Corresponding elastic-sheet activities have also been employed in teaching about black holes \citep{Whitlock2001,Turner2013a,Turner2013b,TranEtAl2018,TranRussell2018a,TranRussell2018b}. 

\begin{figure}[htbp]
\begin{center}
\includegraphics[width=0.75\textwidth]{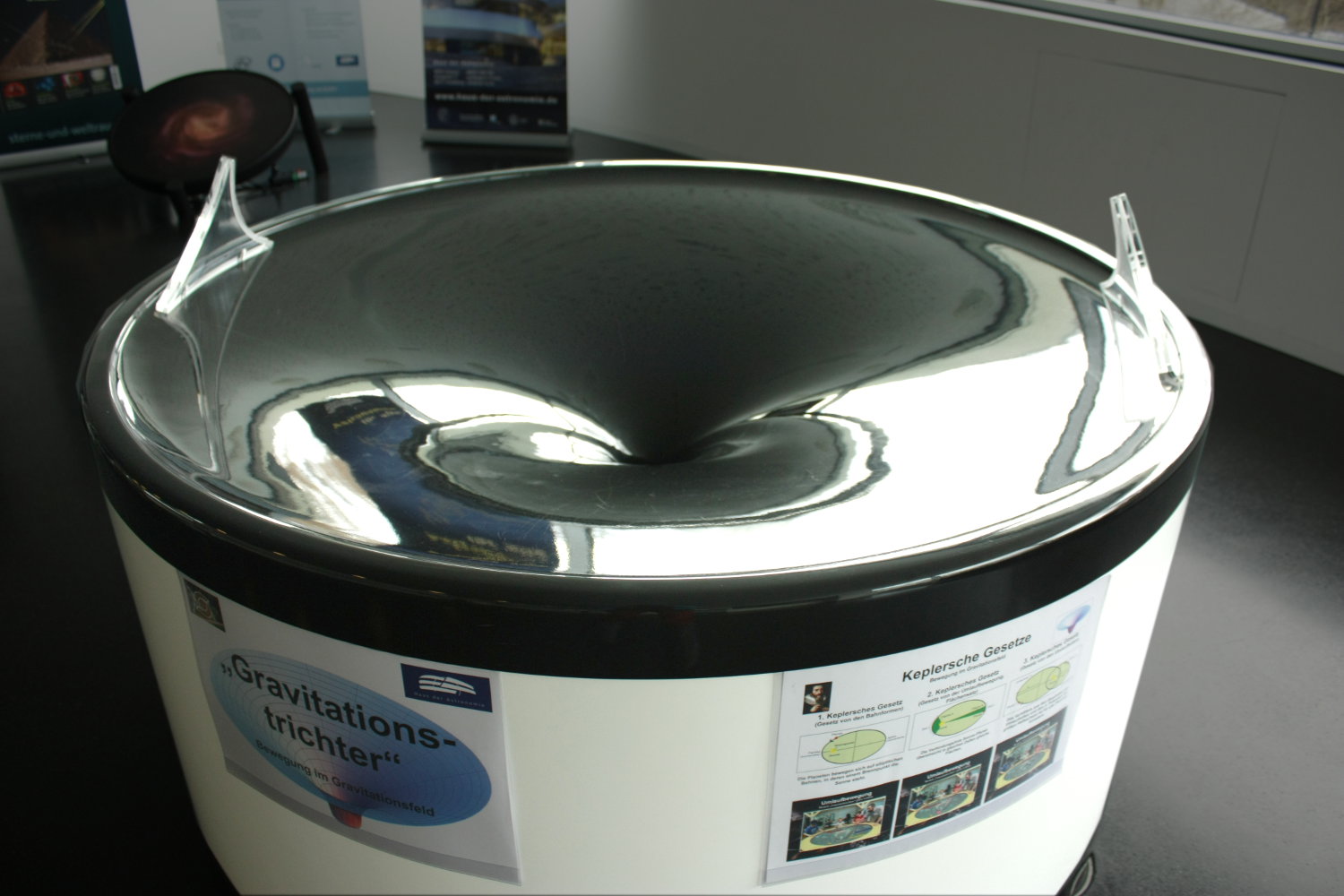}
\caption{Gravitational well for use with coins or small metal/wooden balls at Haus der Astronomie in Heidelberg. Image: C. Liefke (HdA)}
\label{Gravitationstrichter}
\end{center}
\end{figure}

A non-deformable version of the funnel, sometimes explicitly called a black hole, sometimes more soberly represented as a gravitational potential well, for use with either small balls or, more commonly, rolling coins, can be found in numerous planetariums and science centers. An example is shown in Fig.\ \ref{Gravitationstrichter}.

The objections against the elastic sheet model listed in section \ref{ElasticSheetProblems} apply to the elastic sheet as a model for a black hole as well. But due to the expression ``black hole'', there is an additional potential misunderstanding. In the elastic sheet model, there is a hole, after all: the hole that is part of the funnel. This hole should not be confused with the black hole itself. It is a part of the embedding space only, not part of the two-dimensional sheet representing the three-dimensional universe within this model.

\begin{figure}[htbp]
\begin{center}
\includegraphics[width=0.75\textwidth]{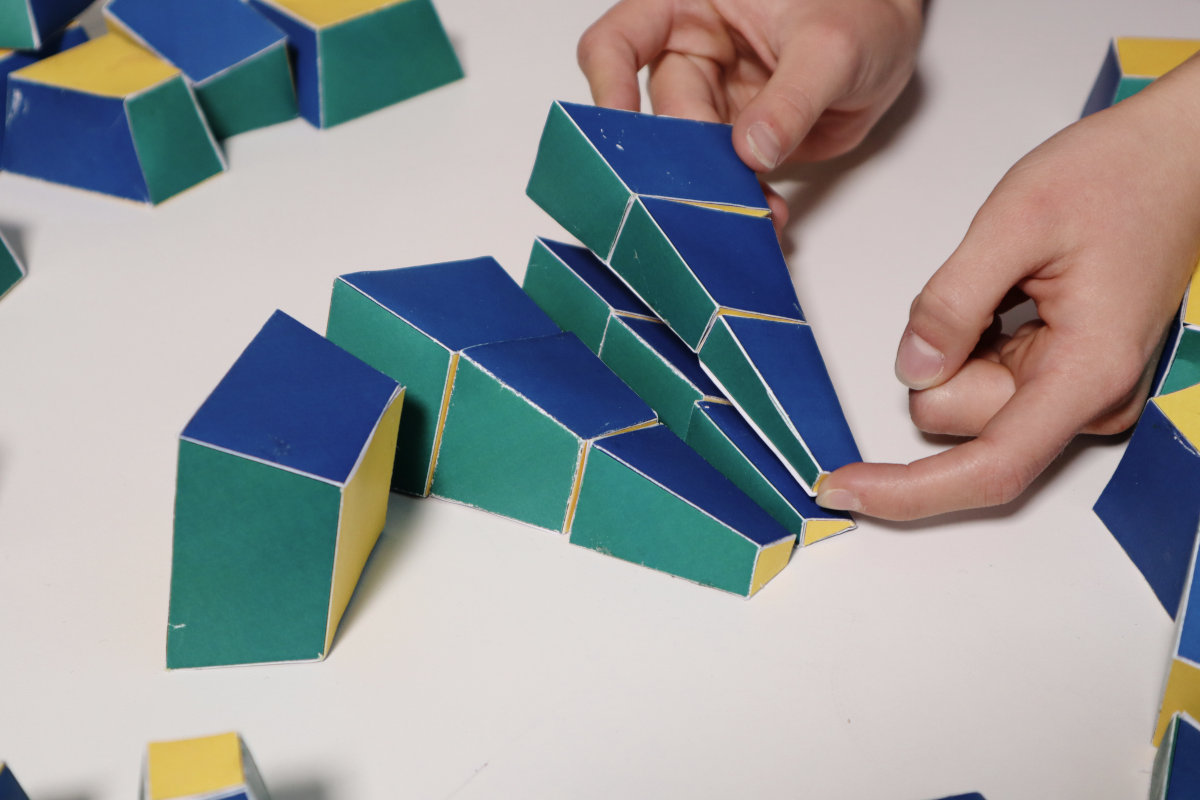}
\caption{Sector model of spatial geometry near a black hole, with building blocks constructed using the worksheet from \citet{KrausZahn2005}. In the space around a black hole with Schwarzschild radius 2.7 cm, the building blocks would fit together snugly. In flat space (which is all we have in most teaching situations), building blocks fit together locally, but cannot be made to fit globally
 }
\label{SectorModelFigure}
\end{center}
\end{figure}

In a different class are those elastic sheet models that dispense with the mechanism for gravity acting on passing bodies, and simply use the elastic sheet's curvature as an analogy for the distortion of space around the black hole, with a curved two-dimensional surface embedded in three-dimensional space standing in for curved three-dimensional space \citep{Thorne2003}.

The three-dimensional spatial geometry around a black hole or a more general mass in three dimensions can be modelled using simple polyhedra. Such models have been developed in detail over the past decade and more by Ute Kraus and Corvin Zahn, make use of {\em sector models}: decompositions of three-dimensional space into adjoining polyhedra \citep{KrausZahn2005,Zahn2010,Zahn2014,Kraus2016b,Kraus2018,Zahn2018}.\footnote{The idea on which these models are based was described earlier by \citet{diSessa1981} under the name of ``wedgie calculus,'' a play-on-words on Regge calculus as the mathematical technique of 
conceptually cutting up curved space into such sections.} In flat three-dimensional space, a space-filling set of polyhedra representing flat space can be packed together without any overlap and without any holes. For a space-filling set of polyhedra faithfully representing positively curved space, negatively curved space, or more complex curved spaces, there will always be overlaps, or holes, when trying to join the polyhedra together in flat three-dimensional space (which, at least approximately, is all we have in everyday teaching situations). Some of the polyhedra can be seen in figure \ref{SectorModelFigure}.

The properties of unavoidable holes and unavoidable overlaps contain information about how the curved geometry for which the polyhedra have been designed diverges from that of flat space, allowing for an instructive, hands-on exploration of the properties of such curved spaces. The concept can readily be extended to curved spacetimes. Both in space and in spacetime, the model allows for an easy (approximate) construction of geodesics. Locally, adjacent polyhedra can {\em always} be fit together snugly, reflecting the fact that locally, curved space is always flat. A local piece of a geodesic -- locally: a straight line on one of the polyhedra -- can always be continued in a straight way onto an adjacent polyhedron. In this way, working one's way from one adjacent polyhedron to the next, a geodesic can be continued through the whole of the represented region of space, similar to the geometric visualisation in section \ref{Inelastic2D}. In the spacetime case, that allows for a straightforward demonstration of light deflection by a compact object.

Tracing light-like trajectories in the vicinity of a black hole, and noting which of them can escape to infinity and which remain ``stuck'' in a finite region, is another way of gaining an intuitive understanding of how a black hole, in particular: its horizon, is defined \citep{Mielke1997,Kraus2005b}.

\begin{figure}[htbp]
\begin{center}
\includegraphics[width=0.6\textwidth]{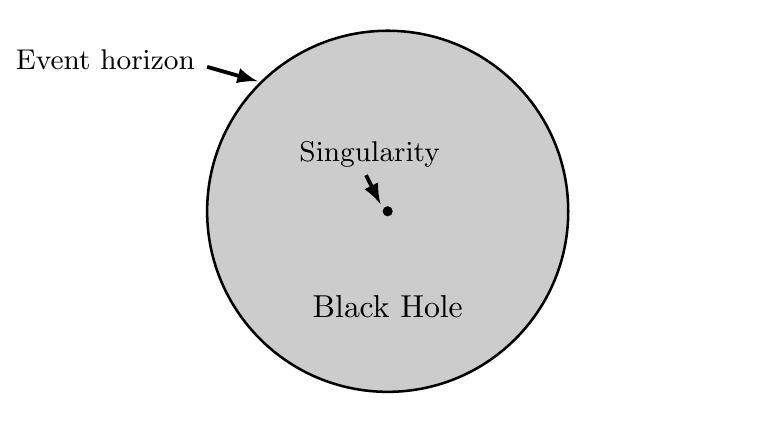}
\caption{Diagram showing a cross section of the ``surface'' (= horizon) and ``center'' (= singularity) of a black hole}
\label{BHStructure}
\end{center}
\end{figure}

When it comes to the interior of a black hole, not only are everyday notions of geometry and of spatial relations of little help, they can be downright misleading. Consider a very simple diagram purporting to show the structure of a black hole, as in figure \ref{BHStructure} (e.g. Fig.\ 24-3 in \citeauthor{Kaufmann1985} [\citeyear{Kaufmann1985}] or Fig.\ 5.6 in \citealt{Mueller2017}),
which is described as showing the cross section of a black hole. Our everyday intuition of reading this diagram as a snapshot of spatial relations is wrong. Inside the black hole's horizon, what used to be the radial spatial direction becomes a time direction, including such directions' compulsory one-way character for particle world lines. The singularity is a stretched-out ``boundary of time'' rather than a well-behaved center point of the surrounding space.\footnote{
I have tried to produce a simple spacetime visualisation of this set-up in \citet{Poessel2010}; more details, in particular beyond simple spherical symmetry, can be found in \citet{Droz1996}.
}

\subsection{River model}
\label{RiverModelBlackHolesSection}

\begin{figure}[htbp]
\begin{center}
\includegraphics[width=\textwidth]{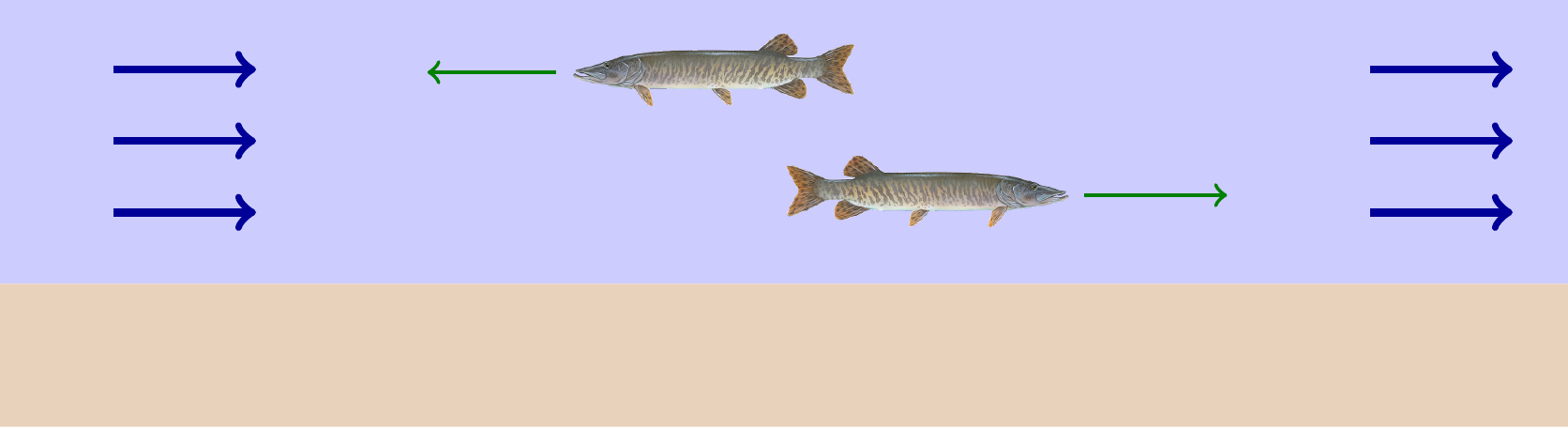}
\caption{River with fish}
\label{RiverModel}
\end{center}
\end{figure}

There is another perspective on black holes, with a model based on a completely different set of simplifications: the river model, described in \citet{Hamilton2008}. Imagine you are standing on the banks of an idealized river, which has the cross section, as viewed from the side, shown in figure \ref{RiverModel}. 
The blue (thick) arrows indicate the flow of the river, and there are also a few fish. The fish can go with the flow, or swim against the flow. Fish speed, denoted by the green (thin) arrows, is limited -- we assume that the fish can swim at any speed up to a well-defined ``maximal speed of fish.''

\begin{figure}[htbp]
\begin{center}
\includegraphics[width=\textwidth]{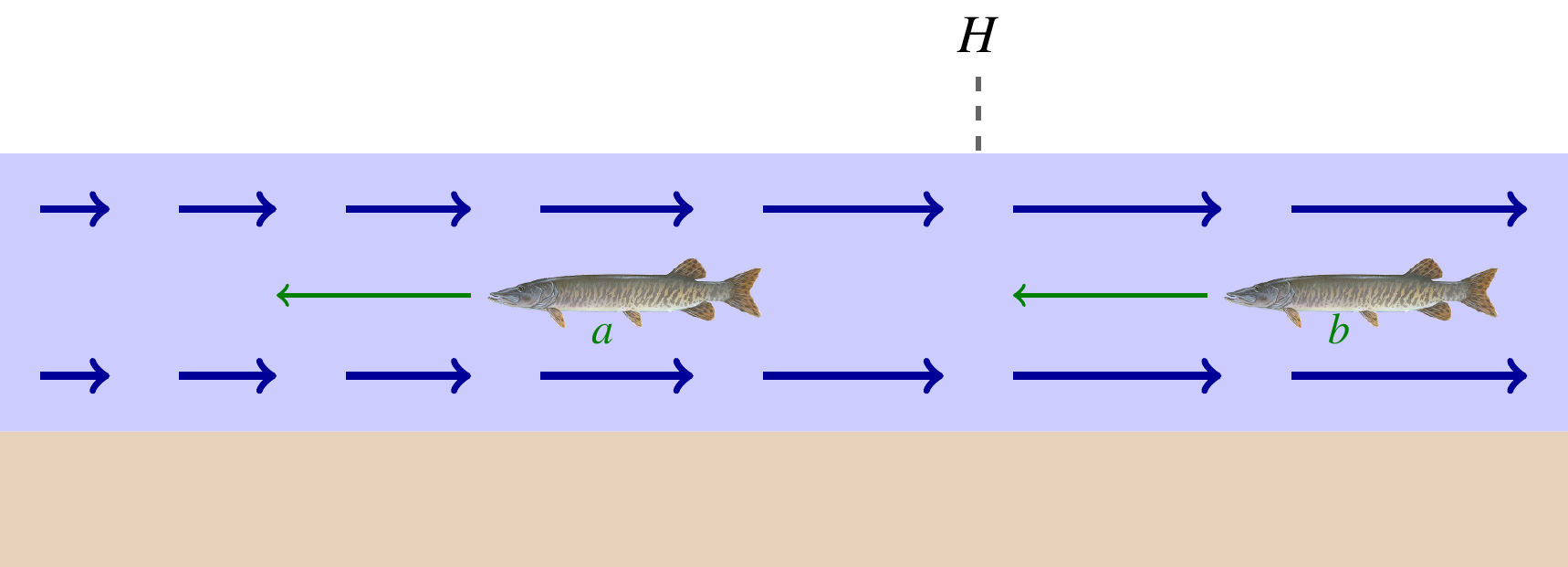}
\caption{River with variable speed of flow, and horizon $H$}
\label{RiverHorizon}
\end{center}
\end{figure}

The flow speed can vary along the river -- physically, where the river becomes more narrow, the water will flow faster; where the river becomes wider, water will flow more slowly. This can lead to the following interesting situation. Imagine that flow speed is increasing from the left to the right, as indicated by the increasing length of the thick blue arrows in figure \ref{RiverHorizon}.
At the point marked $H$, the river flow speed exceeds the maximal speed of fish. For us, who are watching from the river bank, that property makes $H$ special. We will never see a fish to the right of $H$ make any progress towards the left. After all, even if such a fish were to swim upriver, against the flow, with the maximal speed of fish, the flow would carry it downriver at a greater speed than that, so the net movement of the fish would be to the right. The fish marked b is in such a situation. For us river-bank observers, fish b is drawn inexorably towards the right. Nevermore will we see it to the left of $H$.

By contrast, fish that start out to the left of $H$, such as the fish marked ``a'', have a choice. Depending on how they choose to swim, river-bank observers will see such fish move upstream or downstream, towards the right or towards the left. 

The black hole analogy is fairly straightforward: The river represents a radial portion of space around a black hole. The flow of the water models the effects of the black hole's gravitational attraction. The fish are particles moving through space. At most, these particles can move at the speed of light, corresponding to the maximal speed of fish. The pattern in figure \ref{RiverHorizon} represents the increasing strength of gravity closer to the black hole. $H$ is the black hole horizon; to the right of $H$ is the inside of the black hole, to the left, its outside. No particle that has fallen in, that is, finds itself on the right of $H$, can ever get to the outside again. In doing so, it would need to go faster than the speed of light.

The model nicely reproduces the role of the equivalence principle in this specific situation. After all, at least in classical general relativity,\footnote{In particular: outside of ongoing discussions about possible firewalls at the horizon, created by quantum fields \citep{Polchinski2015}.} an observer falling into the black hole will not notice anything special when she crosses the black hole's event horizon. It is only when we take a global view, analysing the light-like world lines at various locations, and following up on where they lead, that we separate the inside from the outside, identify the area of no escape, and map its boundary: the event horizon. In the river model, fish drifting along will not notice anything special when crossing the event horizon, either. As far as these fish are concerned, they are able to swim freely in every possible direction in the surrounding water. It is only from the outside that we notice the point $H$ of no return, the horizon, and the separation into inside and outside. 

One feature of black holes needs to be made clear, though, when using the river model: students tend to think of the river as somewhat symmetric, continuing left and right. But there is an asymmetry in the case of a black hole: The outside (in our images: to the left) is practically infinite. The inside (to the right of the horizon $H$) is finite: Free fall into a black hole comes to an abrupt stop at the black hole's central singularity, after a finite amount of proper time has elapsed on the falling clock.

An exciting property of the river model of black holes is that it {\em can be made exact}: Choose an appropriate radial velocity field; interpret this field as describing the trajectories of observers falling inwards from infinity; apply the equivalence principle, namely that for each such observer, the metric must be locally Minkowskian; transform this Minkowskian metric back to your external coordinates, and you end up with the so-called {\em Gullstrand-Painlev\'e metric} \citep{Martel2001}. With an appropriate transformation of the time coordinate, you can retrieve the usual Schwarzschild metric. In fact, river models for the Reissner-Nordstr\"om metric and even for the Kerr metric have been formulated as well. These, too, can be found in \citet{Hamilton2008}.

Riverine black holes are more than just a model for teaching. They have interesting applications in {\em analogue gravity}: experimental physicists actually producing fluids that flow inwards, in order to simulate spacetime around a black hole -- and possibly answer questions about the behaviour of quantum fields in black hole space times, including the properties of Hawking radiation \citep{Unruh1981,Barcelo2011,Steinhauer2014,Steinhauer2016,Leonhardt2018}.

That said, the main disadvantages of the river models are fairly obvious. When introducing special relativity, didn't we spend a significant amount of time and effort to persuade students that there is no material ether, no all-pervading medium? In the river model, there is. And while all the local rest frames defined by the different regions of water are examples for how the equivalence principle works, in reality, but not in the model, that principle holds for all rest frames that are in constant motion relative to the water, as well. We're singling out a set of local reference frames, and defining a medium pervading all of space.

Then, of course, there is friction; a considerable amount, as the fish move through the water. Add to that the inconvenient fact that these fish propagate by pushing water behind them, which again has no counterpart in real space (at least outside of science fiction).     

\subsection{Visualising black holes}
\label{BHVisualisation}

In the immediate vicinity of black holes, the influence of curved space-time on the propagation of light is particularly pronounced. While the black hole itself is, by definition, not visible directly, the resulting distortions of what an observer would see provide evidence of its presence. 
\begin{figure}[htbp]
\begin{center}
\includegraphics[width=0.9\textwidth]{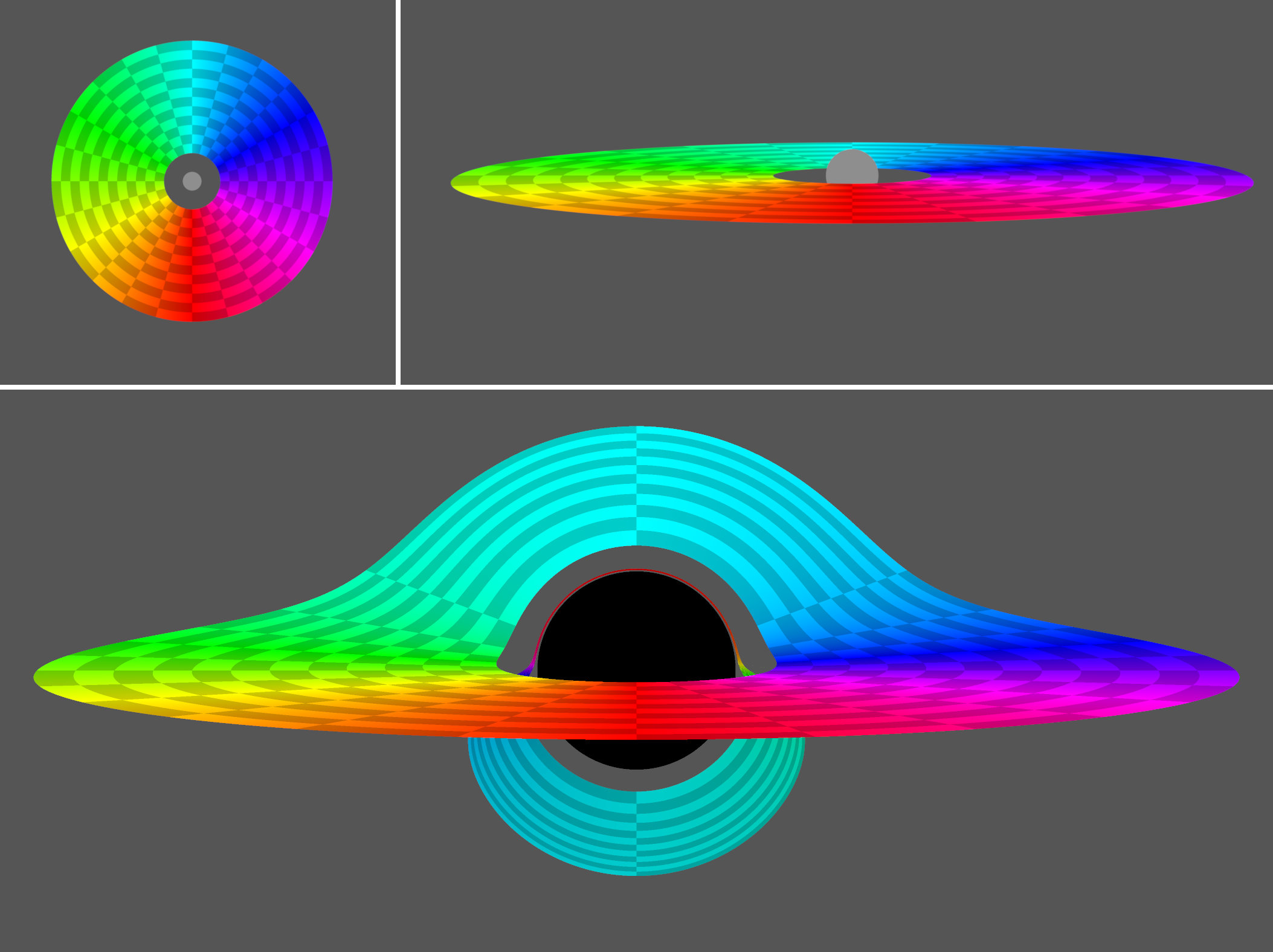}
\caption{Top left: Colored disk with a checker pattern. The grey circle corresponds to the Schwarzschild radius of the black hole whose influence is simulated in the bottom image. Top right: the same disk, seen at an inclination of 84.3 degrees in ordinary flat space with light propagating along straight lines. Bottom: Visual appearance of the disk under the same geometric conditions, taking into account light deflection by the black hole. Image: Thomas M{\"u}ller (HdA/MPIA) }
\label{BHVisuFigure}
\end{center}
\end{figure}
An early example of a visualisation of such distortions is that of a thin disk surrounding the black hole, relevant for the physical situation of a black hole accretion disk, by \citet{Luminet1979}. The most prominent recent example is the visualisation of ``Gargantua,'' the black hole in the movie {\em Interstellar}, released in 2014 \citep{Thorne2014,James2015}. Incidentally, the {\em Event Horizon Telescope} project is currently working on using radio interferometry to obtain a real-life version of this kind of image for our home galaxy's central black hole.\footnote{\href{https://eventhorizontelescope.org/}{https://eventhorizontelescope.org/}}

The way that the black hole's influence distorts the image can be made more understandable using a simple color model for the disk, which makes it easier to identify which parts of the disk an observer will see in different parts of the image, cf. figure \ref{BHVisuFigure}.\footnote{For Android devices, the app ``Accretion Disk'' by Thomas M\"uller enables interactive simulations, see \href{https://play.google.com/store/apps/details?id=tauzero7.android.relavis.blackhole.accretiondisk}{https://play.google.com/store/apps/details?id=tauzero7.android.relavis.blackhole.accretiondisk}.} Other auxiliary structures for helping the viewer comprehend the situation include a map of the Earth \citep{Nemiroff1993} or a checkerboard surface pattern \citep{Kraus1998}. Visualisations also allow viewers to explore a first-person view of falling into \citep{Hamilton2009,Hamilton2010}, or orbiting \citep{MuellerBoblest2011}, a black hole. Similar visualisations can be created for a wormhole \citep{Mueller2004}.

\begin{figure}[htbp]
\begin{center}
\includegraphics[width=0.7\textwidth]{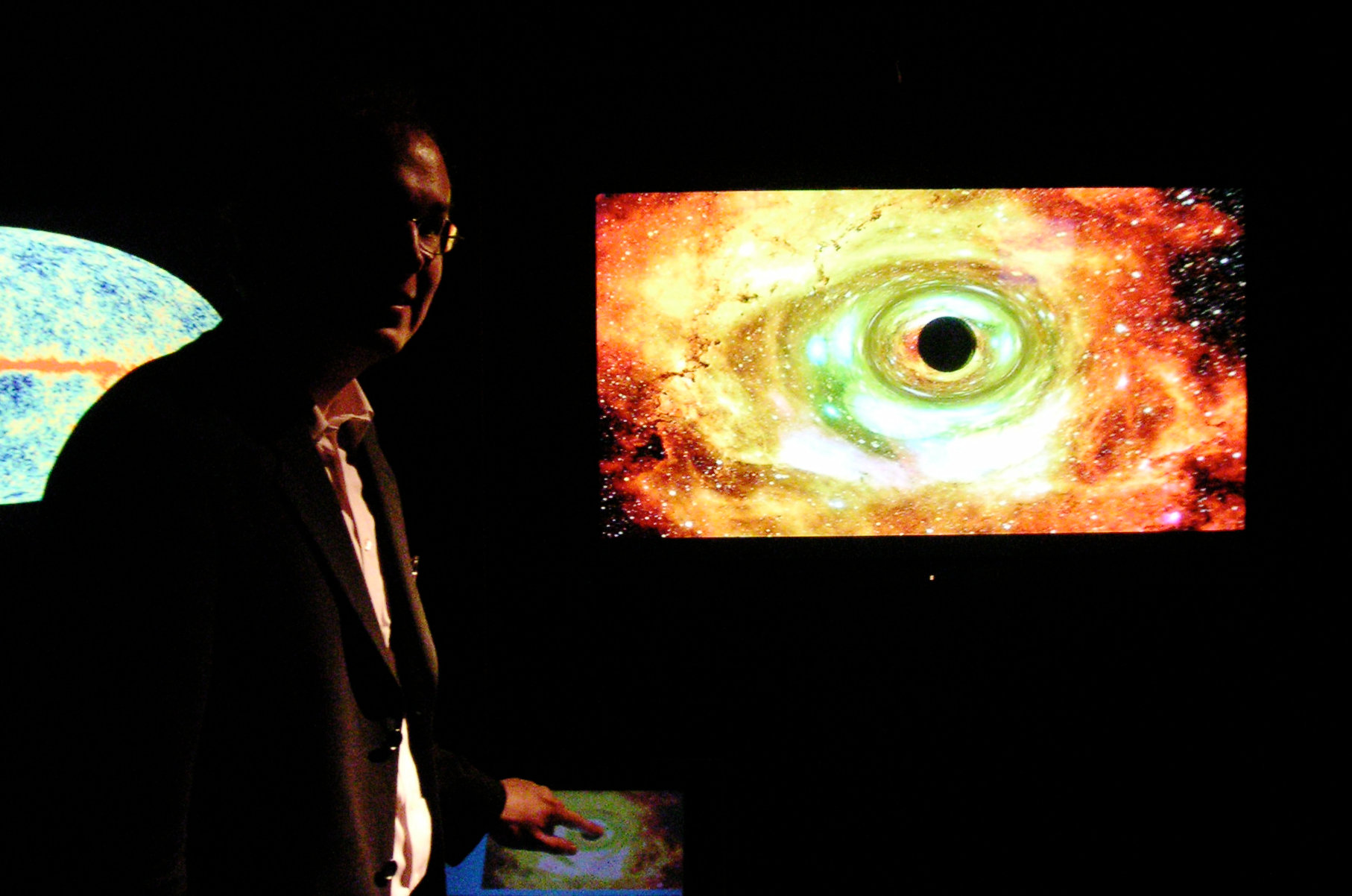}
\caption{Interactive exhibit of a black hole in front of an astronomical background, created by the T\"ubingen relativistic astrophysics group of Hanns Ruder for the exhibition ``Albert Einstein --- Engineer of the Universe'' in Berlin, 2005. The black hole can be moved around on a touch screen (bottom); the result is also displayed on the larger upper screen }
\label{InteractiveBH}
\end{center}
\end{figure}

The simpler situation of a black hole in front of a distant background \citep{Kraus2005} lends itself to interactive implementations \citep{Mueller2010}. Figure \ref{InteractiveBH} shows an example from the exhibition ``Albert Einstein ---Engineer of the Universe'' in Berlin \citep{Renn2005}. 

The "Black Hole (Light)'' app by Thomas M\"uller, available for Android devices, allows users to upload images, and view their distorted versions.\footnote{The Black Hole (Light) app is available for Android devices in the Google Play store at \href{https://play.google.com/store/apps/details?id=tauzero7.android.relavis.blackhole.fastview_free}{https://play.google.com/store/apps/details?id=tauzero7.android.relavis.blackhole.fastview\_free}}
An interactive first-person perspective is provided by the {\em Pocket Black Hole} app developed by Laser Labs for the University of Birmingham gravitational wave group. The app shows the tablet's camera view as it would look if there were a black hole between you and what the camera sees, distorting the view.\footnote{The app is available from \href{https://www.laserlabs.org/pocketblackhole.php}{https://www.laserlabs.org/pocketblackhole.php} (last access 11 May 2018).}

\begin{figure}[htbp]
\begin{center}
\includegraphics[width=0.85\textwidth]{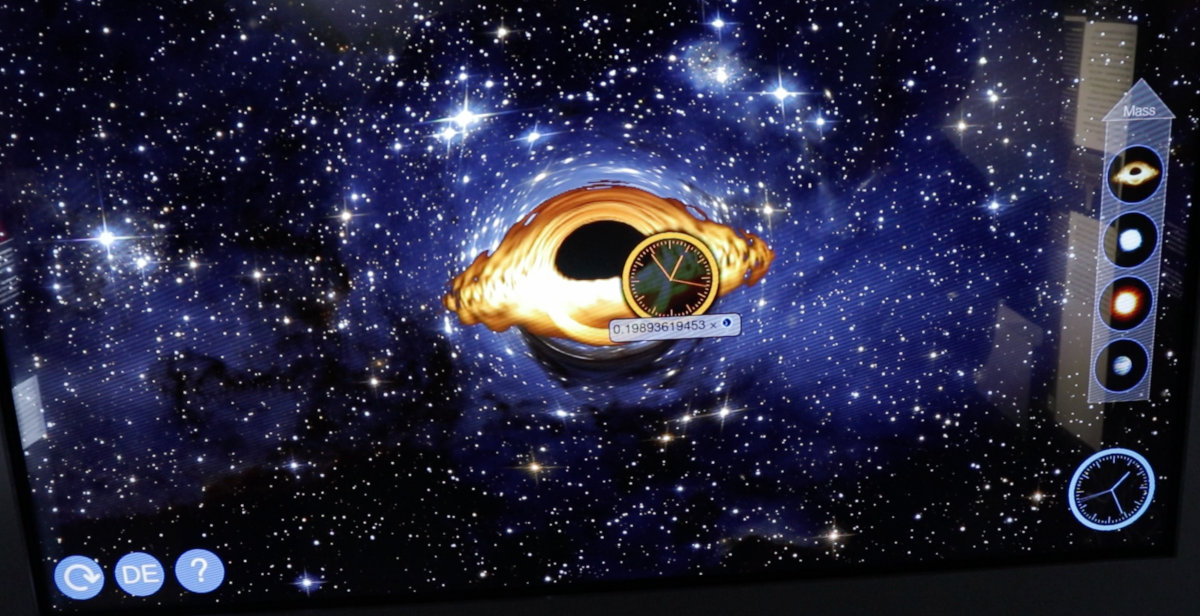}
\caption{Interactive black hole exhibit showing gravitational time dilation, developed by the Heidelberg Institute for Theoretical Studies (Dorotea Duda\v{s} and Volker Gaibler) for the exhibition of the ESO supernova. The exhibit can be seen in (inter-)action in the YouTube video \href{https://youtu.be/GVvcJAtCPF4?t=47}{https://youtu.be/GVvcJAtCPF4} (time mark 00:47)}
\label{blackhole-time}
\end{center}
\end{figure}

Gravitational time dilation can be incorporated into animated visualisations as well, notably in the shape of clocks that run more slowly closer to the black hole, and faster at greater distances. In such models, the visitor's concept of simultaneity is aligned with the model's global time coordinate, so observers can compare clock rates using their everyday notion of time. Fig.\ \ref{blackhole-time} shows an example for an interactive exhibit which demonstrates this effect.

\subsection{Hawking Radiation}

Hawking radiation, derived by Stephen Hawking in the mid-1970s \citep{Hawking1975}, was a breakthrough in semi-classical gravity, specifically the study of quantum fields in curved space-time. Hawking's calculations showed that black holes emit thermal radiation, corresponding to a temperature that is inversely proportional to the black hole mass. Hawking radiation neatly completes what is known as {\em black hole mechanics}, a simple set of rules that was thought to be an analogue of the fundamental laws of thermodynamics. With the thermal radiation supplied in the form of Hawking radiation, the laws of black hole mechanics could in fact be shown to be identical to (a generalisation of) the laws of thermodynamics. 

It is sometimes said that black holes are not, in fact, black. This demonstrates the tension between everyday usage and technical usage of the word: In everyday terms, a black object is one that emits and reflects little to no light. In physics, such behaviour is impossible; instead, a black body is one that absorbs all incident radiation, and itself emits thermal radiation as determined by its own temperature. For everyday objects, the emission is predominantly in the infrared region of the spectrum, so that such an object indeed appears to radiate little or no visible light. To a physicist, black holes became black in the usual sense once Hawking showed that they indeed have a 
(non-zero) temperature, and radiate in the way of all physical black bodies.

\begin{figure}[htbp]
\begin{center}
\includegraphics[width=0.65\textwidth]{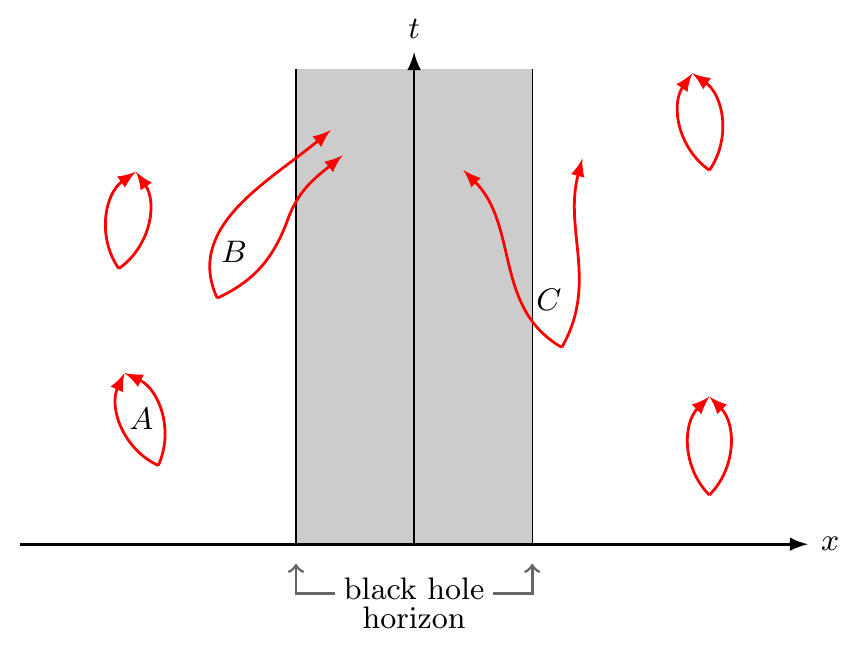}
\caption{Simple model for Hawking radiation: virtual particles near the horizon of a black hole. The situation $C$, in which the negative-energy particle of a pair falls into the black h ole, and the positive-energy particle escapes, is taken to be the production mechanism for Hawking radiation}
\label{Fig:HawkingRadiation}
\end{center}
\end{figure}

Hawking himself has used a simplified model for the radiation named after him (chapter 7 in \citealt{HawkingBrief}, and chapter 4 in \citealt{Hawking2001}), which is sketched in Fig.\ \ref{Fig:HawkingRadiation}.
The model makes use of a common picture for explaining the quantum properties of a vacuum state. A quantum field cannot be identically zero (corresponding to the complete absence of the associated particles). That, after all, would mean that both the current state of the field and its rate of change would be known precisely, which is impossible in quantum theory due to the applicable uncertainty relations. Thus, even a quantum vacuum state, corresponding to energy zero, must have fluctuations. One can picture these as pairs of virtual particles, each consisting of a particle and the corresponding antiparticle. In a fluctuating quantum field, such pairs are created continually, each vanishing again after a short while. Energy conservation demands that one of the two have negative energy, so that the total energy is and remains zero. While classically, particles with negative energy are forbidden, in a quantum vacuum, the argument goes, such particles can exist at least for a short time. In this situation, there are several outcomes, cf.\ Fig.\ \ref{Fig:HawkingRadiation}: Usually, as in case $A$, the negative-energy particle and its positive-energy partner will annihilate after a short time, leaving behind zero total energy. But if there is a black hole horizon present, other outcomes are possible. One is comparatively boring: In case $B$, both particles of a pair fall into the black hole. More interesting is case $C$, where the negative-energy particle falls into the black hole, and the positive-energy partner escapes. 

Why only the positive-energy partner? In Hawking's picture, that is explained as follows: Close to a massive body, a particle has less energy than at a greater distance from that body, since one would need to invest energy to lift the particle to a greater distance (in physics terms, due to the potential energy associated with the attracting mass). Inside the black hole, gravity is so strong that negative-energy particles are actually {\em not} forbidden even according to the laws of classical theory --- after all, the negative energy state is produced by a well-known mechanism, that of potential energy. Thus, once it is inside a black hole, the negative-energy particle of a virtual pair is not forbidden any more, and can just go on existing, without the need to vanish after a short while as would be required of an ordinary virtual particle. The situation where the negative-energy particle falls in, and the positive-energy particle escapes, is thus allowed! The converse situation, though, with the negative-energy particle outside the horizon, remains forbidden --- outside the horizon, any negative-energy particle can only be virtual, doomed to re-uniting in short order with its partner for a total energy of zero. For someone who observes the situation from afar, the net effect will be one of positive-energy particles (and positive-energy antiparticles) escaping from the black hole.\footnote{An alternative explanation can be given once it is realised that, behind the horizon, radial coordinate and time coordinate switch places, cf. \cite{Poessel2010}. Under those conditions, a classically forbidden negative energy value outside the horizon becomes a classically allowed negative momentum component.}

The advantage of this model is that it meshes well with models of quantum fluctuations given in elementary particle physics; the model is thus a part of a broader explanatory landscape. The restriction of classically forbidden energy values, in this case negative energies, meshes well with the relations linking the limited lifetimes of excited quantum states with the energies of those states. 

The model has been criticised on several grounds by \citet{Freistetter2018}, who mentions the problematic concept of negative energy, asks why only the negative-energy particle falls in, and criticises that the model predicts that Hawking radiation should only be created right at the horizon and not, as calculations show, in a finite region (``atmosphere'') surrounding the horizon. Freistetter then goes on to describe a simplified scenario with initial and final quantum states that is closer to Hawking's original set-up.\footnote{While I believe that Freistetter does not adequately take into account Hawking's argument for why the negative-energy state can only exist indefinitely inside the horizon, this resolution requires, admittedly, accepting an unusual property of black holes. From looking at Fig.\ \ref{Fig:HawkingRadiation}, I cannot follow Freistetter's argument about radiation production only directly at the horizon; after all, the relevant particle pairs will generally come into being a bit outside the horizon.} It should be noted that this model can, in fact, be made more exact and used for a derivation of the properties of Hawking radiation \citep{Parikh2000,Parikh2004}.

Fluid models in the context of analog gravity, with the aim of answering questions about the behaviour of quantum fields in black hole space times, had already been mentioned in section \ref{RiverModelBlackHolesSection} in the context of the river model of black holes \citep{Unruh1981,Barcelo2011,Steinhauer2014,Steinhauer2016,Leonhardt2018}; these, too, are of course hoped to include a model analog of Hawking radiation.

\section{Cosmology}
\label{Cosmology}

For more than a decade now, cosmology has been an area of particularly lively debate about models for teaching general relativity. It is characteristic for these discussions that the foundations of modern cosmology, which are based on the Friedmann-Lema{\^\i}tre-Robertson-Walker (FLRW) solutions, is not in dispute, and in most cases, neither are the calculations performed using these models. But different authors differ, sometimes significantly, about the meaning of ``cosmic expansion,'' about the appropriateness (or not) of different models of the expanding universe, their relativistic implications and their advantages or disadvantages as teaching tools \cite[and references therein]{Davis2004,DavisLineweaver2004,Liebscher2007,BunnHogg2009,FrancisEtAl2007,AbramowiczEtAl2009,Chodorowski2011,Rebhan2012b}. I will comment on some aspects of this debate directly related to models, but not all (leaving out, for instance, questions about cosmic horizons).

Before I come to that debate in sections \ref{ExpandingSubstrate}--\ref{CosmoRedshift}, I will give an overview of other kinds of models related to cosmology. For instance, trees in a forest and the way that, given a sufficiently large forest, we will see tree trunks everywhere, have long been a helpful model for understanding Olbers' paradox \citep{Lotze1995a}. The astronomical distance measurements that are crucial for cosmology can be explained using models, as well: Light sources in the lab or outside (street lights!) can be used to understand (and even to reproduce quantitatively!) the standard candle method \citep{Poessel2016a}, and there are various models for demonstrating the parallax method \citep{DeJong1972,Deutschman1977,Ferguson1977,Poessel2017b}. 

Specific aspects of cosmology can be conveyed by interactive exhibits. An example is the relation between the color of simple light waves and their wavelengths, and the way both of these change in the course of a Doppler redshift or cosmological redshift. 
\begin{figure}[htbp]
\begin{center}
\includegraphics[width=0.7\textwidth]{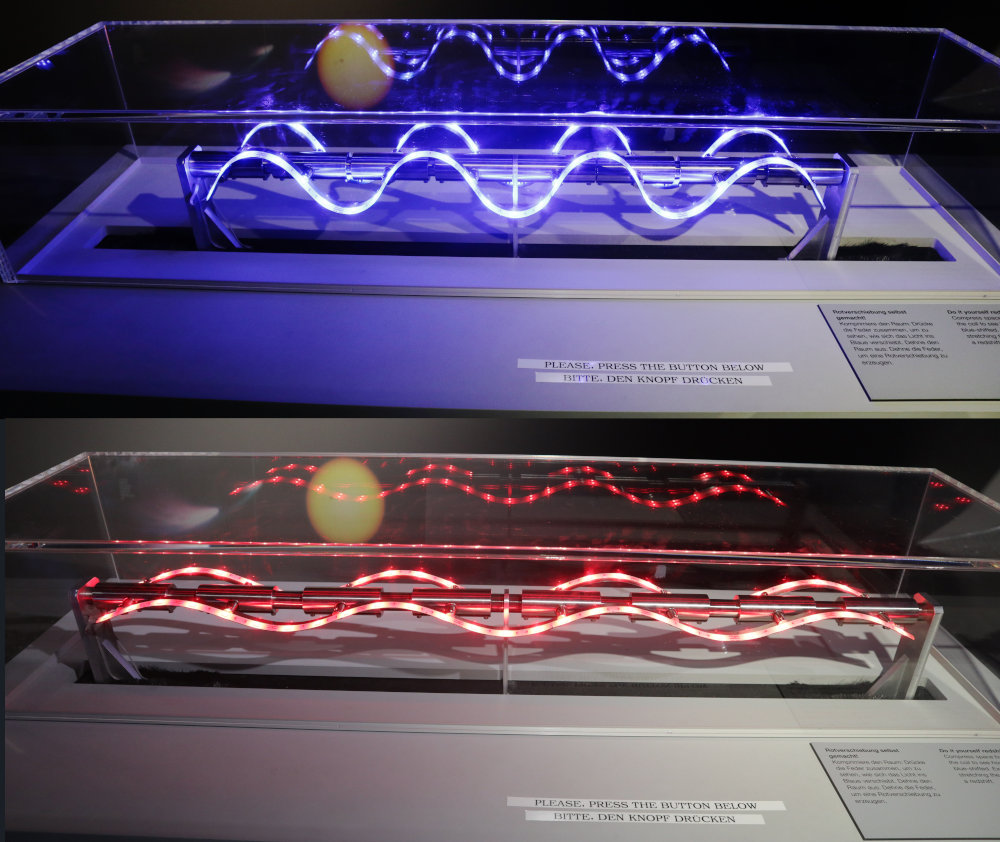}
\caption{Doppler shift exhibit in the cosmology section of the ESO Supernova in Garching, demonstrating the relationship between wavelength and color. The exhibit can be seen in (inter-)action on the YouTube video \href{https://youtu.be/GVvcJAtCPF4}{https://youtu.be/GVvcJAtCPF4}  }
\label{ESORedshift}
\end{center}
\end{figure}
One example is the cosmological redshift exhibit in the ESO Supernova in Garching, Germany, developed by the European Southern Observatory. Here visitors can turn a crank to deform the ``wave'' from shorter wavelengths (top), lit in blue and accompanied by a higher-pitched sound, to longer wavelengths (bottom) with red light and a lower-pitched sound.
(The exhibit does not go through all the intermediate spectral colours, however, but instead interpolates between blue and red via magenta.) Another example is ``The Expanding Universe'' at the Adler Planetarium in Chicago; in that exhibit, a slinky can be compressed or stretched; in the first case, the slinky is illuminated blue to indicate blueshift, while a stretched slinky is illuminated in red to indicate redshift.\footnote{Mark SubbaRao, personal communication, January 2, 2018.}

Inhomogeneities in the cosmic background radiation, that is, the radiation left over from the hot and dense initial Big Bang phase, are the seeds for structure formation over the billions of years of cosmic history. As a visualisation of these inhomogeneities, as a suitable telescope would see them in the night sky, these inhomogeneities have been 3D printed as a physical model of our universe's early state \citep{Clements2017}.

Regarding the evolutionary history of the universe, there are various models that map the history of the universe to other linear structures. A famous example is the ``Cosmic Calendar'' pioneered by Carl Sagan, which maps the history of the universe onto the 365 days of a terrestrial year \citep{Sagan1977,CosmosEp1}. Cosmic history can also be mapped to a one-dimensional {\em spatial structure.} An example is the Harriet and Robert Heilbrunn Cosmic Pathway along the spiral ramp underneath the Hayden Planetarium dome in the American Museum of Natural History in New York, which maps cosmic history onto a path 360 feet (110 meters) in length, including displays of artefacts from different periods of history and several media stations which provide additional information. Yet another model relates cosmic history to the phases of childhood development \citep{Fisher2018}. 

The long time scales of cosmic evolution require specific research techniques: Astronomers are given a ``cosmic snapshot'' (more precisely: information from our own past light cone) showing many different objects in different stages of development. They need to combine these bits of information about separate objects so as to deduce the stages of evolution for each class of object, inferring, for instance, the stages of stellar evolution from the different kinds of stars we can currently observe. A model for this task is to imagine what a short-lived fruit fly can deduce from the humans in various stages of development encountered during its short life. Could such a fruit fly deduce the evolution from birth and childhood to adulthood, old age, and eventual death \citep{Kippenhahn1984,Kippenhahn1995}? An alternative model is that of visiting a football game, taking a picture of all the humans in the audience, small people and large people, then trying to deduce the stages of human development (``how do small people become large people?'') from the snapshot \citep{Mather2012}.

Software packages for digital planetariums such as Uniview\footnote{Uniview by SCISS, \href{http://www.scalingtheuniverse.com}{http://www.scalingtheuniverse.com}} or DigitalSky,\footnote{DigitalSky by Sky-
Skan, \href{https://www.skyskan.com/products/ds}{https://www.skyskan.com/products/ds}} or their counterparts for individual users such as Celestia\footnote{\href{https://celestia.space/}{https://celestia.space/}} or the World Wide Telescope\footnote{American Astronomical Society, \href{http://www.worldwidetelescope.org}{http://www.worldwidetelescope.org}} \citep{McCool2009b}, provide immersive models for the large-scale structure of the universe. When those model universes are based on astronomical data, they effectively amount to travelling along our past light cone --- in the extreme case, back to the origin of the cosmic microwave background, and thus to the Big Bang phase of the universe.

At the tasty end of cosmological visualisations are food-based models that have been used to visualise not only black holes as ``space-time tortillas'', cosmic evolution in the shape of a cocktail with ever less dense layers from bottom to top, the large-scale structure of the cosmos using a parmesan tuille, and even the multiverse, using different kinds of chocolate praline universes \citep{Trotta2018}.

\subsection{Expanding substrate models}
\label{ExpandingSubstrate}

Various models for cosmic expansion fall under the umbrella term ``expanding substrate models.'' The oldest example of which I am aware (thanks to \citeauthor{Harrison1993} [\citeyear{Harrison1993}]) is a technical article by Arthur Eddington.\footnote{Incidentally, this is the article that showed the instability of Einstein's 1917 static cosmological solution.} After having introduced the formulae for an expanding universe, Eddington remarks ``It is as though they were embedded in the surface of a rubber balloon which is being steadily inflated'' \citep{Eddington1930}. Willem de Sitter followed suit directly with an expanding ``india-rubber ball'', and the galaxies represented by ``specks of dust'' on the ball's surface \citep{deSitter1931a,deSitter1931b}, and Eddington revisits the rubber balloon in his lecture and book {\em The Expanding Universe} a few years later \citep{Eddington1933}.\footnote{Earlier on, in the context of his static cosmological model with space shaped like a 3-sphere, Einstein himself had introduced a curved cloth for comparison in a letter to Willem de Sitter. Specifically, he compared the spherically curved space with a spherically curved region of a cloth floating in the air, noting our inability to deduce from the geometry that region whether or not, globally, the whole space would be closed in on itself, or end in a boundary, or go on to infinity \citep{Einstein1917b}.}

\begin{figure}[htbp]
\begin{center}
\includegraphics[width=0.75\textwidth]{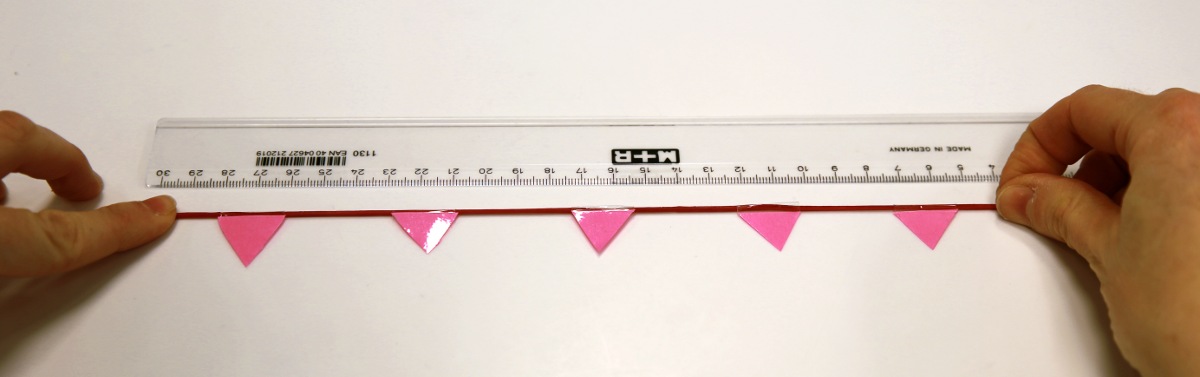}
\includegraphics[width=0.75\textwidth]{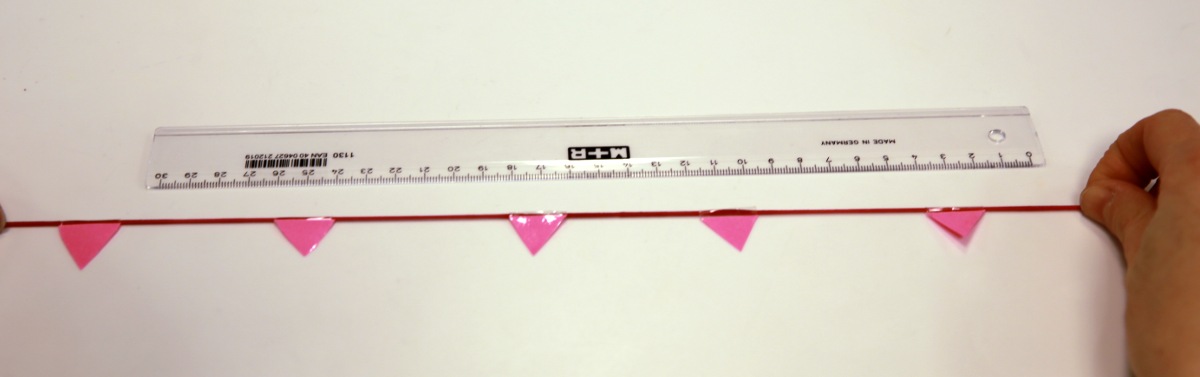}
\caption{One-dimensional expanding substrate model: rubber band with pasted-on cardboard triangles to mark the positions of galaxies}
\label{cosmicGummiband}
\end{center}
\end{figure}

In one dimension, a rubber band or latex strip makes for a simple demonstration of cosmic expansion: as the rubber band is stretched, marks on the strip -- applied with a marker, or pasted on -- or knots represent galaxies in the Hubble flow, which move apart in simple scale factor expansion \citep{Lightman1991,Poessel2005}. Figure \ref{cosmicGummiband} shows an example. The simple linear setup makes it easy to measure the changing distances quantitatively (\citealt{Klimishin1991,Lotze1997,Hogan1998,Price2001,Lotze2002c} and activity H--9, ``Modeling the Expanding Universe'' in \citealt{UAYF2}). This means the model can be used directly to let students derive the Hubble-Lema{\^\i}tre law.  Alternatively, at least as a Gedankenexperiment or as part of an animation, a rubber ruler with tick marks for galaxies will do the trick just as well \citep{Plait2015}. Disadvantages of the rubber band are that it has two ends and thus a well-defined center, in contrast with the boundary-less homogeneous universe we inhabit; also, we are observing the rubber band from the outside, whereas for an inhabitant of the modelled universe, that rubber band would be all there is \citep{Lotze2002c}. Another possibility is a collapsible radio antenna, with a little flag representing a galaxy attached to each segment \citep{Bauman1980}; in practice, though, I have found that it is not at all easy to extend the antenna so uniformly that the distances between the little flags undergo scale-factor expansion.

Considering two-dimensional substrates, I already described some aspects of the balloon model for cosmic expansion in section \ref{GoodBad}.
\begin{figure}[htbp]
\begin{center}
\includegraphics[width=0.9\textwidth]{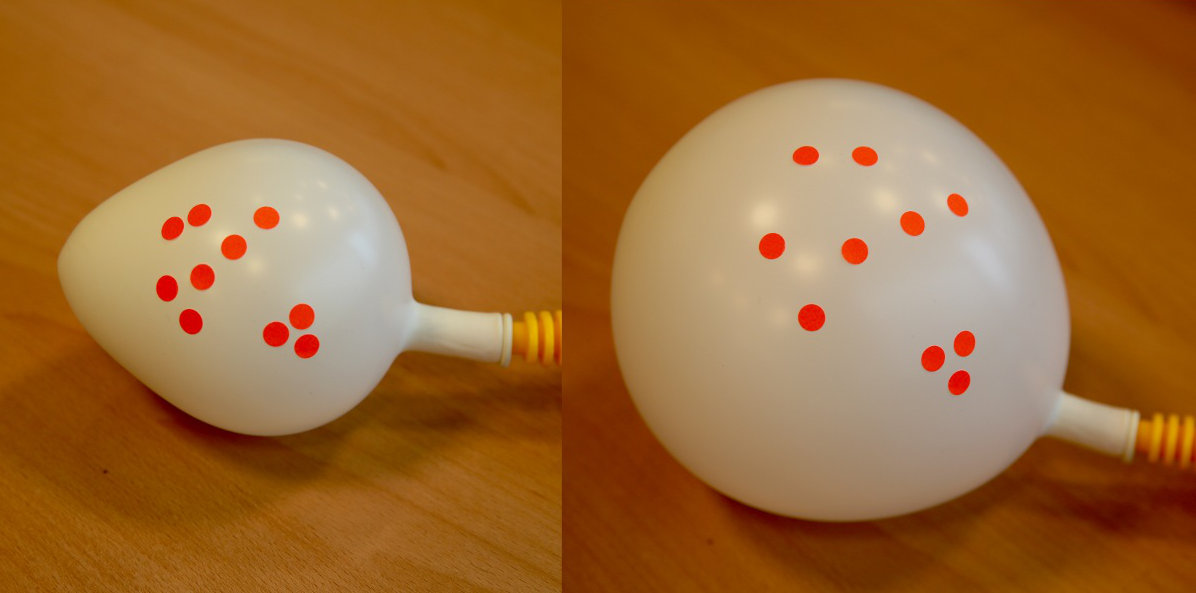}
\caption{The rubber balloon universe in two different states of expansion. The red stickers on the balloon's surface represent galaxies in the Hubble flow}
\label{BalloonUniverse}
\end{center}
\end{figure}
A rubber balloon provides a suitable two-dimensional surface; again, marks painted on or in the form of stickers represent galaxies in the Hubble flow, as shown in figure \ref{BalloonUniverse}. As you inflate the balloon, distances between the model galaxies, measured as shortest connecting lines on the balloon surface, increase proportional to a cosmic scale factor \citep{Lotze1995a}. This model is common in popular science books \citep{Gamow1940,Barnett1948,Trefil1983,Asimov1984,Nicolson1981,Nicolson1985,Wheeler1990,Barrow1994,Hawking1996,Hawking2005,Bennett2014,Vaas2018} and is included in a number of text books \citep{MTW,Kaufmann1988,Ingham1997,Taylor2000,Foster2006,Ellwanger2008,Hanslmeier2013,Bennett2017} as well as in the cosmological philosophy book by \citet{Kanitschneider1984}. It is possible to use a tape measure to measure intergalactic distances on the balloon as a hands-on activity \cite[activity H-4 and H-7 respectively, ``A ballooning universe'']{UAYF,UAYF2}.\footnote{In those instances where I have tried this with a group of students, the deviations of the balloon from a perfectly spherical shape made it difficult to get a halfway decent Hubble diagram out of this, though.} Instead of at a balloon, one can look at an expanding rubber sheet universe, such as the one introduced (with careful attention to its model character, and the limits of the model) in \citet[chapter 14]{Harrison2000}, cf.\ \citet{Harwit1988} and \citet{Lotze1997}.

\begin{figure}[htbp]
\begin{center}
\includegraphics[width=0.7\textwidth]{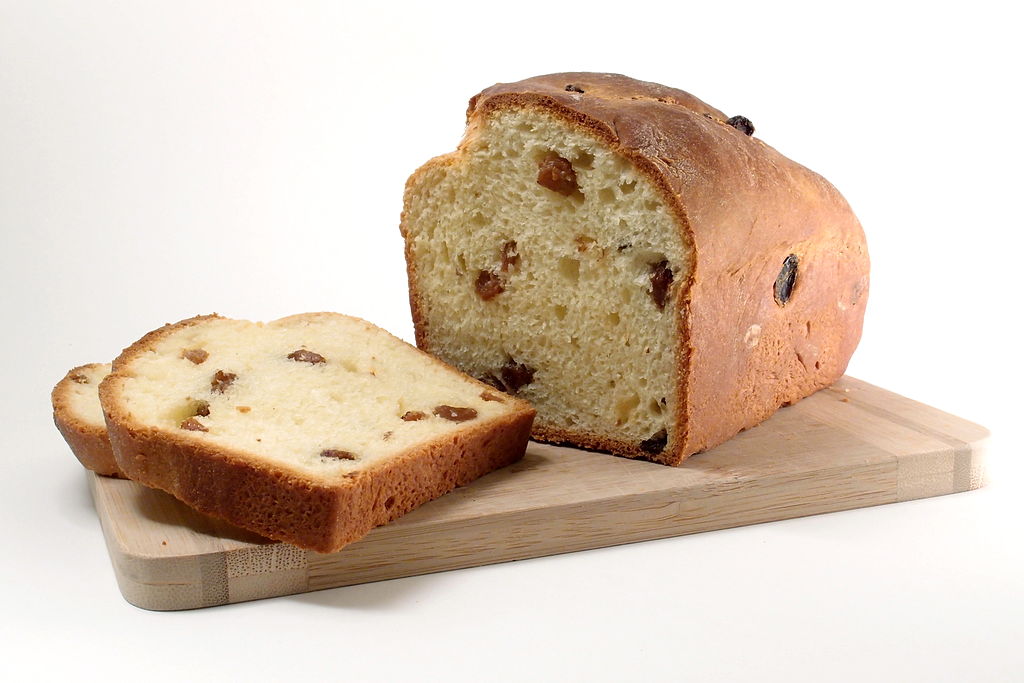}
\caption{Raisin cake, specifically: Sweet bread with raisins, made from a rich yeast dough. Image by User SKopp via \href{https://commons.wikimedia.org/wiki/File:Rosinenstuten_with_slices.jpg}{Wikimedia Commons} under license \href{https://creativecommons.org/licenses/by-sa/3.0/deed.en}{CC BY-SA 3.0}}
\label{RaisinCake}
\end{center}
\end{figure}
A raisin cake, figure \ref{RaisinCake}, can serve as a three-dimensional expanding universe: as the cake rises, the raisins move apart. In practice, this is a thought experiment rather than a demonstration. We can see through the oven door and observe the cake getting larger as it rises, but we cannot see directly how the raisins move apart --- in visualizations, the cake is shown with a cut-away view, or as transparent; real raisin cakes in the process of being baked are neither. Readers with access to an MRI machine and an idea of how to construct an oven without metallic parts are encouraged to try the experiment of monitoring the expansion of the cake in real time. For the rest of us, it is still  instructive to have students, or readers, think about how the raisins move apart, as a mental picture, aided by illustrations showing the raisin bread in different states of expansion \citep{Shipman1976,Abell1984,Kaufmann1985,Lederman1989,Pasachoff1992,Kippenhahn2003, Hetznecker2007,Bennett2014,Strauss2016}. As some practitioners have noted, bringing real cake into the classroom may also be a welcome change from ordinary, low-calorie teaching. An alternative to the cake is a three-dimensional cubic grid, with nodes connected by straight links --- whether a famous version by M. C. Escher \citep{Rees1999}, a plain version \citep{Trefil1983,CosmosEp10}, or a jungle gym with telescoping pipes, populated by children representing galaxies (attributed to George Gamow in \citealt{Shipman1976}). In all these cases, you are meant to picture the links lengthening or stretching, as if they were telescopic, or made of rubber. Alternatively, an exhibit can show the cubic grid at different times, corresponding to different values of the cosmic scale factor. An example can be found in the ``Evolution of the Universe'' exhibition in Deutsches Museum in Munich, where the grid is realised using white LEDs as galaxies in a glas cube with five mirrored walls producing an illusion of a much larger universe \citep{Wankerl2010}.

Neither the raisin cake nor the more abstract cubic grid provides a physical demonstration model of cosmic expansion in progress. Another three-dimensional model fares better in that respect: the boiling milk model of the universe's early inflationary expansion can be demonstrated in the classroom, and contains some interesting analogies (such as phase transitions) with the original \citep{Yusofi2010}.

In addition to stretchable substrate models, there are unstretchable ones, which serve to give a physical basis to the size of space. One possibility is that of a fully populated chessboard, which doubles in size while each chess figure stays the same size \citep{Fath1955}, another that of a cinema in which three persons initially sit next to each other, then have one empty seat magically inserted in between each pair, allowing for an explanation of the Hubble-Lema{\^\i}tre law \citep{Bauer2017}.

\subsection{Explosion models and the Milne universe}

A different family of models of cosmic expansion makes use of the familiar process of an explosion. In the simplest case, this is depicted as an ordinary explosion, with particles moving away from an identified central point of the explosion \cite[at 18:23--18:53]{CosmosDark2014}. Several problems with an ordinary explosion model are well-known. For one, it encourages the notion of a definite center of cosmic expansion -- namely the starting point of the explosion -- as the location ``where the Big Bang happened.'' This contradicts the cosmological principle, which states that we live in a homogeneous universe, with all locations on an equal footing. In the FLRW models, space is homogeneous, and the Big Bang can be said to have happened {\em everywhere} -- in the sense that all world lines of the Hubble flow can be traced back to the initial Big Bang singularity and that, since observers in the Hubble flow can consider themselves to be at rest, they can all claim that the Big Bang (also) happened at their own location.

The second misleading conclusion that can be drawn from the ordinary explosion model is that there is empty space, into which galaxies move as they follow the Hubble flow --- an expanding universe that pushes its (spherical, expanding) boundary ever further out into previously empty space.

Both problems are solved by the (special-)relativistic version of an explosion model: the {\em Milne universe}, first proposed by Edward Arthur Milne in the 1930s as an alternative to the cosmological models of general relativity, but long since been recognised as a special case of a FLRW spacetime with particular value as an educational tool \citep{Ellis2000,Rindler2001,Mukhanov2005,Liebscher2005,GronHervik2007,Rebhan2012a}. The Milne universe is a mathematical model, although a comparatively accessible one. Since a two-dimensional version (time direction, one space direction) can be easily drawn, it also provides an example for spacetime visualizations: Simple diagrams, often two-dimensional, sometimes three-dimensional, tracing light cones and particle worldlines in order to visualize physics in curved spacetime. (A systematic exposition of general relativity, based on such visualizations, is \citeauthor{Geroch1981} [\citeyear{Geroch1981}].)

\begin{figure}[htbp]
\begin{center}
\includegraphics{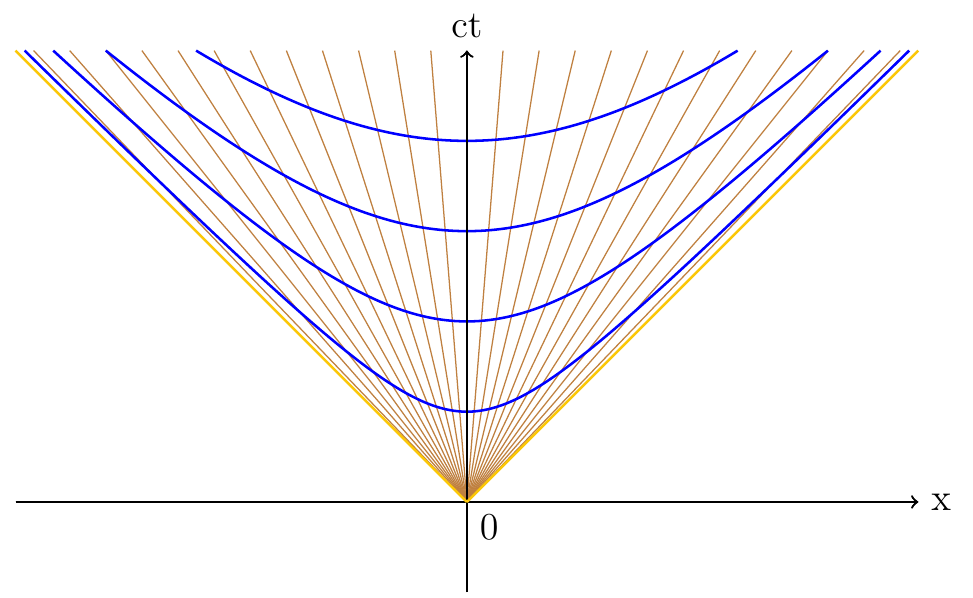}
\caption{Milne model spacetime: test particles moving away at constant speed from a common origin. Shown in blue are the lines of constant cosmic time.}
\label{MilneModel}
\end{center}
\end{figure}

The basic setup for the Milne model is shown in figure \ref{MilneModel}: test particles, which correspond to galaxies in the Hubble flow, leave a common location ($t=0,x=0$) at constant speeds. A cosmic time coordinate $\bar{t}$ is defined so that time intervals along the world lines of these test particles correspond to proper time intervals for these particles. A cosmic space coordinate $\bar{x}$ is defined as the proper distance at time $\bar{t}$ (found by integrating the metric along the appropriate slice $\bar{t}=\mbox{const.}$). Thus, we have two sets of coordinates: the private coordinates $x,t$ which are the usual inertial-frame coordinates for the inertial observer at rest at $x=0$, and unified, public coordinates $\bar{t}, \bar{x}$, in which cosmic time is defined by the proper time of galaxies in the Hubble flow (with the Big Bang event $t=0, x=0$ defining the zero point for all those proper time measurements), and distances measured along the surfaces of constant cosmic time. These coordinates correspond to the usual definitions of cosmic time and proper distances in a FLRW universe. Public and private coordinates are linked by
\begin{eqnarray}
ct &=& c\bar{t} \cdot\cosh(\bar{x}/c\bar{t})\\[0.5em]
x &=&  c\bar{t} \cdot\sinh(\bar{x}/c\bar{t}).
\end{eqnarray}

Number densities of galaxies can be chosen so as to yield a universe of constant co-moving galaxy density. A drawback of the model is that it is not complete, in the following sense: as shown in Fig.~\ref{MilneModel}, it only encompasses the forward lightcone of the Big Bang event. In principle, influences from outside that lightcone can propagate into the Milne universe (although nothin within the universe can propagate to the outside region). This feature vanishes with the transition to a universe with non-zero mass density: in the private coordinate representation shown in Fig.~\ref{MilneModel}, number density and thus mass density diverges as we approach the boundary light cone. The boundary itself becomes part of the initial singularity, sealing off what used to be the regions outside the light cone of the zero-density Milne model.

While the usual presentation requires at least some idea of the concept of a metric, the main points can be captured by a toy model that requires no more than an understanding of the usual Lorentz transformations between inertial systems \citep{Poessel2018}. In the following, we compare and contrast several aspects of the expanding substrate model and the Milne model.

\subsection{Geometry of (expanding) space}

The immediate and considerable advantage of expanding substrates is that they provide simple, widely accessible images. As the balloon inflates, you can see the whole pattern of dots change; as the rubber band is stretched, you see scale-factor expansion in action. In consequence, expanding substrate models have a wide range of applications. They are as suitable for the general public as they are for pupils of all ages, and can even serve to illustrate scale factor expansion for student audiences 

The Milne model, by contrast, is mathematical in nature, and requires knowledge at least of special relativity. This restricts the audience to more advanced high school students, university students, and those members of the general population with the required knowledge and a suitably serious interest in the subject. 

On a related note, at least the concrete one-and two-dimensional substrate models allow for direct measurements --- their correspondence with true scale factor expansion limited by possible inhomogeneities of the set-up --- which can then be used to reconstruct the relative change in universe scale factor and the Hubble relation. Working with the Milne model requires calculations; on the plus side, this results directly in simple, closed-form formulae.

Both models hold interesting lessons about the geometry of space. The balloon model gives a direct demonstration of a spherically curve surface, as an analogue for a spherically curved space; this illustrates nicely that geometry in modern cosmology need not be flat. This is a valuable demonstration of possibilities, although with the current measurements of cosmological parameters, we cannot distinguish between the universe's actual spatial geometry and a perfectly flat space. The balloon also demonstrates a possibility that intrigued Einstein in his original 1917 article: that space can be finite even in the absence of a boundary. Again, though, we have no indication that this is true for our universe.

The Milne universe gives a more advanced demonstration. Private space has Minkowski geometry, and is perfectly flat. Yet space in the Milne universe, defined by slices of constant cosmic time, is hyperbolic. Three-geometry is not invariant, but depends on the choice of time coordinate; four-curvature, the key criterion for deciding whether or not we are dealing with the gravitational effects of masses/energy, is markedly different from three-curvature. In particular, an infinite space, as determined by splitting spacetime into space and time using the cosmic time coordinate, can emerge from a finite region, in the Milne model: from the point-like Big Bang event. This is a helpful model when trying to comprehend similar transitions in the bubble universes of inflationary multiverse scenarios \citep[ch.\ 3]{Greene2011}.

\subsection{Dragging effects and bound systems}

The biggest drawback of expanding substrate models is that they are suggestive of objects getting dragged along with the Hubble flow. This is seen most directly in the latex strip version of the rubber band model, with galaxies represented by coins on the strip: If you place an additional coin on the latex strip while expansion is in progress, that coin will remain where it was placed, instantaneously joining the Hubble flow. 

You can even play out the so-called {\em tethered galaxy problem} in this setting: let coin A be glued to the latex strip ({\em definitely} part of the Hubble flow!), and let coin B be joined to coin A using a thin strip of paper (affixed to each coin with sticky tape or glue). Stretch the latex strip to simulate expansion; the coins, of course, will remain at a constant distance to each other; coin B will slide along the stretching latex to keep its constant distance from coin A.

But if you cut the paper strip while simulated expansion is going on, coin B will instantly join the Hubble flow. As soon as the paper strip is cut, it will lie still on the latex strip, following whatever simulated expansion might follow. 

This, of course, is completely different from what would happen to a tethered galaxy in FLRW models \citep{Peacock2002,DavisLineweaverWebb2003,Whiting2004,Clavering2006,GronElgaroy2007,BarnesEtAl2006}. The behaviour of such a galaxy is not influenced at all by the immediate Hubble flow -- only by the accelerations encoded in the second time derivative $\ddot{a}$ of the cosmic scale factor: the deceleration caused by the universal gravitational attraction of ordinary matter, or the acceleration caused by a cosmological constant/dark energy. In a universe dominated not by dark energy, but by matter and/or radiation, the galaxy represented by coin B would begin to move {\em towards} coin A!\footnote{This is readily understood from the point of view of Newtonian gravity, cf. section \ref{SectionNewtonianCalculations}: The gravitational force of any homogeneous spherical mass shell on a test particle inside the shell is zero; the gravitational force of such a shell on a test particle outside the shell is the same as if all the mass were concentrated in a point particle at the center of the sphere. Thus,
mass within the sphere that is centered on galaxy A and whose surface contains galaxy B acts as if all the mass were concentrated at the location of galaxy A; mass outside that sphere does not make a net contribution at all.}

In the Milne model, cosmic expansion is a consequence of initial conditions. A galaxy with different initial conditions is not going to immediately join the Hubble flow. At best, there is going to be what might be termed ``kinematic sorting'': a galaxy that is not part of the Hubble flow will overtake those Hubble flow galaxies that are slower, but not catch up with galaxies that are faster, and thus end up in the company of galaxies with the same velocity, going in the same direction, at the same speed. Conceptually, it is not a big leap from this purely kinematic situation, with galaxies moving as free test particles, to a Milne universe with the addition of Newtonian gravity. In that case, too, intuition from classical physics would posit that a tethered galaxy has different initial conditions from galaxies in the Hubble flow. Where it ends up will be a combination of kinematic sorting and additional acceleration caused by Newton's gravitational force (possibly modified to include the pressure term that allows for effects like those of dark energy).

Closely related to the tethered galaxy problem is the question of bound systems in an expanding universe. Expanding substrate models imply that space is some kind of a substance -- it is certainly modelled as such! -- and that this substance is expanding. Related descriptions sometimes talk about space expanding, about space being the only thing affected by cosmic expansion and matter being ``just along for the ride,'' of the expansion of space as something fundamentally different from galaxies moving {\em through} space \citep{Kanitschneider1984,Kaufmann1988,Liddle1988,Lederman1989,Winnenburg1998,BeyversKrusch2009,Freistetter2013a,Muller2014,Higbie2014,Siegel2015,Plait2015,Bahr2016,DLRnextUrknall,ChamWhiteson2017,StarTalk2017,Bauer2017}, or, even more strongly worded, about new space being created in the expansion process \citep{Gassner2016}.

But if expansion is a property of space itself, one could, with some justification, expect that for {\em all} objects that are separated by some intervening space, their distance must increase. If there is space in between the atomic nucleus, and the atom, and if that space, like all space, expands, atoms would get larger. Similarly, if space between the Sun and the Earth expands, the distance between the two should increase. It is one thing to say that strong forces somehow keep objects together by counteracting the expansion of space.\footnote{E.g. the Minutephysics explanation \citep{Minutephysics2013b}, which directly uses the problematic analogy of objects getting dragged along with an expanding, stretching surface.} But in the extreme case of the ``new space getting created'' picture, it is not even clear how forces and the creation of new space in between two objects can, or should, interact. 

\begin{figure}[htbp]
\begin{center}
\includegraphics[width=\textwidth]{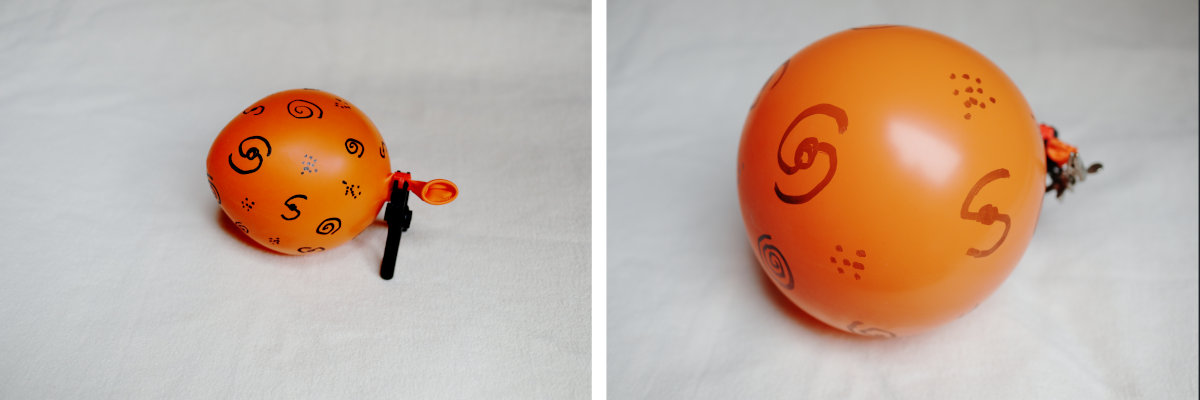}
\caption{Galaxies drawn onto a rubber balloon expand as the rubber balloon universe expands, even though, as bound systems, they should keep their size}
\label{GrowingGalaxies}
\end{center}
\end{figure}

One particular feature of an expanding substrate model can aggravate the problem of expanding bound systems: Whether on the latex strip or the rubber balloon's surface, when you paint or draw your galaxies onto the surface, they {\em are} going to expand. An example can be seen in Fig. \ref{GrowingGalaxies}. That is why glueing objects to the elastic substrate, or using small stickers, is advisable.

In the Milne model, with its emphasis on cosmic expansion as defined by initial conditions, not by some unusual dragging-along mechanism, bound objects are as unproblematic as in ordinary, classical physics. The statement that, say, the Solar system is bound implies suitable initial conditions, and in particular a certain initial velocity for the planets, at speeds below the system's escape velocity. The statement that galaxies are participating in the Hubble flow implies a different set of initial conditions. 

Analyses of bound systems in the full framework of FLRW spacetimes is more similar to the explanation within the Milne model than that of expanding substrate models. The simplest such calculations, choosing an almost Newtonian coordinate system tied to a galaxy in the Hubble flow, and analyzing the consequences, show that what happens can indeed be approximated as an equilibrium of forces -- but just as in the case of the tethered galaxy, the influence of cosmic expansion comes purely from the accelerating or decelerating part ($\ddot{t}$); there is no ``dragging along,'' no contribution that comes directly from a first-order change in the cosmic scale factor \citep{CooperstockEtAl1998,Price2012,CarreraGiulini2006,FaraoniJacques2007,Giulini2014}.

\subsection{Cosmological redshift}
\label{CosmoRedshift}

\begin{figure}[htbp]
\begin{center}
\includegraphics[width=\textwidth]{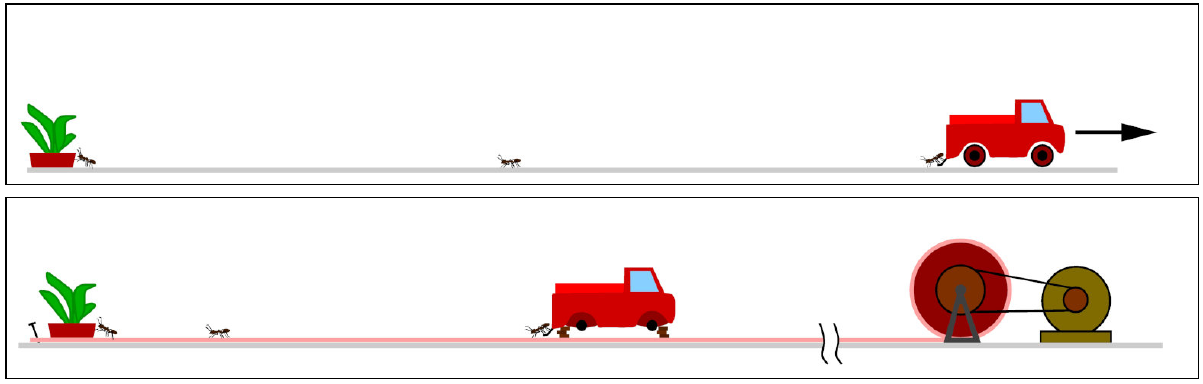}
\caption{Ants dropping out of a moving miniature car (top) at fixed frequency as a model of the Doppler effect, and ants dropping out of a stationary car fixed to an expanding rubber strip (bottom) as a model for the cosmological redshift. Images from \citet{Poessel2005} }
\label{CosmicAnts}
\end{center}
\end{figure}

Finally, there is the interpretation of the cosmological redshift in the two models. In the rubber strip model and the balloon model, light propagation can be pictured, thought-experiment style, by considering small creatures such as ants walking along the surface at constant speed as in Fig. \ref{CosmicAnts} \citep{Abell1984,Price2001,Poessel2005,Bahr2016}.

\begin{figure}[htbp]
\begin{center}
\includegraphics[width=\textwidth]{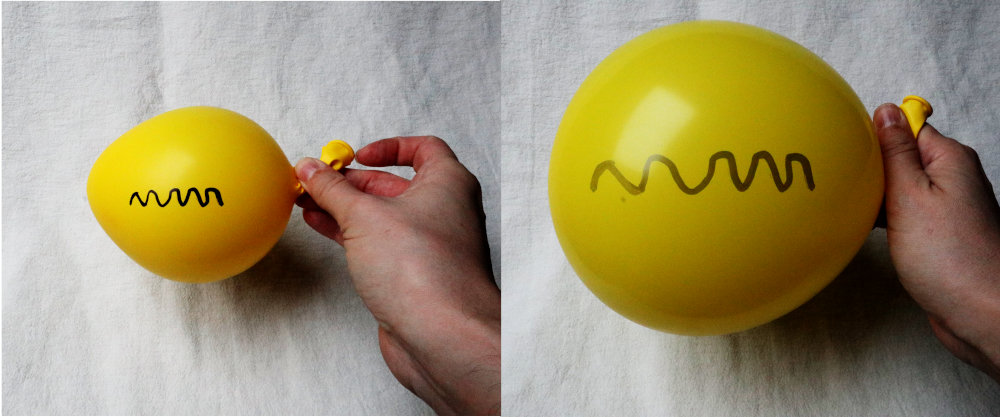}
\caption{A light wave that was drawn on a rubber balloon with a felt pen expands as the rubber balloon is inflated}
\label{Fig:LightWaveStretched}
\end{center}
\end{figure}

For emphasizing interpretations in which light is ``stretched'' by cosmic expansion, by light waves drawn or otherwise applied to a rubber balloon surface \citep{Winnenburg1998,Lotze1999,Scorza2008,SixtySymbolsRedshift,Moche2009}, as in Fig.~\ref{Fig:LightWaveStretched}. The appropriateness of this part of the model can be understood using the equivalence principle: Each infinitesimal surface region on the strip or the balloon represents locally flat spacetime, in which light propagates at the usual constant speed $c$. What changes is the way that the infinitely many locally flat regions are stitched together to form expanding space time, with its possibly curved space. This can be taken as far as calculating the propagation of light in a homogeneous, expanding universe, cf.\ section 6.3 in \cite{Poessel2017a}. A disadvantage of the balloon model is that the amplitude of the light-wave is stretched, as well. 

Both models suggest that distances between two consecutive light pulses -- two ants following each other, two adjacent maxima of the drawn-on light-wave -- increase in the same way as distances between galaxies in the Hubble flow: in proportion to the universal cosmic scale factor. That is indeed the case in FLRW models.

In the Milne model, the most direct interpretation of the cosmological redshift is as a special-relativistic Doppler shift. The relative velocity between two galaxies, which determines the redshift, is readily calculated by using private space (in which the Milne universe is a subset of Minkowski spacetime). This, too, mirrors a property of FLRW models, where the cosmological redshift has a well-defined interpretation in terms of a special-relativistic Doppler shift. In general, though, there is no globally flat reference frame in which four-vectors can be compared directly. There is, however, the general relativistic generalisation of a direct comparison: parallel transport of the distant galaxy's four-velocity along the light-like geodesic of the distant galaxy's light to our own galaxy. The relativistic relative speed obtained in this way, plugged into the special-relativistic Doppler formula, does indeed give the proper cosmological redshift \citep{Narlikar1994,Liebscher2007,BunnHogg2009,CookBurns2009,Kaya2011}.

In terms of redshift, then, the two different models lead to two alternative, but equally valid interpretations of the general cosmological redshift in FLRW models. The expanding substrate interpretation has the advantage of being directly visually accessible. For pupils who know about special relativity, the Doppler shift interpretation has the advantage of connecting directly to that simpler theory.

\subsection{Expansion on transparencies}

Another interesting physical model is specifically tailored to show how cosmic expansion does not single out any observer. Just because we in the Milky Way galaxy see all other galaxies moving away from us, their speeds proportional to their distances as per the Hubble relation, that does not mean we are special --- all other observers in the Hubble flow make similar observations. The model uses two transparencies and an overhead projector (activity H-6, ``Visualizing the expansion of space'' in \citealp{UAYF}, \citealp{Lotze1997}, \citealp{Lotze2002c} and activity H--9, ``Modeling the Expanding Universe'' in \citealp{UAYF2}), and has also been used in a demonstration video \citep{Minutephysics2013a}.

\begin{figure}[htbp]
\begin{center}
\includegraphics[width=0.4\textwidth]{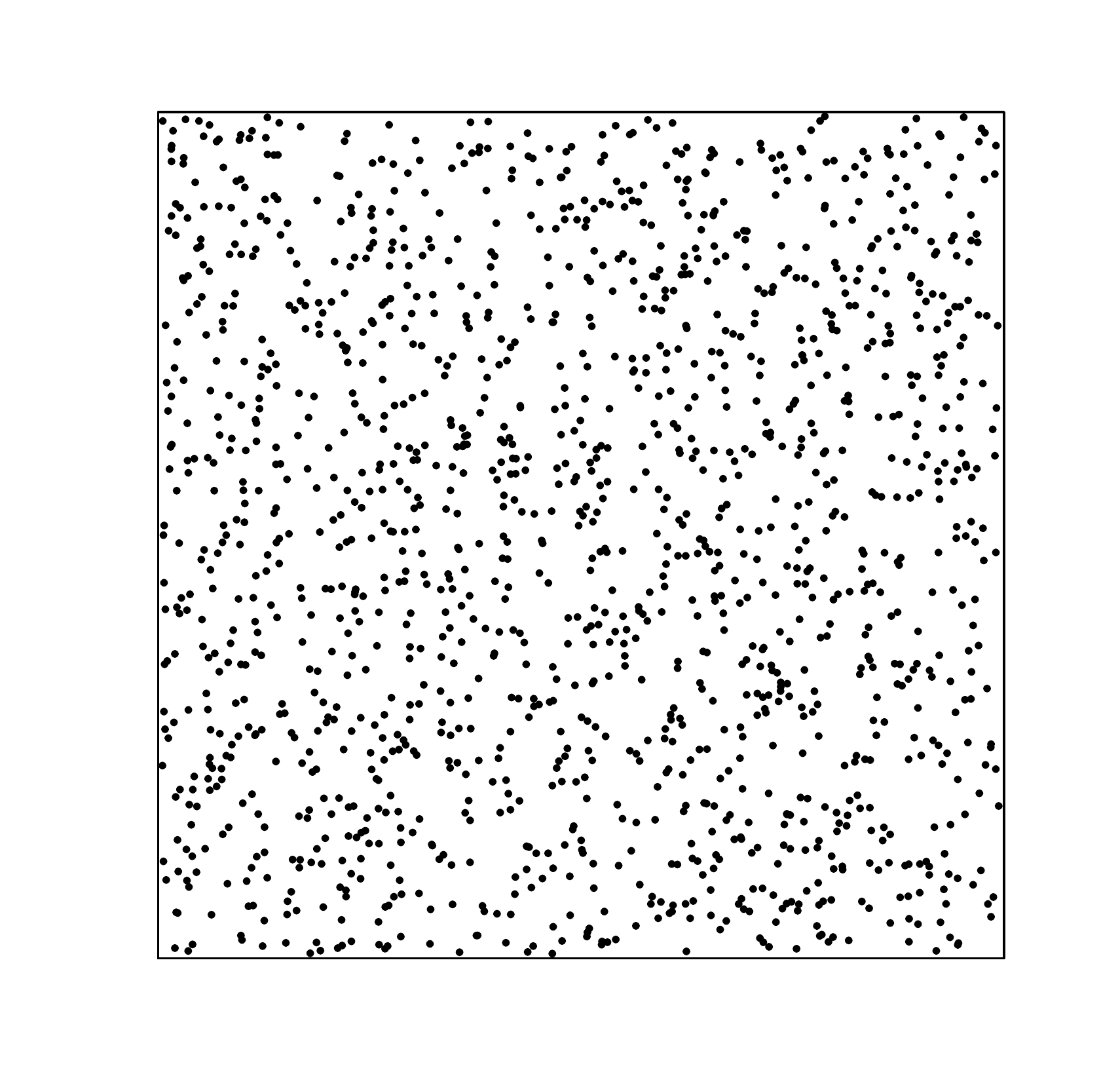} \hspace*{1.5em}
\includegraphics[width=0.4\textwidth]{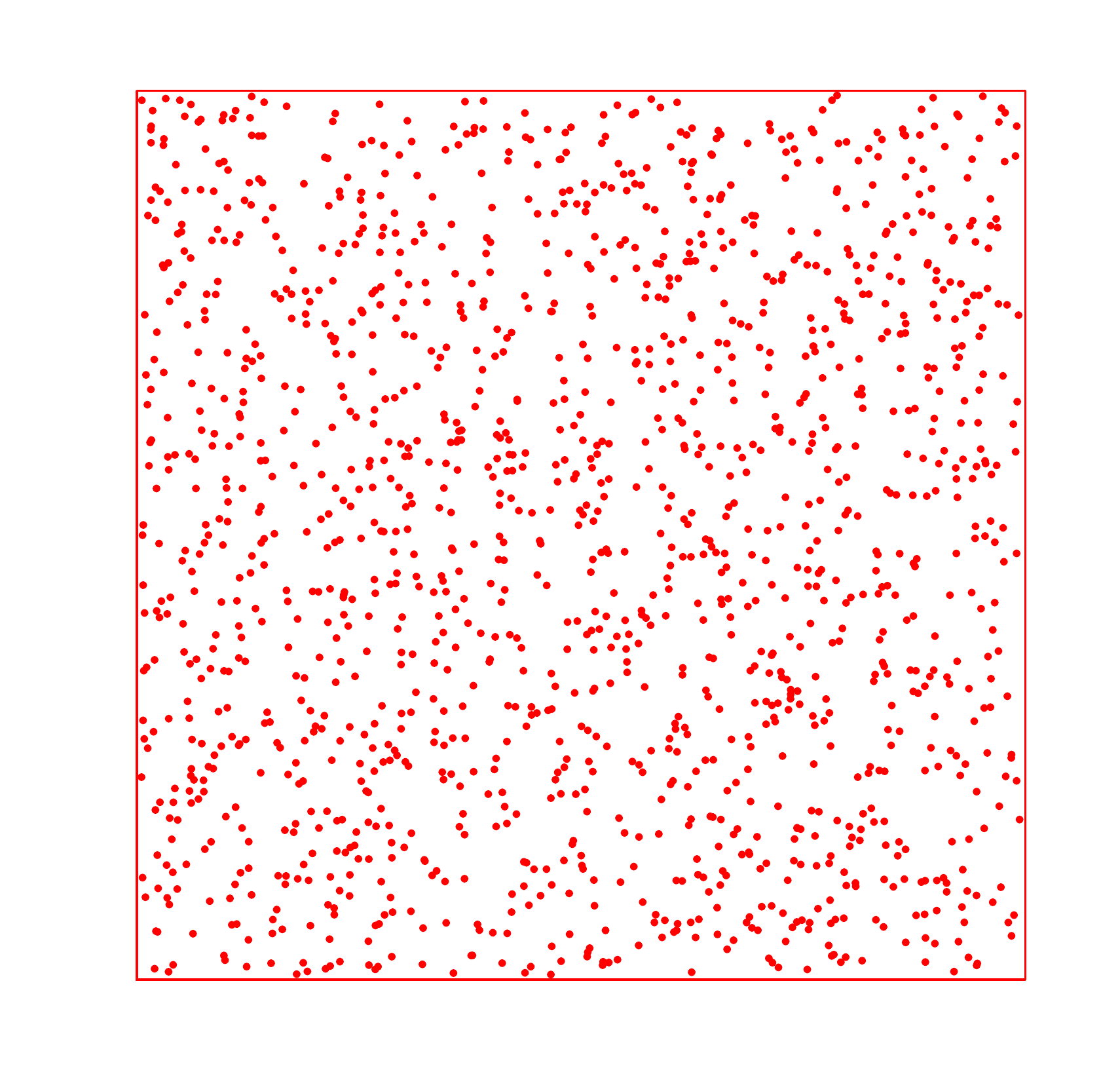}
\caption{The same pattern of 1500 points, printed on two transparencies. For the right transparency, the pattern is magnified by the factor 1.05.}
\label{Transparencies1}
\end{center}
\end{figure}

To begin with, create a two dimensional random pattern of points. Each point represents a galaxy. Print this pattern out on a transparency (left part of figure \ref{Transparencies1}). Then, print the same pattern out again, but at slightly different magnification, say with all distances changed by a factor 1.05 (right part of figure \ref{Transparencies1}).
The transparencies can either be viewed directly, on a white background, or projected using an overhead projector. Next, shift the transparencies relative to each other, changing the location where one galaxy-dot comes to lie directly below its counterpart on the other slide. In that particular configuration, the two slides represent distances from that particular galaxy at two different moments in time. As can be seen in figure \ref{Transparencies2}, the Hubble relation will be immediately visible:
\begin{figure}[htbp]
\begin{center}
\includegraphics[width=0.4\textwidth]{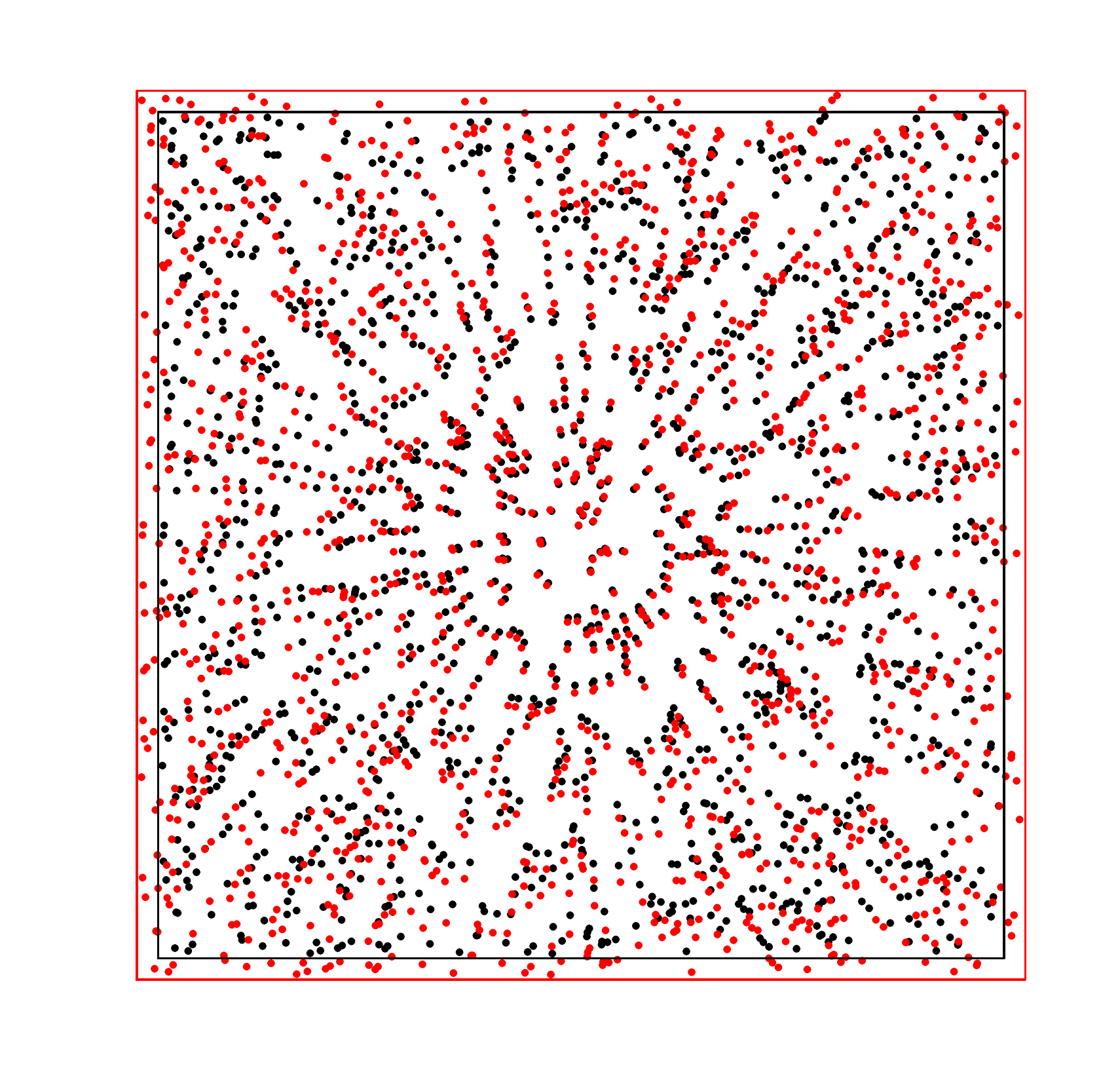} \hspace*{1.5em}
\includegraphics[width=0.4\textwidth]{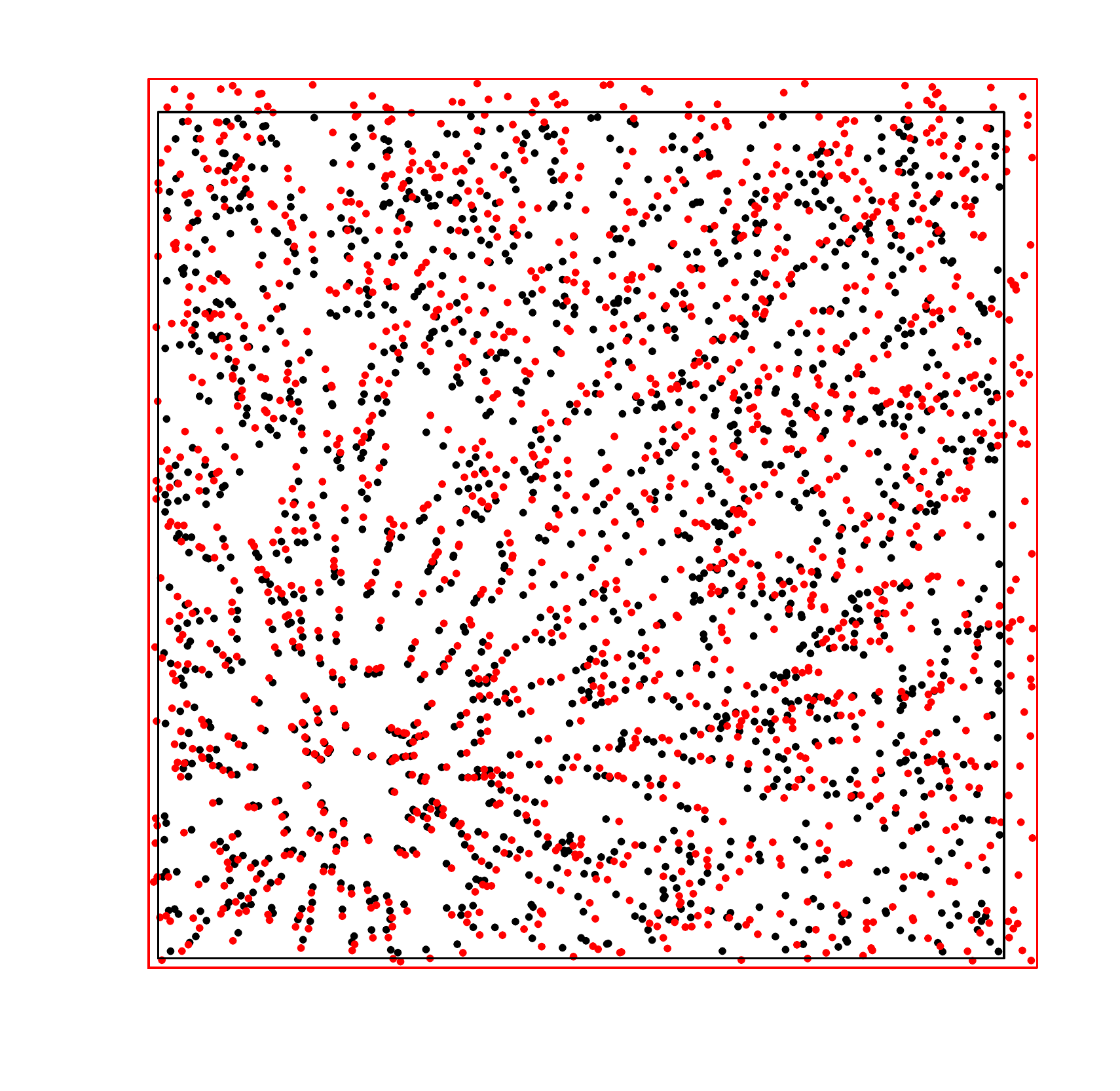}
\caption{The two transparencies from figure \ref{Transparencies1}, placed on top of each other, with different relative shifts}
\label{Transparencies2}
\end{center}
\end{figure}
at the coincidence point, there will be no relative shift between red and black dots by definition; the farther away one goes from the coincidence point, the greater the shift between black and red points --- corresponding to increasing shifts in distance between the first and second snapshots. 

By moving the coincidence point around, it is readily demonstrated that the Hubble relation holds for all observers in this expanding universe, not just for some privileged special observer. This can be a hands-on activity, where participants can explore for themselves how moving one of the transparencies changes the ``focus'' of the pattern, vividly demonstrating that, while the distances between us and all the distant galaxies are changing according to the Hubble-Lema\^itre relation, this does not make us special in any way.

\subsection{Newtonian calculations}
\label{SectionNewtonianCalculations}
\begin{figure}[htbp]
\begin{center}
\includegraphics[width=0.4\textwidth]{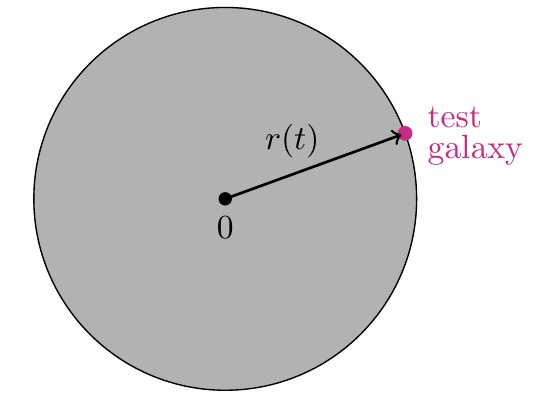}
\caption{Basic setup for analysing cosmic dynamics using a Newtonian approach. For analysing the acceleration of the test galaxy relative to the origin $0$, only the mass inside the grey sphere contributes}
\label{NewtonSphere}
\end{center}
\end{figure}

When it comes to cosmic dynamics, that is, the evolution of the cosmic scale factor $a(t)$ over time, Newtonian calculations provide a powerful and useful model. In Newtonian gravity, the gravitational force of a homogeneous spherical shell is zero for a test particle inside the shell, and the same as for a point particle in the shell's centre, with the shell's total mass, for a test particle outside. On this basis, glancing over possible problems with infinitely large shells, one can argue that in a homogeneous universe, the only force acting on a test galaxy when considering that galaxy's motion relative to the spatial origin is the gravitational attraction by the matter in a spherical volume around the origin, whose surface includes the location of the test galaxy, a scenario that is shown in figure \ref{NewtonSphere}.

In a homogeneous universe, expansion is governed by the scale factor on all scales. Hence, the Newtonian analysis can take place in an infinitesimal region around the spatial origin, which is customarily chosen to coincide with our own location in the cosmos. In this region, all galaxy speeds are much slower than the speed of light, and the Newtonian approximation provides an arbitrarily precise description of scale factor dynamics. The only contribution missing in this model is the pressure contribution to the active gravitational mass. If this additional term is inserted by hand, the full Friedmann equations can be derived.

Historically, some of the first such calculations were viewed as a possible replacement for the general-relativistic cosmological models \citep{Milne1934,McCreaMilne1934,McCrea1955}. Such models, in an appropriate generalised formalism, have also been proposed as a generalisation of Friedmann cosmologies \citep{Tipler1996a,Tipler1996b}. In an educational context, they are typically used either to provide a derivation of the Friedmann equations without using the full Einstein field equations, as a helpful simplification or approximation, or as an intermediate state in the development of fully relativistic cosmic dynamics \citep{Lemaitre1934,CallanEtAl1965,Harrison1965,Sciama1971,Weinberg1972,Berry1976,Burke1980,Liddle1988,Coles1995,Lotze1998,Lotze2002a,Lotze2002b,Jordan2005,CarrollOstlie2007a,CarrollOstlie2007b,Huggins2013,Wells2014,Poessel2017a}.
                                             
\section{Gravitational waves}
\label{GWSection}

With the first direct detection of gravitational waves, that particular subfield of general relativity has become much more prominent in the public eye as well as in teaching at various levels. In the following, I give an overview of models for the properties of gravitational waves in section \ref{GWProperties}, of different kinds of models for the basic principles of gravitational wave detection in sections \ref{InterferometricDetector}--\ref{LightMoving}, before addressing the issue of noise in such detections in section \ref{GWDetection}.

\subsection{Gravitational wave properties}
\label{GWProperties}

The most elementary text book illustration of the action of gravitational waves depicts a ring of test particles, and traces how the distances of the ring particles from the center (or center particle) change over time as a gravitational wave passes orthogonally through the plane of the ring, creating the characteristic quadrupole pattern where stretched distances in one direction coincide with shrunk distances in the orthogonal direction within the plane.

\begin{figure}[htbp]
\begin{center}
\includegraphics[width=0.8\textwidth]{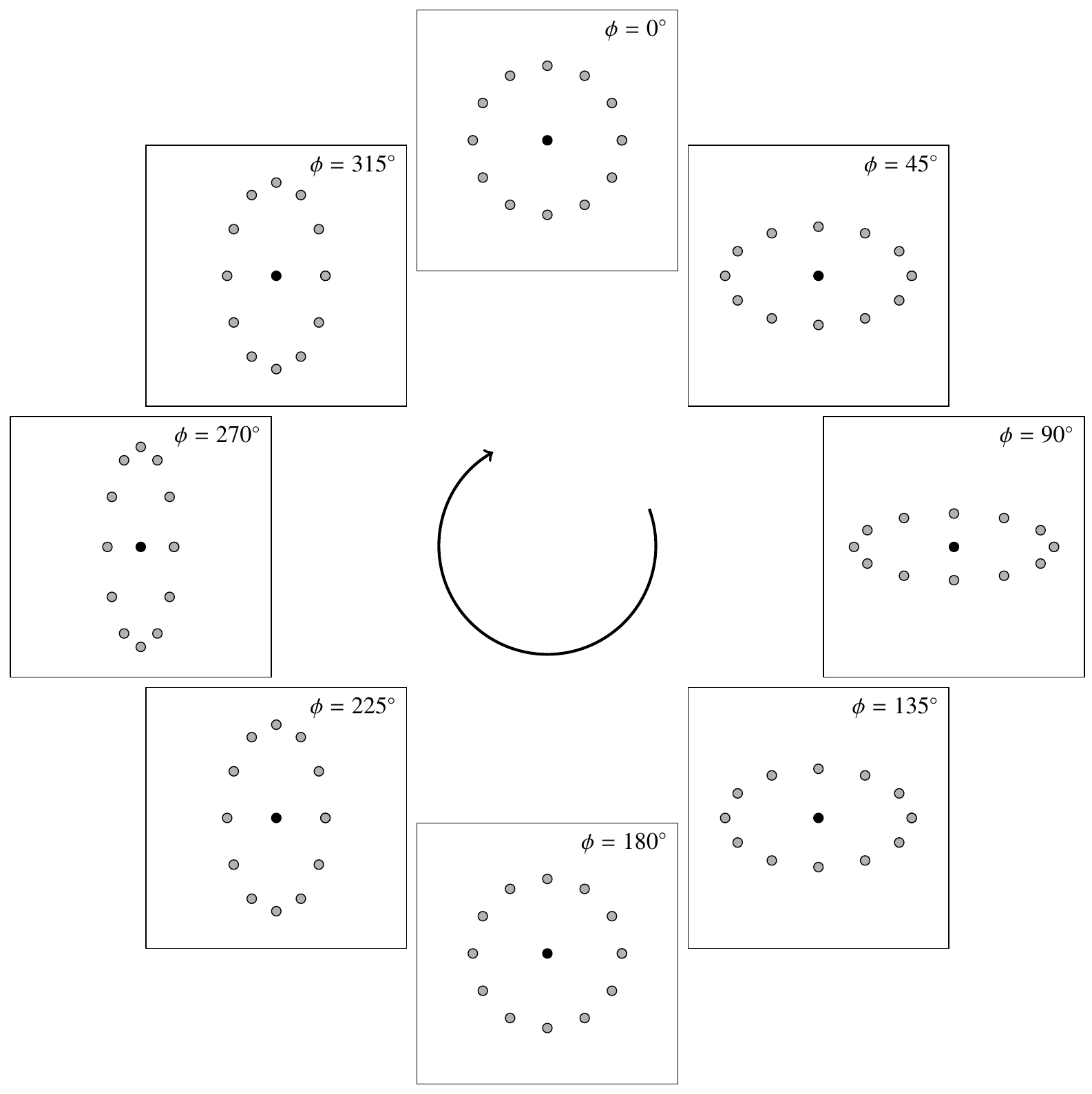}
\caption{Quadrupole distortion pattern for a linearly polarized gravitational wave passing through a ring of freely floating test particles. Fig.\ 5 from \citet{Poessel2016b} }
\label{GWDistortionPattern}
\end{center}
\end{figure}

The quadrupole distortion pattern, stretching distances in one direction while squeezing them in an orthogonal direction, is one of the most common images used to illustrate the action of gravitational waves on free-floating test particles. It is most commonly illustrated using a ring of such test particles, by showing proper distance at constant times as measured by the standard time coordinate for a linearized gravitational wave described in TT gauge. An example is shown in figure \ref{GWDistortionPattern}.\footnote{An animated version can be found as part of the animation \href{https://youtu.be/AXgljRvI_Tg}{https://youtu.be/AXgljRvI\_Tg} }
The center of the ring, here marked by the test particle shown as a black dot, is kept fixed, and the changing distances of the ring particles from the center particle are shown.

\begin{figure}[htbp]
\begin{center}
\includegraphics[width=0.8\textwidth]{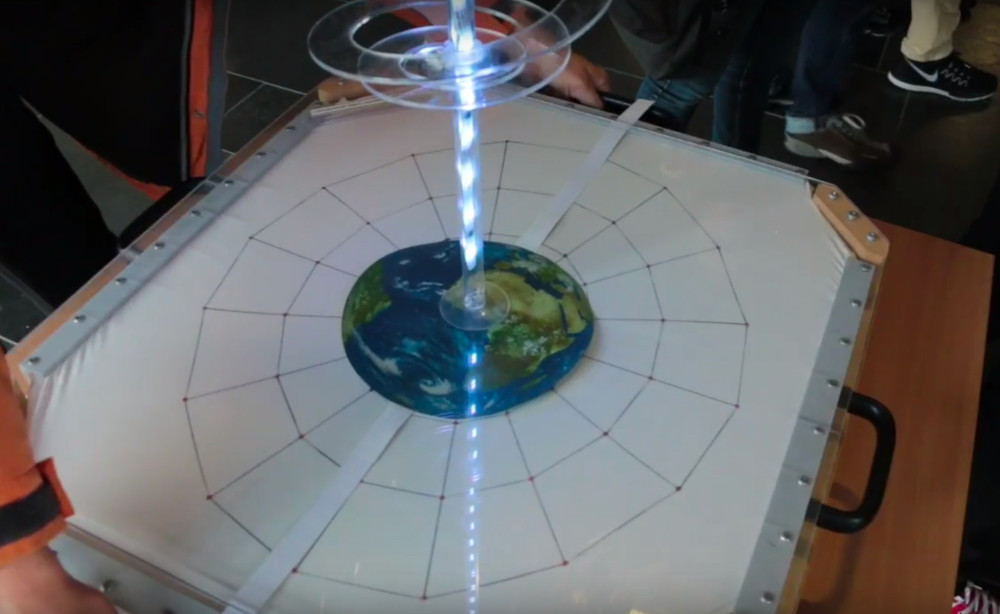}
\caption{Hands-on elastic sheet model of the action of a gravitational wave on a ring of test particles surrounding the Earth (below) coupled with a symbolic representation of the propagating wave itself. The rubber sheet can be stretched or shrunk using the handles on the side; an underlying mechanism ensures that the distortion will follow a quadrupole pattern. A brief video showing the exhibit in action can be seen on \href{https://youtu.be/siwMvNQAr38}{https://youtu.be/siwMvNQAr38}. The model is part of the exhibition {\em Einstein inside}, \href{http://www.einstein-inside.de/}{http://www.einstein-inside.de/}}
\label{gw-einstein-inside}
\end{center}
\end{figure}

There are several physical realisations of this scenario. One of them again represents space by an elastic sheet (as in section \ref{ElasticSheet}) and then applies the appropriate quadrupole distortions. An example from an exhibition is shown in figure \ref{gw-einstein-inside}.
A variety of hands-on demonstrations of quadrupole distortion is presented by \citet{Kraus2016a}, and shown in figure \ref{KrausZahnFigure}:
\begin{figure}[htbp]
\begin{center}
\includegraphics[width=0.9\textwidth]{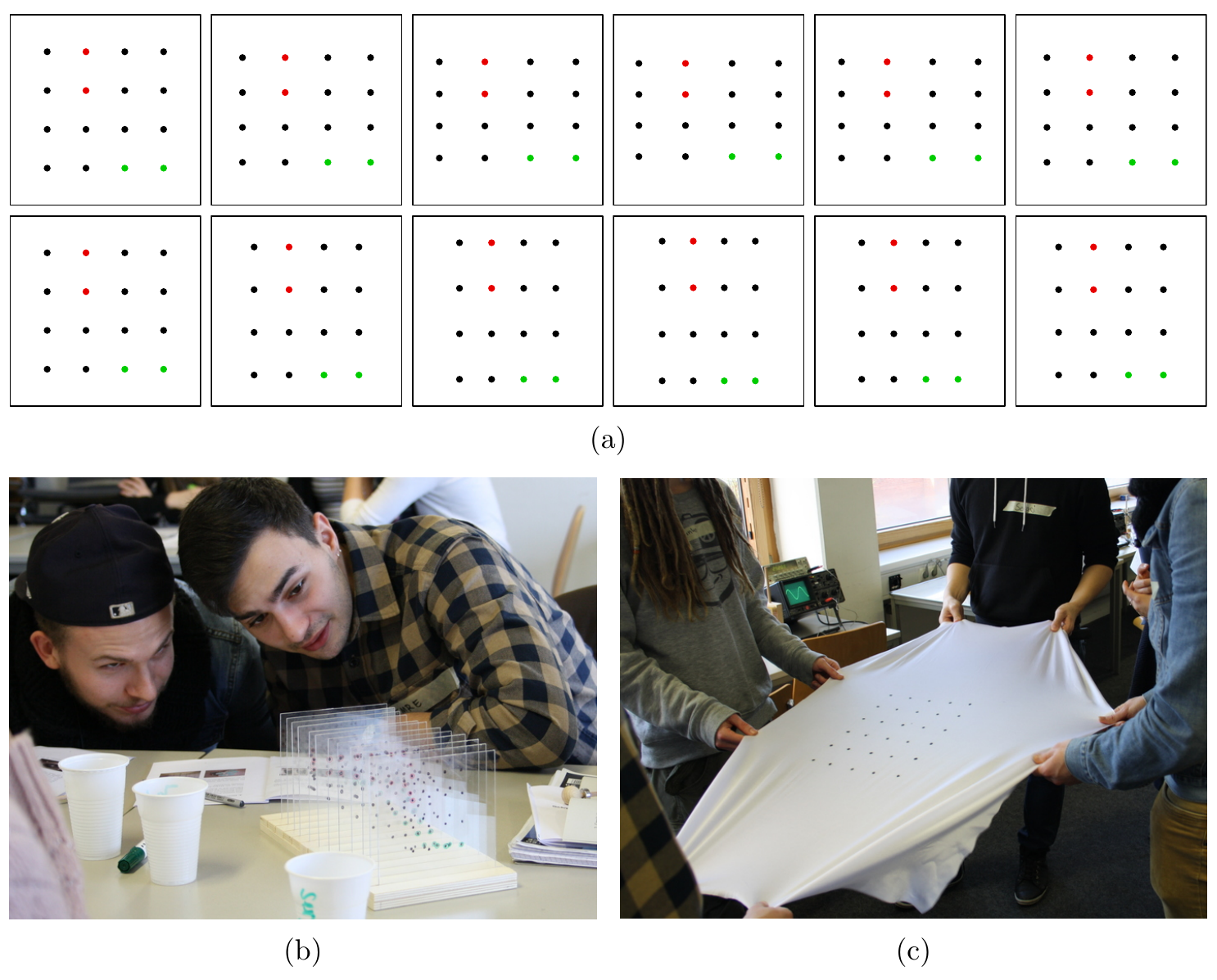}
\caption{Visualizations of quadrupole distortion as used in the ``Sch\"ulerlabor  Raumzeitwerkstat''  (student  lab  on  relativistic
physics) at Hildesheim University (Germany), used to teach pupils in grades 9 to 13 (ages 15 to 19 years) about gravitational waves. Fig.\ 3 in \citet{Kraus2016a}, used by permission.
}
\label{KrausZahnFigure}
\end{center}
\end{figure}
In this case, the test particles are arranged into a square pattern (with some test particles marked in color for ease of identification). The sequence of distortions can be shown in the form of snapshot images as in figure \ref{KrausZahnFigure}a. Transparent versions of these images on perspex can be arranged in sequence to show the wave nature of the effect, as in figure \ref{KrausZahnFigure}b. Finally, by recreating the point pattern on an elastic sheet and having pupils pull and release in a coordinated manner as in figure \ref{KrausZahnFigure}c, the time evolution of the distortion can be recreated. 

Several visualisations are provided by the {\em Stretch and Squash} app, developed by the non-profit company Laser Lab for the gravitational wave group at the University of Birmingham. The app allows users to show the typical distortions caused by sinusoidal gravitational waves with various polarisations, or by a chirp signal, for a variety of background images, from the classic ring of free-floating particles to representations of detectors and even a live deformation of the current image seen by the computer's or tablet's own camera.\footnote{The app is available from \href{https://www.laserlabs.org/stretchandsquash.php}{https://www.laserlabs.org/stretchandsquash.php} (last accessed on May 11, 2018.} A caveat is that, in reality, only freely floating particles will exhibit that kind of distortion. For bound systems, the distortions will be reduced severely in most cases, although the modification is frequency-dependent and there might even be resonance at certain frequencies (an effect that the so-called resonant detectors seek to exploit). Showing the same kind of distortion for a ring of freely floating particles, for a schematic detector whose components are suspended as pendulums in part to allow them to react to the gravitational wave as free particles would (at least in the directions in which deformations can be measured by the detection mechanism), and for bound systems (such as most of what the live camera will show) blurs this distinction.

\begin{figure}[htbp]
\begin{center}
\includegraphics[width=\textwidth]{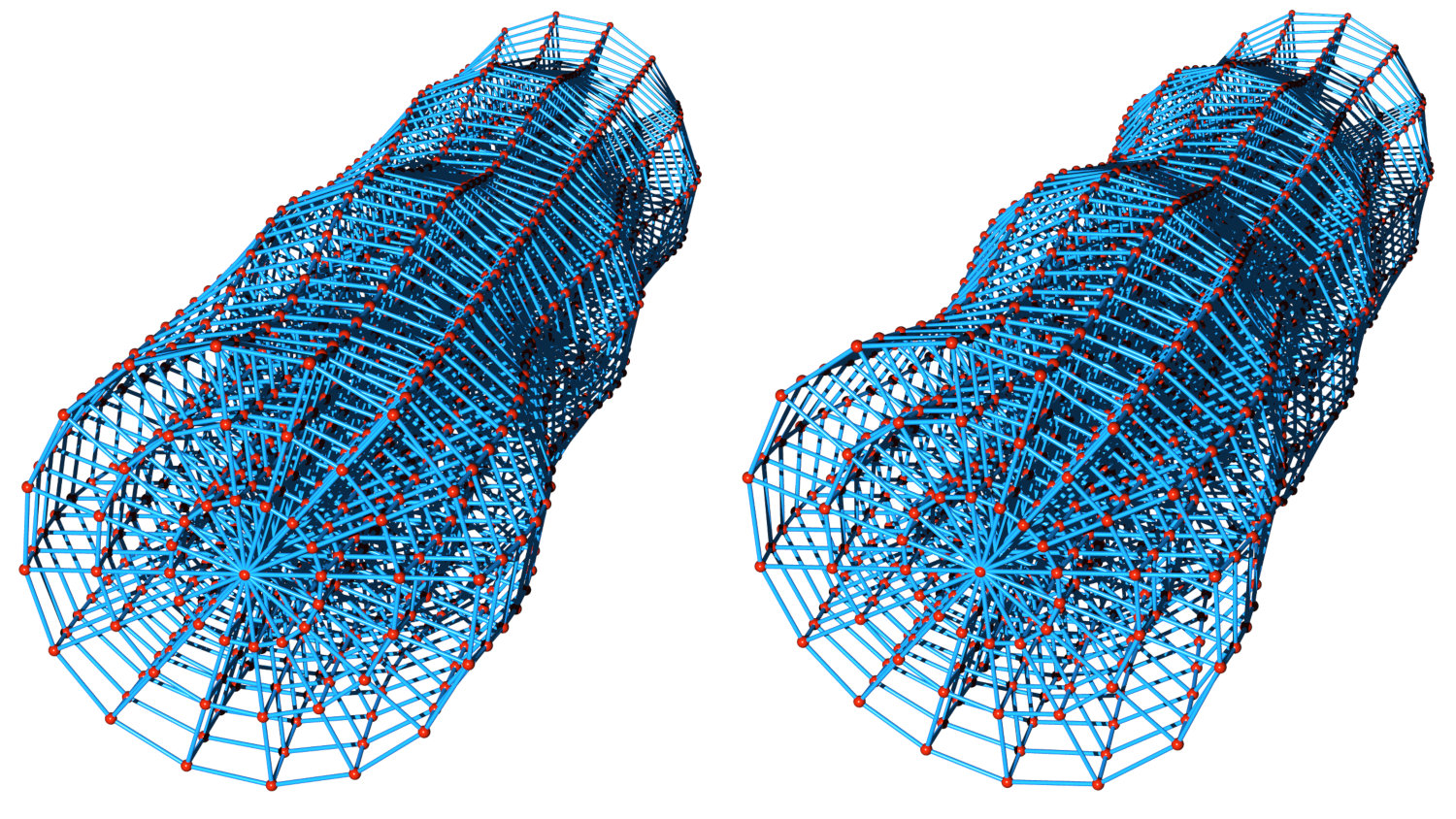}
\caption{Snapshot of animations of a linearly polarized (left) and a circularly polarized (right) gravitational wave, showing local quadrupole distortions lining up to make a wave that propagates through space. The animation itself can be seen at \href{https://youtu.be/AXgljRvI_Tg}{https://youtu.be/AXgljRvI\_Tg}}
\label{GWWaveAnimation}
\end{center}
\end{figure}

Another visualisation shows the wave nature of gravitational waves, that is, the link between oscillation and spatial structure, as follows: Consider a collection of rings lined up in the direction of propagation of the wave. As the wave passes, each ring will show a different phase of quadrupole distortion. Taken together, the distorted rings show how the distortion states are propagating through space, cf. figure \ref{GWWaveAnimation}.

The surface defined by the lined-up distorted rings can also be  turned into a physical, three-dimensional model. Files for 3D-printing the surfaces corresponding to the first detected gravitational wave signals can be found on the popular 3D model site Thingiverse, although their twisted shape means they are not that easy to print.\footnote{The models were contributed by Thingiverse user Paul Klinger (user name: Almoturg) and can be downloaded under \href{https://www.thingiverse.com/thing:2886889}{https://www.thingiverse.com/thing:2886889}}

A different application for an elastic sheet model is to demonstrate not the quadrupole distortion, but the outwards travel of the distortion phases (most notably, the distortion maxima) for a binary system as one of the simplest gravitational wave sources, creating a characteristic expanding spiral pattern. However, given the typical speeds of propagation of disturbances on an elastic sheet, the rotation frequency of the binary system will need to be rather high, and the spiral wave pattern will not be visible with the naked eye, only via a high-speed or stroboscopic recording \citep{Overduin2018}.

\begin{figure}[htbp]
\begin{center}
\includegraphics[width=0.9\textwidth]{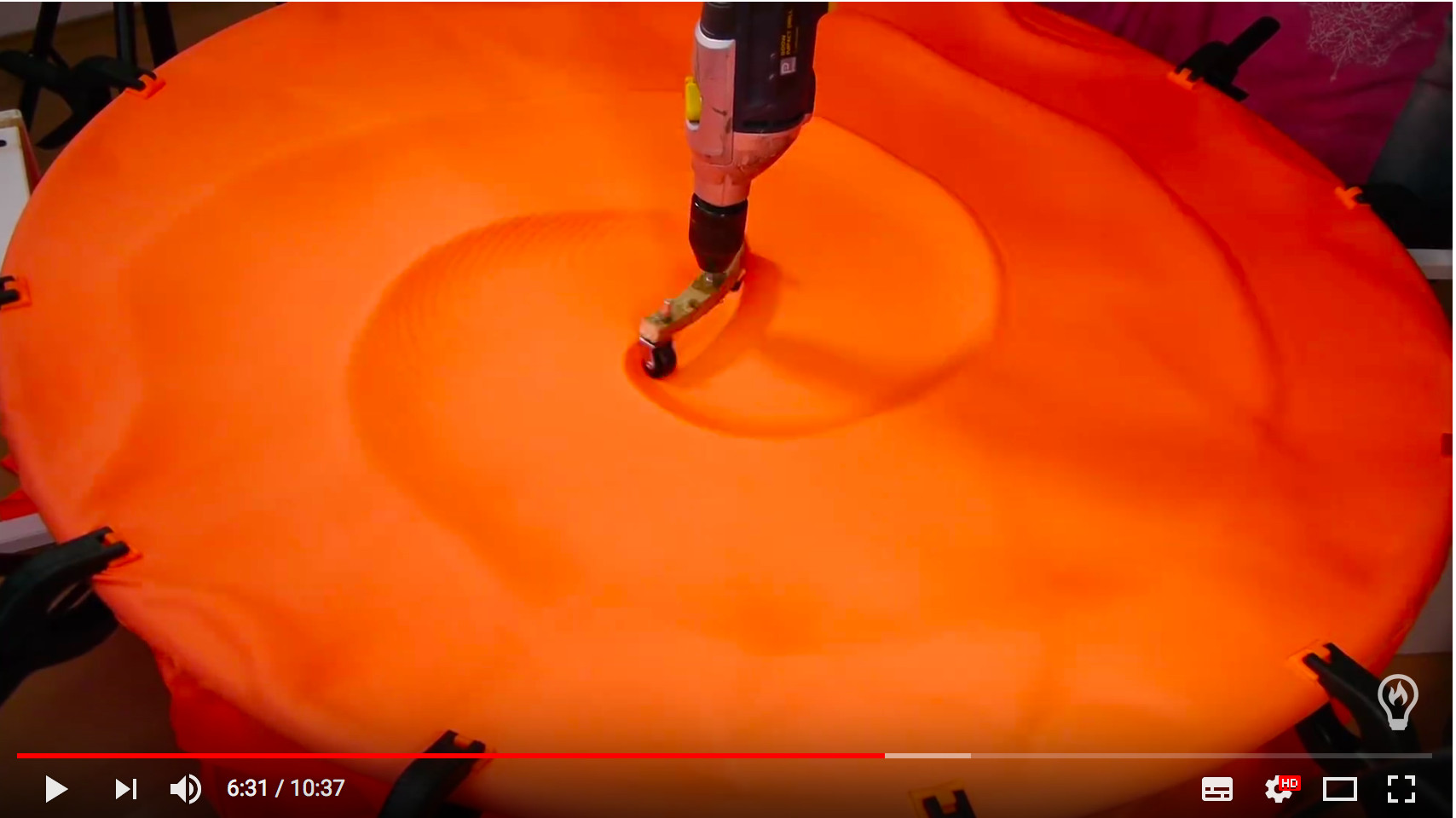}
\caption{Snapshot from the video ``Visualizing gravitational waves'' by Steven Mould, \href{https://youtu.be/dw7U3BYMs4U}{https://youtu.be/dw7U3BYMs4U}}
\label{MouldFigure}
\end{center}
\end{figure}

Fig.\ \ref{MouldFigure} shows a still from an impressive realisation of this setup by science communicator Steve Mould.\footnote{Several  different realisations can be watched on YouTube:\\[0.5em]
\bgroup
\renewcommand{\arraystretch}{1.2}
\begin{tabular}{|l|r|l|}
\hline
Steve Mould & 9 Aug 2016 &  \href{https://youtu.be/dw7U3BYMs4U}{https://youtu.be/dw7U3BYMs4U}\\\hline
LIGO-Caltech & 26 Aug 2016 & \href{https://www.youtube.com/watch?v=YfSyhcFu_MM}{https://www.youtube.com/watch?v=YfSyhcFu\_MM}\\\hline
Gottlieb Planetarium & 15 Oct 2016 & \href{https://www.youtube.com/watch?v=wnWmGr_523s}{https://www.youtube.com/watch?v=wnWmGr\_523s}\\\hline
Royal Obs. Edinburgh & 20 Jan 2017 & \href{https://www.youtube.com/watch?v=T6B1U-5oAp4}{https://www.youtube.com/watch?v=T6B1U-5oAp4}\\\hline
\end{tabular}
\egroup
\vspace*{0.5em}
} 

On a more elementary level, aiming at a qualitative understanding, there are simpler models that are particularly aimed at audiences unfamiliar with the properties of physical waves (beyond everyday knowledge). For such an audience, one needs to introduce the concept of a wave in general, as well as gravitational waves as a specific case. One possible analogy that can serve to illustrate wave properties, wave propagation, and noise is with water waves.\footnote{Cf. this outreach video produced by Cardiff University in and around a swimming pool, \href{https://youtu.be/Lcxt097G4Ps}{https://youtu.be/Lcxt097G4Ps}} Wave properties such as wavelength and period of oscillation, and the difference between longitudinal and transversal waves, can be demonstrated using standard tools used in physics teaching, such as a giant slinky \citep{Stuver2008}.

Another model for gravitational waves, which has been employed both in a popular-science setting \citep{Bartusiak2000,Levin2011,Poessel2012,Bartusiak2017} and in an educational setting \citep{Kwon2017}, is based on the analogy between sound waves and gravitational waves. While the model does not reproduce certain basic wave properties of gravitational waves --- namely the action on test particles and the transversality ---, it does reproduce key features of the way gravitational waves are produced, and for the information astronomers hope to extract from them. When astronomers observe light and other forms of electromagnetic radiation from distant objects, each minute surface region of an object produces its own light, and astronomical observations (given suitable spatial resolution) can resolve the resulting radiation into a picture of the object's visible surface, showing structure and details. 

The wave form of a sound wave produced, say, by a violin, on the other hand, is a result of the way the whole of the violin's body oscillates. We hear the resounding sound as a whole; we do not hear separate sound components that would allow us to distinguish between small regions of an oscillating violin. Analogously, gravitational waves are the overall result of large-scale accelerated motion of mass. Their properties tell us about the large-scale structure of that motion (more concretely, about the quadrupole moment, and how it changes). And similar to microphones, or our ears, in those cases that are of current astronomical interest at least, such as merging black holes, gravitational wave detectors do not give us a detailed picture of separate partial gravitational wave sources. They record the overall wave pattern, although, similar to our two ears, and to an arrangement of multiple microphones, they can reconstruct the direction the waves are coming from. 

An additional feature of this model is that the frequencies of gravitational waves for typical source events for today's ground-based detectors, namely for merging black holes and neutron stars, lie in the same range as the sound wave frequencies that are audible for human ears. These gravitational wave signals --- as well as the associated noise --- can be sonified, made audible, illustrating both their generic ``chirp'' form of rising frequency and amplitude as well as the deleterious influence of detector noise.\footnote{A collection of such sonifications, as well as the associated explanations, can be found on \href{https://www.soundsofspacetime.org/}{https://www.soundsofspacetime.org/}, hosted by Montclair State University, NJ.}

Finally, electromagnetic waves themselves can be used as a model for gravitational waves, or alternatively as a model with which to compare and contrast gravitational waves. Naturally, this works particularly well for audiences that already know about electromagnetism, but the analogy can also be found in books aimed at a more general audience (ch.\ 5 in \citealt{Blair1997}, \citealt{Giulini2017}). Together with versions of Newtonian gravity modified to include finite propagation speed, electromagnetism provides a basis for {\em mathematical} models of gravitational waves, both through comparing and contrasting the two physical phenomena and through taking the equations of electromagnetism as a role model for gravitational equations \citep{KalckarUlfbeck1974,Berry1976,Campbell1976a,Campbell1976b,Davies1980,Schutz1984,Lotze2000a,Lotze2016,Hilborn2018}. These models allow for actual calculations as well as the derivation of some of the relations linking the properties of a gravitational wave source with its gravitational wave luminosity. A sub-genre are calculations aimed at a simplified derivation of the quadrupole formula, which combine dimensional analysis with some assumptions in analogy with electromagnetic waves such as assumptions about the multipole nature of the radiation \citep{Bracco2009,Mathur2017}. An interesting variation of the gravitational/electromagnetic wave analogy involves an exploration of a visual model, namely force field lines, and analogous concepts for gravitational fields \citep{Price2013}. For the detection of gravitational wave using pulsar timing arrays, the basic principle has been explained using either electromagnetic or sound waves as a model \citep{Jenet2015}.

\subsection{Interferometric gravitational wave detectors: the basics}
\label{InterferometricDetector}

The groundbreaking first direct detection of gravitational waves in 2015 made use of an interferometric gravitational wave detector. In this section, I will focus on conceptual models and visualizations of the basic detection principle of detectors such as LIGO or Virgo; model experiments and models for the noise in such detectors are treated further on in section \ref{GWDetection}. A simple conceptual model of an interferometric gravitational wave detector is the {\em Michelson interferometer} \citep{Spetz1984}. In its most basic form, this optical setup can be found in the curriculum of introductory physics classes. A schematic set-up is shown in figure \ref{MichelsonBasic}:
\begin{figure}[htbp]
\begin{center}
\includegraphics[width=0.5\textwidth]{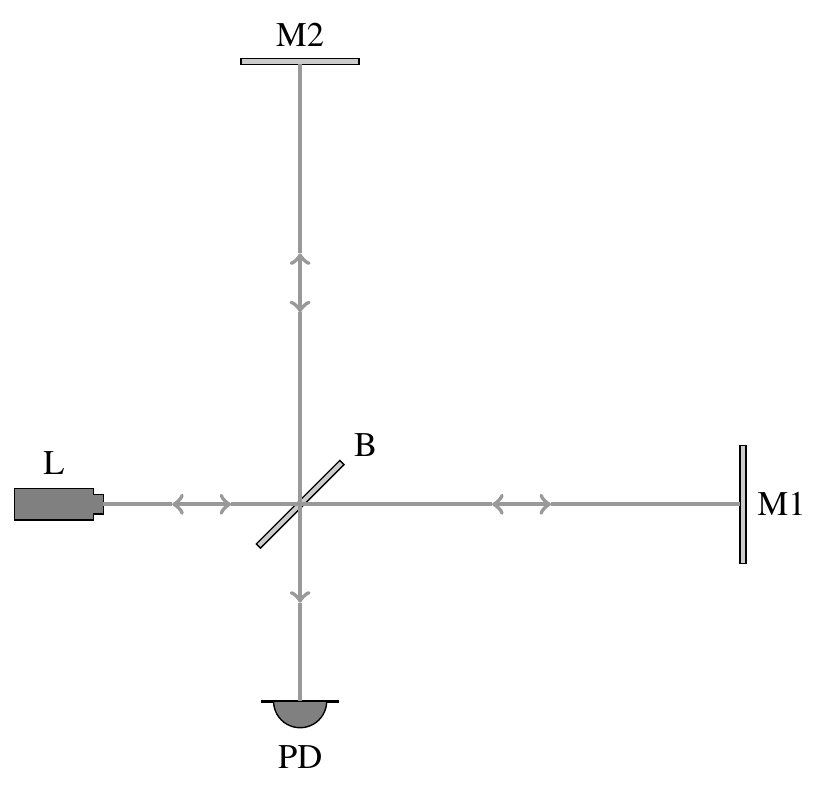}
\caption{Basic elements of a Michelson interferometer. Fig.\ 6 from \citet{Poessel2016b} }
\label{MichelsonBasic}
\end{center}
\end{figure}
Monochromatic light from a laser, modelled as a plane wave, travels to the beam-splitter B. Half the light travels on towards the mirror M1 on the right, the other half to the mirror M2 on top. After reflection, both components arrive back at B. Half of each component travels back to L, while the other half travels on to the photodetector PD. 

The key part is that the light travelling towards PD has two components, linearly superimposed: light that has travelled to M1 and back, and light that has travelled to M2 and back. The component waves have the same amplitude, but if the distance $L_1\equiv \overline{BM_1}$ is different from the distance $L_2\equiv \overline{BM_2}$, there will be a phase difference between the two as they arrive at PD, which means that the amplitude of the light arriving at PD is modulated by a factor
\be
\cos(2\pi [L_2-L_1]/\lambda_L),
\ee
with $\lambda_L$ the wavelength of the light. Clearly, a relative change in the length of the two arms will lead to a change in amplitude for light reaching the photodetector. Since the wavelength for visible or near-infrared light, as commonly used in such interferometers, is rather small, such an interferometer reacts sensitively to changes in $L_2-L_1$.

The starting point for understanding the basic working principle of an interferometric gravitational detector is to imagine that the beam splitter and the two mirrors M1 and M2 (and possibly the other components, as well), react to a gravitational wave in the same manner as free test particles. Let us take the beam splitter as our spatial origin. Then a gravitational wave passing through the detector perpendicularly to the plane of the basic set-up shown in \ref{MichelsonBasic} would change the distances between the beam splitter and M1 and M2 respectively just like it changes the distances in the diagram \ref{GWDistortionPattern}, with alternate stretching and squeezing of distances. 

A modern interferometric detector is adjusted so that (almost) no light reaches the photodetector in the absence of a gravitational wave. To that end, the length difference $L_2-L_1$ needs to be chosen in just the right way for destructive interference to occur for the light waves travelling from the beam splitter in the direction of the photodetector. The simplest version of the detection principle is, then, that a passing gravitational wave will change the two lengths $L_1$ and $L_2$. In the most straightforward case, one length is stretched while the other is shortened. It is straightforward to see that such a change would upset the specialised arrangement that kept the photodetector dark. In this sense, light falling on the photodetector indicates the presence of a gravitational wave.

\begin{figure}[htbp]
\begin{center}
\includegraphics[width=0.9\textwidth]{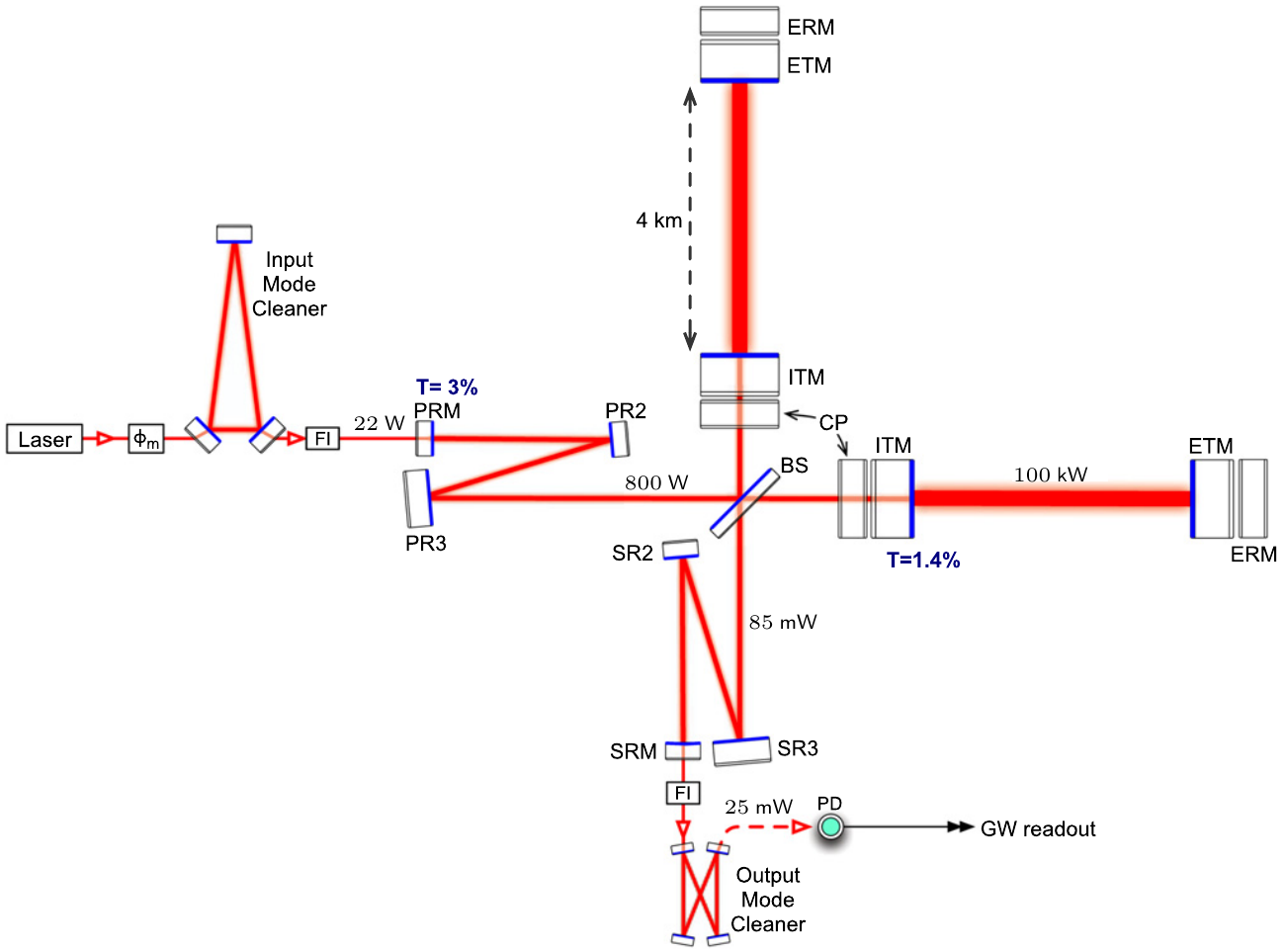}
\caption{Optical layout of advanced LIGO in the configuration that detected the very first gravitational wave signal. Image: LIGO Laboratory, used by permission.}
\label{LIGOLayout}
\end{center}
\end{figure}

I will go into more detail about how an interferometric detector is influenced by a gravitational wave in sections \ref{LightRuler} and \ref{LightMoving}. In this section, I will talk about the more general aspects of the simple Michelson interferometer as a model for an interferometric gravitational wave detector. A number of simplifications which are necessary for the transition from a real interferometric gravitational wave detector such as LIGO to a basic Michelson interferometer are readily visible by comparing figure \ref{MichelsonBasic} with the optical layout of Advanced LIGO at the time of the first direct gravitational wave detection, in late 2015 \citep{LSC2015,Martynov2016}, which is shown in figure \ref{LIGOLayout}.
Pedagogical introductions to the optical systems involved can be found in \cite{Bond2016} and \cite{Saulson2017}. Even that layout is, of course, a simplification --- there are control and calibration systems with their own optical components that are left out. Also, the laser alone, marked here only by a box, consists of more than twice as many optical elements than those shown in figure \ref{LIGOLayout}. How do these additional elements affect the faithfulness of the simplified model?

The triangular contraptions marked as ``input mode cleaner'' and ``output mode cleaner'' in fact help make the simplified model more accurate: The input mode cleaner ensures that what goes into the Michelson interferometer is as close as possible to an idealized plane (TEM${}_{00}$ mode) electromagnetic wave. The output mode cleaner ensures the same for light exiting the interferometer towards the photodetector PD. In both cases, this means the simplified model that treats light as a plane wave is close to the actual description. Higher modes only need to be included in a much more comprehensive description, adding small corrections to the result. 

The inner mirrors ITM (input test mass) are tuned so as to make each of the arms a resonant Fabry-Perot interferometer between each ITM and the outlying ETM (end test mass). This introduces additional resonance conditions that need to be fulfilled to make each Fabry-Perot cavity work, and it leads to a substantial increase in the number of photons interacting with each test mass mirror, leading to a reduction of the effects of photon shot noise. But if we are interested in understanding merely the detection principle, not the noise budget, the net effect of the cavity on the detector sensitivity can be accounted for in a straightforward way: by increasing the arm length of the simple Michelson interferometer meant to represent the interferometric detector. 

In a Michelson interferometer adjusted so as to keep light from reaching the photodetector PD, energy conservation demands that as much light as is fed into the interferometer by the laser should also exit the interferometer in the direction of the laser, travelling from B towards L in figure \ref{MichelsonBasic}. A calculation that includes proper modelling of the beam splitter, namely including a suitable relative phase shift between transmitted and reflected light, confirms this \citep[section 2]{Bond2016}.\footnote{There has been an extensive discussion in the physics education literature about proper phase shifts of this kind, and their physical basis. 
Cf. \cite{Degiorgio1980}, the replies by \cite{Zeilinger1981} [noting where Degiorgio's argument only applies to symmetric beam splitters] and by \cite{Lai1985}.

A more general treatment can be found in \cite{Hamilton2000}, with the physics background, namely the generalised scattering of light by dielectric objects, explained in \cite{NietoVesperinas1986}. A more general discussion of symmetric stratifications, which shows the link with the phase shift $\pi$ in the case of total reflection, is given by \cite{Lekner1990}.} The power recycling mirror PRM, assisted by its henchmirrors PR2 and PR3, reflects most of this light back into the interferometer. The motivation behind this is, again, the mitigation of photon shot noise by increasing the laser power within the interferometer. For understanding the basic detection principle, this is of no consequence, and we can safely leave this feature out.

Another feature omitted in the simplified model is signal recycling. The signal recycling mirror SRM, assisted by SR2 and SR3, reflects some of the light trying to leave the interferometer in the direction of the photodetector. Since, in the standard configuration, (almost) no light tries to leave the detector in that way, except when a gravitational wave is passing through, that amounts to reflecting the signal --- the evidence for a gravitational wave --- back into the interferometer. At least in a rough approximation, this can be seen as exposing the light in question to the influence of the gravitational wave several times before it reaches the photodetector.
This turns out to increase detector sensitivity in a specific gravitational wavelength range, while decreasing sensitivity outside that range. This pattern of frequency-dependent sensitivity change is not a part of the simplified model. 

Last but not least, a real interferometric gravitational wave detector has an active control system. If the key elements --- notably the beam splitter and the end mirrors --- were suspended completely freely, then even with no gravitational wave present, the arm lengths would constantly change relative to each other. Lower-frequency disturbances in particular produce changes much larger than any passing gravitational wave. These disturbances are caused by minute vibrations and undulations of the ground, transmitted to the interferometer either through the sophisticated isolation system or directly via the force of gravity, which cannot be shielded at all. 

This is a crucial problem for the following reason. For very small length changes, the response of the interferometer, as measured by the changes of brightness of the light reaching the photodetector, is a linear measure for the change in $L_2-L_1$. Linearity allows researchers to make a Fourier analysis of the measured signal, separating different frequency ranges --- a crucial step for analysing a gravitational wave signal, given that a typical signal is overwhelmed by noise at very low frequencies below a few dozen Hertz, and at high frequencies beyond a few dozen kiloHertz, but detectable in the limited frequency range in between. Letting the low-frequency noise move the mirrors and beamsplitter will take the detector out of the zone of linear response, making Fourier decomposition impossible. 

The solution is to install a feedback system: As soon as a change of the light level at the photodetector indicates a change in $L_2-L_1$, that signal is used to control an actuator which restores the system to its undisturbed state. If that control system worked instantaneously and perfectly, all the information about length changes --- including the evidence for gravitational waves! --- would be contained in the control signal that was used to control the actuator. In reality, the information is contained in both in the control signal and in the error signal, the latter recording the remaining changes in light level at the photodetector. Changes above some specific frequency will be detectable only in the error signal; the specific high-frequency range where this happens will depend on the ``reaction speed'' of the control system.

A suitable calibration helps the physicists to reconstruct the relationship between length shifts, error signal and control signal. (One way for Advanced LIGO to do this is to move the end mirrors ever so slightly using radiation pressure.) The beauty of this setup is that the detector will have behaved linearly throughout, since any deviation from its undisturbed state will have been kept very small. Thanks to this linearity, the changes in $L_2-L_1$ can be reconstructed in a way that allows for Fourier decomposition, and thus for an extraction of the frequency range that is of interest in detecting gravitational waves. The simplified model is missing this feedback system, of course. 

Leaving out noise sources, and the (sophisticated) mitigation measures associated with specific kinds of noise, is a standard way of simplifying an experiment or instrument. Most of the simplifications that lead us from a realistic interferometric gravitational wave detector to a simple, freely suspended Michelson interferometer, are of this sort, and do not influence the basic reaction of the interferometer to a passing gravitational wave that the model is meant to help illustrate. The exception is signal recycling, which introduces a frequency-dependent response the model is lacking. This is where the model departs, on a fundamental level, from the way a real detector works. But even in that case, if you accept the simplified description that signal recycling amounts to exposing the same light to the same gravitational wave, only longer (or, alternatively, several times in a row), the underlying detection principle is the same as for the Michelson interferometer.

\subsection{Light as a ruler}
\label{LightRuler}

So how does a Michelson interferometer detect the influence of a gravitational wave? The simplest model assumes that a gravitational wave acts just as any other influence which causes a length change in such an interferometer: by shifting the positions of one or both mirrors relative to the beamsplitter. 

A great advantage of this model is that a physical version is not too hard to build. A simple and solid Michelson interferometer can be constructed for about \$200, using a laser and simple optical elements, and it can be used to give a vivid demonstration of such an instrument's sensitivity --- e.g. very slight pressure on one of the mirrors, or even just pressure on the plate on which the optical elements are mounted, is enough to cause a significant change in the interference pattern \citep{Farr2012}. The LIGO Scientific Collaboration provides two possible designs for such do-it-yourself interferometers \citep{Douglas2009,Ingram2014}. A particularly simple toy version involves no more than a green laser pointer, aluminium foil mirrors and a CD case (flat plastic sheet) beam-splitter \citep{Choudhary2018}.

It should be noticed that, in contrast with real gravitational wave detectors and also in contrast with the simplest mathematical description in terms of plane waves, light in these simple interferometers is propagating as part of a spherical wave --- the corresponding light beams not in parallel, but slightly divergent. This creates a characteristic ring-shaped interference pattern at the photodetector (which, in the simplest setup, is replaced by a projection screen).

\begin{figure}[htbp]
\begin{center}
\includegraphics[width=0.9\textwidth]{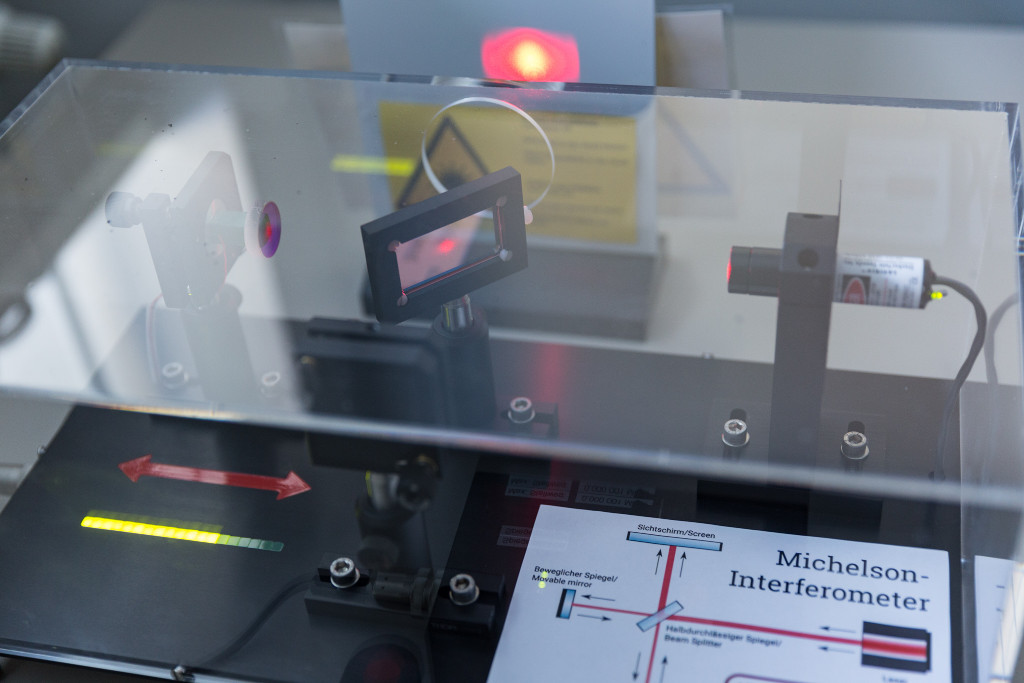}
\caption{Simple Michelson interferometer as an interactive exhibit. When the visitor exerts some pressure to a lever attached to the metal based, labeled ``touch me,'' they produce a minute deformation of the base plate which is sufficient to effect a significant change in the interference pattern. Image credit: B. Knispel, Albert Einstein Institute/Max Planck Institute for Gravitational Physics}
\label{MichelsonExhibit}
\end{center}
\end{figure}

Such Michelson interferometers also make great interactive exhibits. An example for a relatively simple setup can be seen in figure \ref{MichelsonExhibit}. In that version, one of the mirrors can be moved using a piezo-actuator. In another, open version, moderate pressure on the (solid metal) base plate is sufficient to change the distance between mirror and beam-splitter by an amount that causes a clearly visible change in the interference pattern. A much more elaborate version for a Michelson interferometer exhibit is described in \citet{Riles2011}. 

\begin{figure}[htbp]
\begin{center}
\includegraphics[width=0.9\textwidth]{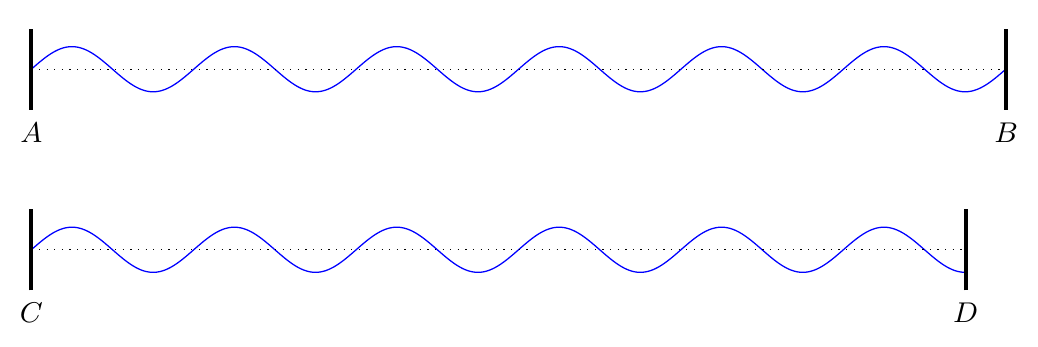}
\caption{Light as a ruler: comparing two distances using a reference light wave. The distance $\overline{AB}$ corresponds to 6 wavelengths, the distance $\overline{CD}$ to 5.75 wavelengths}
\label{BasicRulerMeasurement}
\end{center}
\end{figure}

Implicit in this model is the assumption that the light itself is not changed by the passing gravitational wave. Instead, light acts as a length reference, analogous to a ruler. Basic length measurements are made by comparing the ``marks'' on the ruler --- the periodic phase structure of the light --- with the distance that is to be measured, as sketched in figure \ref{BasicRulerMeasurement}.

In a Michelson interferometer, one ``light ruler'' is used to map the distance between the beam splitter, the first mirror, and back again; a second ''light ruler'' measures the distance between the beam splitter, the second mirror, and back. At the beam splitter, the two light waves are combined, producing an interference pattern. Changes in the pattern signal changes in relative length, as long as those changes are small enough --- relative shifts by multiple integer wave lengths cannot be detected.

\begin{figure}[htbp]
\begin{center}
\includegraphics[width=0.8\textwidth]{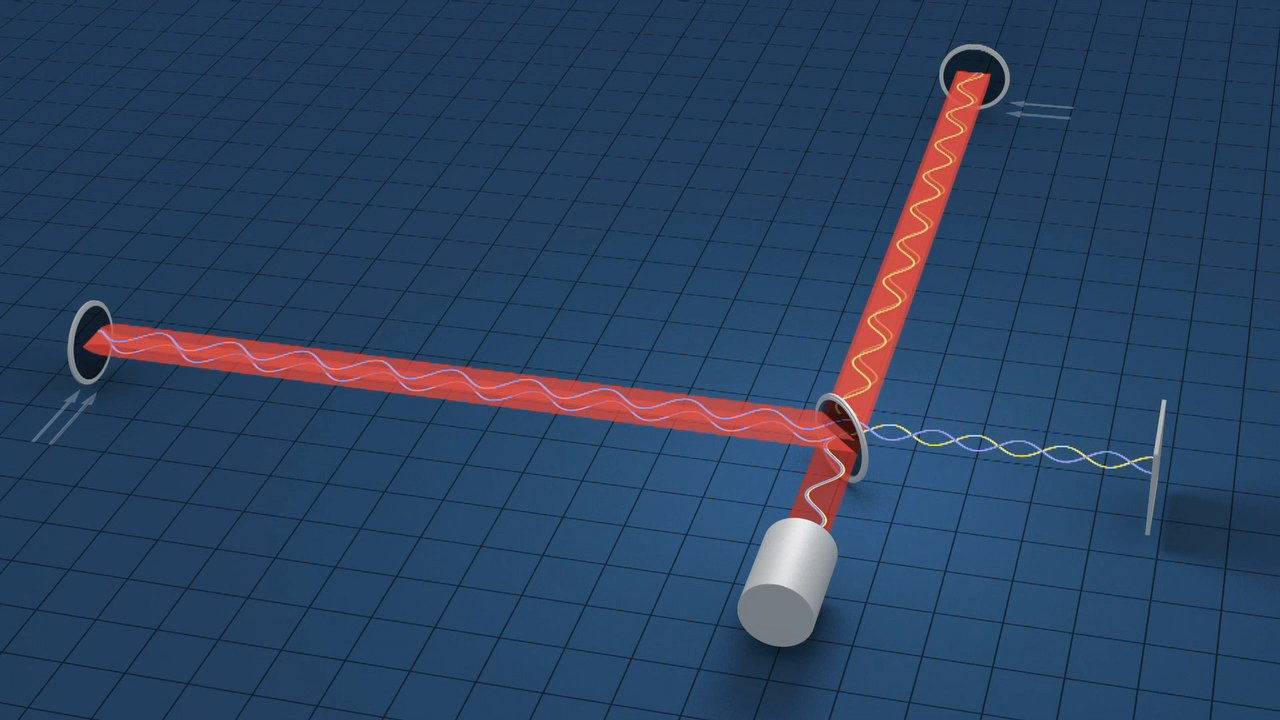}\\
\includegraphics[width=0.8\textwidth]{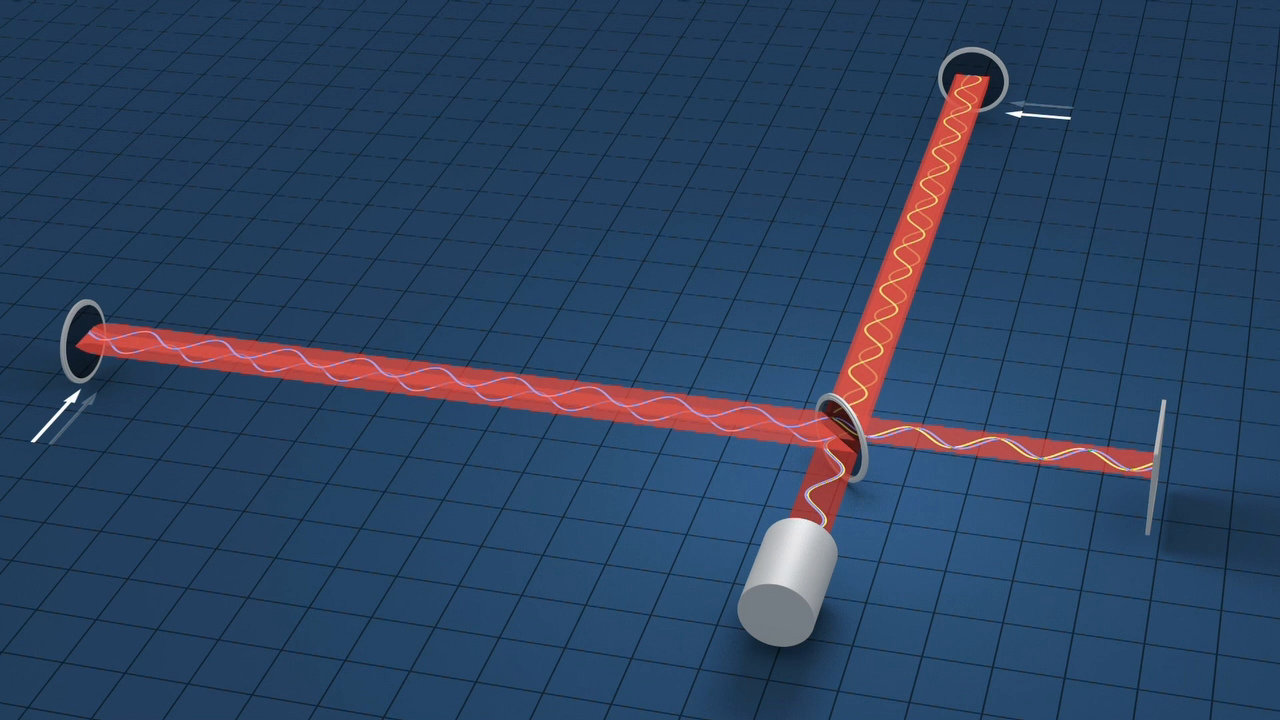}
\caption{Example for the light-as-a-ruler model for interferometric gravitational wave detection (snapshots). The top image shows the ``ground state'' of the detector, in the absence of gravitational waves, with complete destructive interference at the photodetector on the right. The bottom image shows constructive interference as a gravitational wave passes through the system. Animation: LIGO/T. Pyle. The corresponding video can be found online at \href{http://www.ligo.caltech.edu/video/ligo20160211v6}{http://www.ligo.caltech.edu/video/ligo20160211v6}}
\label{LIGOVideo}
\end{center}
\end{figure}

The light-as-a-ruler model is not only used in practical realisations of Michelson interferometers, but also in visualisations of how an interferometric gravitational wave detector works. Two snapshots from an animation created for the LIGO collaboration can be seen in figure \ref{LIGOVideo}. The two images show complete destructive interference (top) resulting in darkness at the photodetector, and perfect constructive interference (bottom) caused by the shift in mirror positions due to the gravitational waves. The animation goes back and forth between those two extreme states, leaving in the light wave as a static ruler (that is, without visible propagation of maxima or minima) as the mirrors shift position.

This is a valid simplified version of the basic detection principle, demonstrating the link between distance changes caused by passing gravitational waves and changing interference patterns. The simplified chain of argument ``gravitational waves create (relative) changes in armlength; Michelson interferometers measure (relative) changes in armlength'' is a valid summary of how modern interferometric gravitational wave detectors work.

The light-as-a-ruler model can become confusing, though, once the question is raised {\em why} we should assume light to remain uninfluenced by the gravitational wave. This question is particularly pertinent if the influence of the gravitational wave has been explained as a ``stretching of space''. Light is propagating through space; if space is stretching, why should light not be stretched along with it? This question is more likely to come up if the audience is familiar with cosmic expansion, and with the associated cosmological redshift --- where light {\em does} stretch along with the ``stretching of space,'' cf. section \ref{CosmoRedshift}.\footnote{In fact, looking at the metric describing an elementary gravitational wave and the metric describing an expanding universe, the situation is exactly analogous. Just as the wavelengths of light in an expanding universe are stretched proportionally with the cosmic scale factor, so a gravitational wave alternately stretches and shrinks the wavelengths of light in an interferometric detector, cf. \citep{Poessel2016b}.}

This is a problem for the static light-as-a-ruler model. If light is influenced by the gravitational wave, how can it serve as a (static) ruler? With a ruler, we can only measure length changes relative to the ruler. If the armlength were stretched in the same way as our ruler, we would not measure any change at all. 

The answer to this apparent paradox can be found in \citet{Saulson1997} and \citet{Faraoni2007}. It is based on the fact that we {\em know} light not to be a static ruler. As the detector operates, light is propagating through the optical system. For one, that means light travel times become important. As the armlength changes, it takes a slightly longer or slightly shorter time for one particular phase of the light wave to propagate through the detector. This shifts the arrival time of wave maxima and minima, and the relative shift for light that has travelled through one vs. light that has travelled through the other arm produces part of the interference pattern. If the time it takes for the light to travel from the light source to the photo detector is short, compared with the period of the gravitational wave, this will be the dominant effect.

In addition, light {\em is} being redshifted and blueshifted by the gravitational wave directly. But since those shifts are, in general, different in the two arms, due to the quadrupole nature of gravitational waves, this effect will still serve to shift the interference pattern away from the dark null state. After all, you cannot create perfect destructive interference by superposing light with two different wavelengths.

The response at least of ground-based gravitational wave detectors is typically calculated in the short-arm limit, or long wavelength limit, assuming that light-travel time through the detector is small compared with the period of the gravitational waves that are to be detected. In this approximation, light propagating through the interferometer will happen so quickly that distances within the detector do not significantly change as, say, one maximum of a light wave travels to the photodetector. In this way, each portion of light entering the detector only probes the present state, producing an interference pattern that is easily calculated. 

However, with its Fabry-Perot cavities amounting to a much longer effective arm length, and with power and signal recycling lengthening the time spent by light inside the detector, deviations from the short-arm approximation can actually become important in LIGO-sized detectors. A more thorough calculation shows that going from the short-arm approximation to a full calculation makes a difference when searches for bursts and for stochastic gravitational waves are extended to higher frequencies of 10 kHz or more (\citealt{Rakhmanov2008}, cf. \citealt{Baskaran2004}). For the third-generation bound to succeed the current LIGO and Virgo detectors, with (physical) armlengths on the order of 10 km, these effects will need to be taken into account routinely \citep{Essick2017}.

An advantage of the more complete description is that it allows for a unified description of ``light-time'' gravitational wave detectors that includes the ground-based interferometers, but also space-based detectors like the planned LISA mission with armlengths of several million kilometers and pulsar timing arrays \citep{Koop2014}.

\subsection{Interferometer animation with tracer particles}
\label{LightMoving}

\begin{figure}[htbp]
\begin{center}
\includegraphics[width=0.6\textwidth]{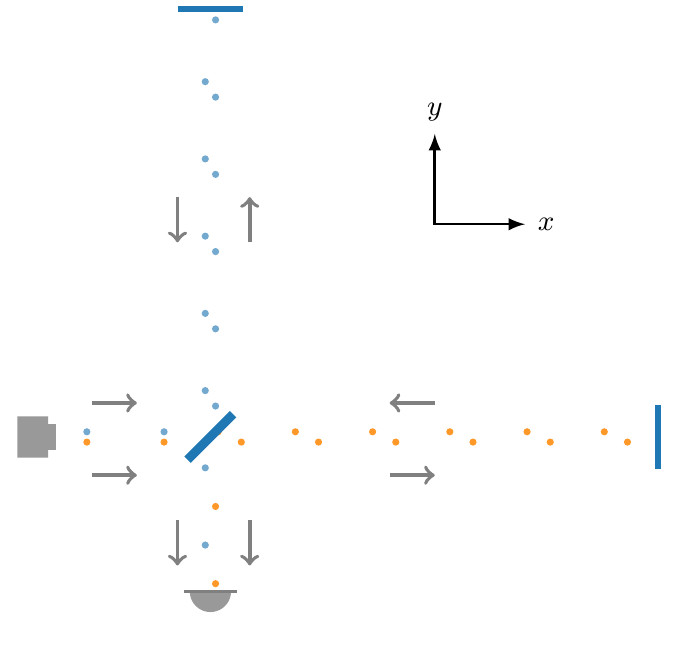}
\caption{Radar ranging model of a Michelson interferometer: snapshot from the animation \href{https://youtu.be/HshG4ueCnY0}{https://youtu.be/HshG4ueCnY0} at a time when no gravitational wave is passing through. Grey arrows have been added to indicate the direction in which the tracer particles are moving, for instance: in the right arm, particles in the top row are moving to the left, particles in the bottom row to the right}
\label{MichelsonIntAnimation1}
\end{center}
\end{figure}

\begin{figure}[htbp]
\begin{center}
\includegraphics[width=0.6\textwidth]{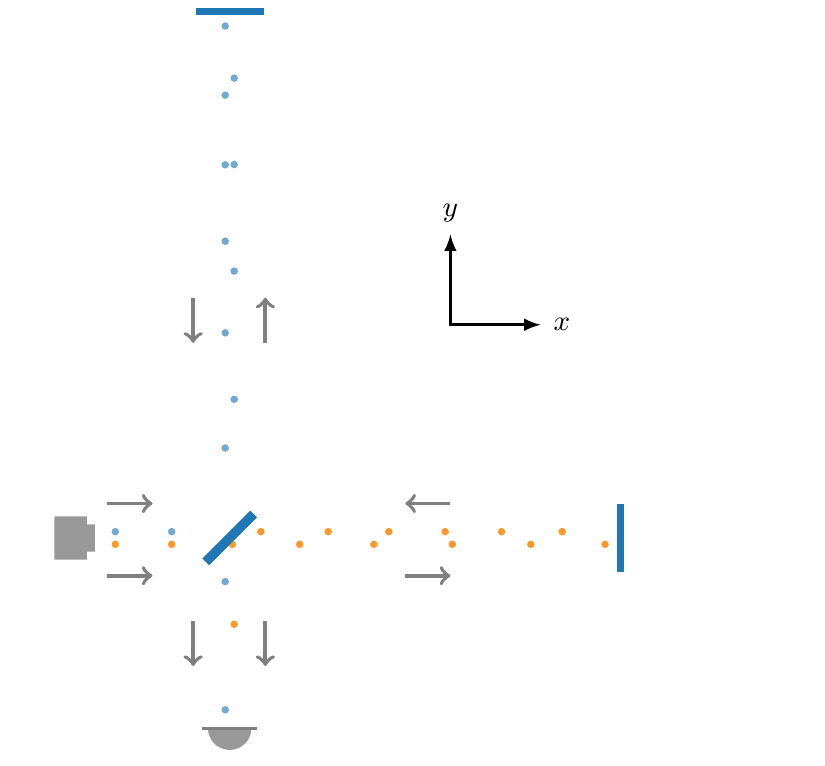}
\caption{Radar ranging model of a Michelson interferometer, with this snapshot taken at a time when a gravitational wave passes perpendicular to the image plane. Grey arrows have been added to indicate the direction in which the tracer particles are moving. Snapshot from the animation \href{https://youtu.be/HshG4ueCnY0}{https://youtu.be/HshG4ueCnY0}}
\label{MichelsonIntAnimation2}
\end{center}
\end{figure}

The key role of light travel time suggests the usefulness of models that model light travelling through an interferometric detector. Since the properties to be modelled involve motion and, more generally, changes over time, the model can be realised in the shape of an animation. Two snapshots from such an animation --- the first showing the detector in the absence of gravitational waves, the second while a gravitational wave is passing through --- are shown in figures  \ref{MichelsonIntAnimation1} and \ref{MichelsonIntAnimation2}.

The animation shows the basic elements of a Michelson interferometer. The travelling blue and yellow dots are photons, modelled as point particles propagating at the speed of light. The usual metric for a linearly polarised linearised gravitational wave propagating in the z direction, written down using the transversal-traceless gauge (TT gauge; one way of choosing a particular coordinate representation of this metric) is
\be
\Dd s^2 = -c^2 \Dd t^2 + [1+h\cos(\omega[t-z/c])]\,\Dd x^2 + [1-h\cos(\omega[t-z/c])]\,\Dd y^2 + \Dd z^2,
\label{TTMetric}
\ee
where $h$ is the gravitational wave amplitude, $\omega = 2\pi f$ the angular frequency of the waves and $f$ their frequency. The coordinates are co-moving: free particles remain at constant coordinate values as the gravitational wave passes, while the metric (\ref{TTMetric}) describes how the proper distances between them change.

In the animation, the position of the beamsplitter is held constant and chosen as the spatial origin. The image plane is the xy plane, with the gravitational wave travelling at right angles to the screen, towards the viewer. The comoving time $t$ is used as the animation time. Laser, end mirrors and photodetector are at constant comoving coordinate locations (corresponding to test particles in free fall). Each frame shows their proper distance from the beam splitter, as given by the metric (\ref{TTMetric}) at constant $t$. 

Light propagation in general relativity is commonly modelled by having light follow null geodesics.\footnote{This is commonly introduced as an axiom when teaching the fundamentals of general relativity, so it is important to keep in mind that
light propagation is not an independent postulate. As soon as we introduce Maxwell's electromagnetic fields, the equation of motion for those field is fixed, in the usual way, by Einstein's equations, more specifically by the vanishing divergence of the energy-momentum tensor that follows from the relevant Bianchi identity. Whatever we want to assume about light propagation must properly be derived from this equation of motion. This is no simple calculation, but one can derive, for instance, that up to terms linear in the Riemann tensor, energy transported by electromagnetic radiation indeed travels on the null cone \citep{Noonan1993}. Incidentally, accurate descriptions of gravitational waves need to take similar effects into account \citep[and references therein]{Blanchet1998}.} Surfaces of constant phase can be traced through an interferometric detector using the eikonal approximation \citep{Rakhmanov2009}.

In the simulation presented here, such phase surfaces are not modelled as surfaces. Instead, we only trace one-dimensional propagation (although the one-dimensional paths do not, of course, lie along a single line) through the detector. The animation shows ``tracer particles,'' or ``tracer photons'' if you will, travelling along null geodesics and represented by small solid circles. Each tracer particle is taken to mark the location of a maximum of a light wave travelling through the detector.\footnote{We do not specify which component of the electric or magnetic field we are talking about and, given the fixed relation between them, do not need to.} Tracer particle propagation can be calculated setting $\Dd s^2=0$, and the coordinate speed of light either in the x or in the y direction, $\Dd x/\Dd t$ and $\Dd y/\Dd t$ respectively, can be read off (\ref{TTMetric}) in a straightforward way, since in our detector set-up, light is propagating either in the x direction or in the y direction. 

The model includes a few artificial changes for added visual clarity, which do not correspond to physical properties: The wave components travelling the two different paths from the laser to the photodetector are represented by blue and yellow tracer particles, respectively. Also, for better visibility, these particles are offset from each other in the transversal direction: From the laser to the beam splitter, yellow and blue particles travel side by side; particles travelling outwards to the end mirrors are offset from the returning particles. This leads to curious jumps as the particles are reflected at the mirrors, but makes it much easier to discern the changing patterns of the inter-particle distances.

As the animation is running, viewers can first see the state of the interferometer in the absence of gravitational waves. At the photodetector, the yellow and blue tracer particles arrive perfectly spaced: yellow, blue, yellow, blue, each pair equidistant. In terms of waves, this means complete destructive interference. 

After some time, a gravitational wave passes through the detector. The period of the passing gravitational wave is chosen to be 10 times the period of the light wave propagating through the detector. As the gravitational wave sets in, the regular pattern changes. Now, yellow and blue tracer particles sometimes arrive close together, sometimes almost evenly spaced. On the simplest level, this transition between a regular blue, yellow, blue, yellow and an irregular pattern is what allows the experimenters to detect the gravitational wave. The fact that the gravitational wave stretches or shrinks the distances between the tracer particles in the same way it stretches and shrinks the armlengths is clearly visible in the animation --- when, say, the horizontal arm is at maximal length, so are the distances between the particles going left, or the particles going right.

\begin{figure}[htbp]
\begin{center}
\includegraphics[width=\textwidth]{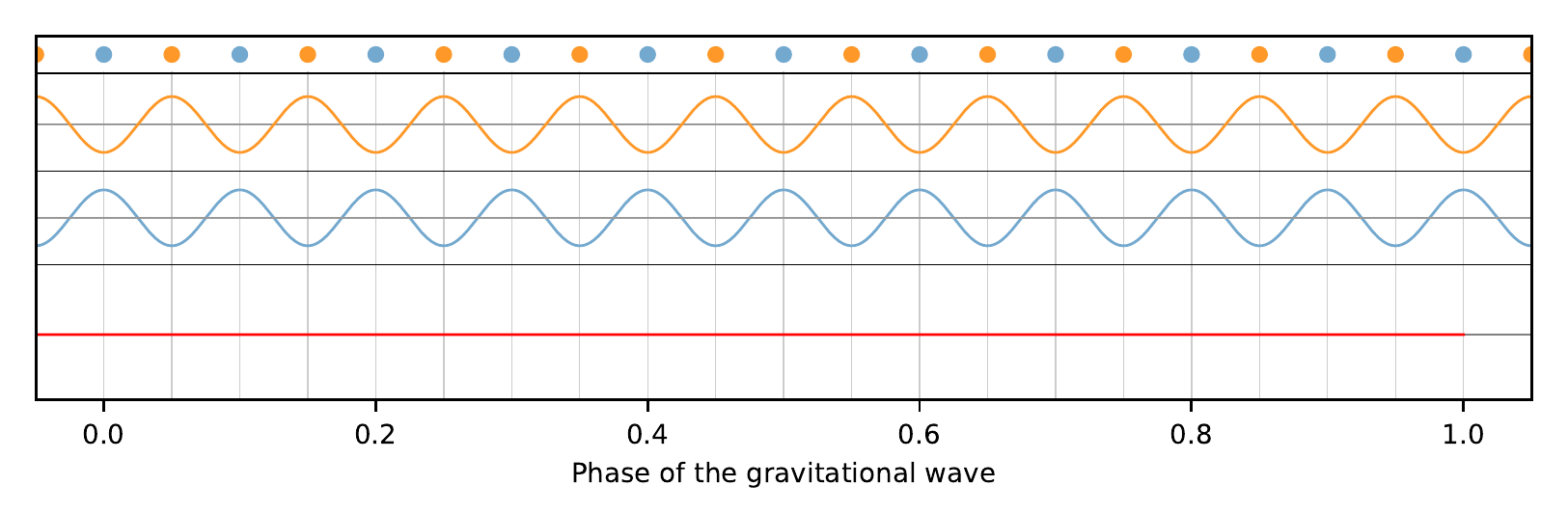}\\

\vspace*{0.25em}

\includegraphics[width=\textwidth]{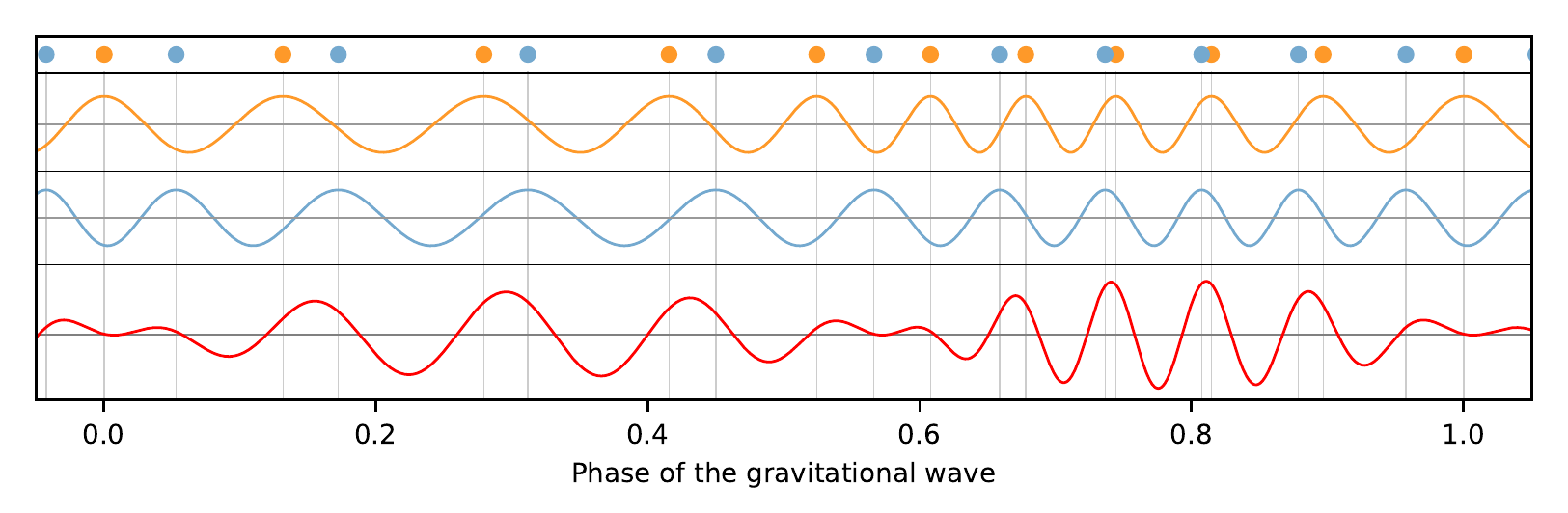}
\caption{Arrival times of tracer particles, and of the corresponding wave components, at the photodetector for the detector animation. Arrival times of the tracer particles are plotted in the top row. Below, the wave components arriving at the photodetector via the vertical arm (blue) and the horizontal arm (yellow), and below that their superposition (red). Time is given in fractions of the period of the gravitational wave. The period of the light wave amounts to one tenth of the gravitational wave period. Top: in the absence of a gravitational wave, the regular pattern of blue and yellow tracer particles leads to complete destructive interference. Bottom: with a gravitational wave passing through, the pattern is perturbed and there is non-zero power at the photodetector}
\label{MovingStrip}
\end{center}
\end{figure}

In the animation (although not in the two snapshots reproduced here), the arrival of particles at the photodetector is recorded on the left-hand edge of a moving strip, creating a real-time record of arrival times and arrival time differences. Two such diagrams are shown in figure \ref{MovingStrip}. Both show the order of arrival of the particles for one full period of the gravitational wave.\footnote{In the animation, the moving strip is moving to the right, chosen as the more natural direction, given Western left-to-right reading habits. In figure \ref{MovingStrip}, the diagram is left-right reversed relative to the moving strip to adapt to the usual way of plotting quantities against time from the left to the right.}

In addition, the diagrams show the wave components arriving at the photodetector via the first and the second arm, as well as their superposition. The changing interference pattern that could actually be measured at the photodetector would correspond to the square of the superposition. In the absence of a gravitational wave, the superposition leads to complete destructive interference, as shown in the top diagram of figure \ref{MovingStrip}. Time is plotted as a fraction of the period of the gravitational wave. At phase 0, the horizontal arm is fully stretched by a factor of 1.35 and the vertical arm is at minimum length at $1/1.35 = 0.74$ of its undisturbed length; at phase 0.5, the situation is reversed.

When, after the onset of the gravitational wave, the situation has settled, the signal at the photodetector has the same periodicity as the gravitational wave. There is unavoidable delay, as the tracer particles in the arm need to travel to the beamsplitter, and thence to the photodetector, before the pattern is registered. But it is clear from the bottom diagram of figure  \ref{MovingStrip} that, within one period of the gravitational wave, there are two maxima and two minima of the envelope of the interference, corresponding to the two maximally distorted states of the detector at phases 0.0 and 0.5. 

That the patterns are not the same in both cases is testament to the fact that the tracer particles are subject to the distortions caused by the gravitational wave on the home stretch, between beam splitter and photodetector, as well. That this effect is so prominent is a consequence of the simplified geometry of the animation, where the arm lengths are at most one order of magnitude longer than the distance between beam splitter and photodetector. In the real LIGO, the difference amounts to two orders of magnitude.\footnote{This can be seen, for example, in the document showing the LIGO nominal optical layout including coordinates that is available at \href{https://dcc.ligo.org/LIGO-D0902838/public}{https://dcc.ligo.org/LIGO-D0902838/public}.} A version of the animation that would, for instance, leave out distortion effects for the whole region between beamsplitter and photodetector would create a more realistic pattern in this respect but would, unrealistically, exempt one part of space from the effect of the gravitational wave, potentially confusing viewers who are trying to make sense of the animation.

Beyond the animation shown here, variations are possible that show different configurations: the detector's insensitivity when the two principal directions of the gravitational wave are at 45 degrees to the detector's L shape, for instance, but also the detector's inability to detect a gravitational wave whose period is an integer multiple of the round-trip light travel time between beam splitter and end mirrors, where stretching and shrinking effects cancel out --- an effect invisible in the static light-as-a-ruler model.

The light travel time model has the advantage of incorporating all basic effects of the gravitational wave on the gravitational wave interferometer. This becomes important, for instance, for the Laser Interferometer Space Antenna LISA, with its million-kilometer arms. The model can also be readily generalised to the detection principle of pulsar timing arrays --- in that case, the tracer particles do not stand in for wave maxima, but for the regular pulses emitted by a distant pulsar. By monitoring the regularity of those pulses, rhythmic changes in the arrival time differences between the pulses caused by a passing low-frequency gravitational wave can be detected. 

The light travel time model's disadvantage is that it has to remain virtual: It can be simulated, and even made into an interactive visualisation, but I can think of no practicable way to build a physical version. In this respect, the model will remain less immediately accessible than a Michelson interferometer, since the latter can be manipulated, hands-on, by the audience.

\subsection{Detecting gravitational waves: noise}
\label{GWDetection}

Some of the practicalities of detecting gravitational waves -- in particular, the various kinds of noise and the corresponding mitigation measures -- can be taught using simple models, as well. A powerful demonstration is that of a pendulum or double pendulum whose point of attachment is moved with a given frequency. Below the pendulum's resonance frequency, the pendulum bob's motion faithfully follows any motion of its attachment point. At the resonance frequency, periodic motion of the attachment point leads to a runaway increase in the pendulum's amplitude of oscillation. Markedly above the resonance frequency, the pendulum bob hardly moves at all. This is the effect used to isolate the test masses and optics of interferometric gravitational wave detectors from ground vibrations, and it is readily demonstrated using a hand-held model pendulum \citep{Hammond2016}.\footnote{A free-hand demonstration on the occasion of a teacher training workshop in late 2015 at Haus der Astronomie, Heidelberg, can be seen in \href{https://youtu.be/AzZbgc40_jE}{https://youtu.be/AzZbgc40\_jE}}

Radiation pressure can also be demonstrated in a toy Michelson interferometer, using a nerf gun to fire ``photons'' at the mirror \citep{Choudhary2018}.

The study of seismic disturbances at the LIGO sites, an important source of noise when detecting gravitational waves at frequencies of the order of 10 Hertz and lower, is the basis for a virtual laboratory, available online.\footnote{{\em LIGO e-lab}, QuarkNet and I2U2, \href{https://www.i2u2.org/elab/ligo/}{https://www.i2u2.org/elab/ligo/} (last accessed 27 April 2018).}

\begin{figure}[htbp]
\begin{center}
\includegraphics[width=0.48\textwidth]{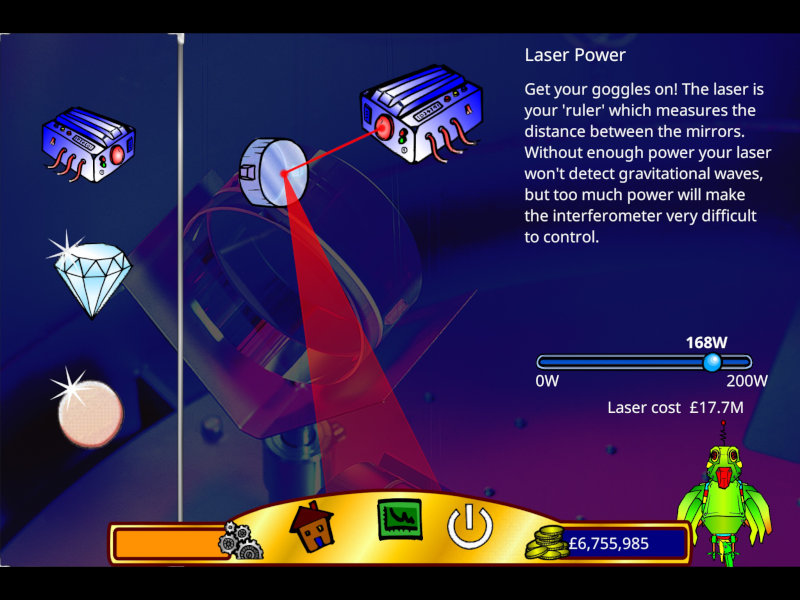} \hspace*{0.5em} \includegraphics[width=0.48\textwidth]{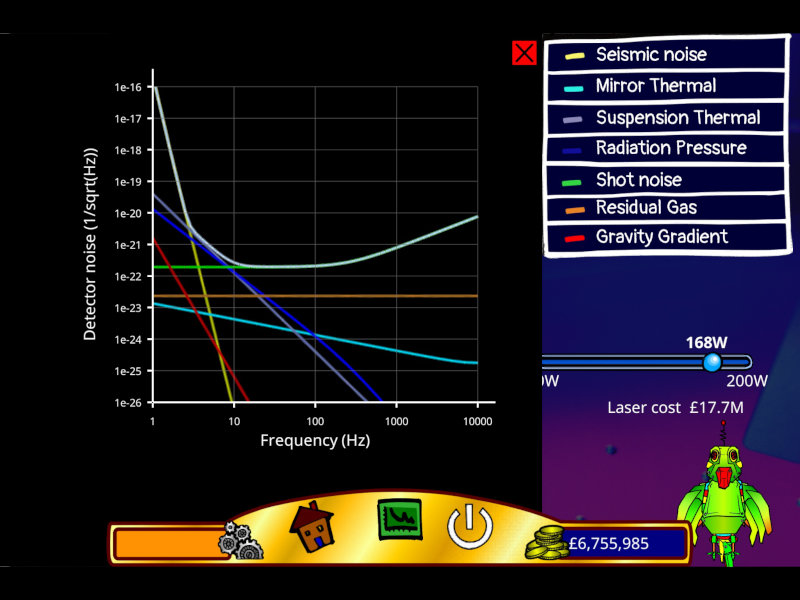}
\caption{Screenshots from the game Space-Time Quest by Laser Labs. Players can change various parameters for their detector (here: choosing laser power using a slider) and are shown directly how their choice effects detector sensitivity (curves on the right). After completing their detector, they are shown its effective range and the probable number of detections they would have made during their science run}
\label{spacetimeQuest}
\end{center}
\end{figure}

Designing a gravitational wave detector, taking into account the various noise mitigation aspects, is an optimisation problem. A simple model for this optimization is at the core of the game {\em Space-Time Quest} developed by the non-profit company Laser Labs for the Gravitational Wave Group at the University of Birmingham,\footnote{ \href{https://www.laserlabs.org/spacetimequest.php}{https://www.laserlabs.org/spacetimequest.php} (last accessed 27 April 2018).} where players are given a limited amount of in-game money, and can make a variety of detector design choices, with the aim of maximising their virtual detector's sensitivity, cf. figure \ref{spacetimeQuest}. An open source version {\em Space Py Quest} exists as well, and allows students to see directly the formulas behind the simulation.

Data analysis is a key part of the direct detection of gravitational waves. For this part of the detection effort, namely for the search of gravitational wave signals in noisy data, there are simple models, as well. The game {\em Black Hole Hunter}\footnote{Cardiff University and LIGO Scientific Collaboration, \href{http://blackholehunter.org/}{http://blackholehunter.org/} (last accessed 27 April 2018)} translates gravitational wave signals into audio signals and lets players search for black hole merger events. Alternative models for the search for gravitational wave signals include simple image pattern matching activities \citep{Larson2006} and simplified analysis based on straightedge-assisted measurements in printouts of simplified simulated signals \citep{Rubbo2007}. 

\section{Conclusion}
\label{Conclusion}

The problem of how to teach the concepts and applications of general relativity to non-specialists, undergraduates and pupils, questions about usefulness of employing models and analogies, but also the pitfalls of that strategy --- in effect, creating only an illusion of understanding --- have been around as long as general relativity itself. Einstein, thirty years after his own attempt at explaining relativity to a general audience \citep{EinsteinAllgemein}, wrote a rather sobering foreword to a popular book on relativity \citep{Barnett1948}, stating:
\begin{quotation}\small
Anyone who has ever tried to present a rather abstract scientific subject in a popular manner knows the great difficulties of such an attempt. Either he succeeds in being intelligible by concealing the core of the problem and by offering the reader only superficial aspects or vague allusions, thus deceiving the reader by arousing in him the deceptive illusion of comprehension; or else he gives an expert account of the problem, but in such a fashion that the untrained reader is unable to follow the exposition and becomes discouraged from reading any further. If these two categories are omitted from to-day's popular scientific literature, surprisingly little remains. But the little that is left is very valuable indeed.
\end{quotation}

Different kinds of models allow us to teach general relativity at many different levels, and to many different kinds of audience. As we make use of images, of narrative structure and previous knowledge, models can help us to reach young school students or aid learners in the setting of an advanced graduate course.

Models can help us understand, and make understand, but they can also create confusion. By their very nature, models for teaching are more simple than the original they are meant, at least in part, to represent. It is particularly unfortunate that the audience members most affected by potentially misleading features of a model are likely to be those who are taking the matter most seriously, and who try hardest to understand the model and its consequences. They are the most likely to worry about the double role of gravity in the elastic sheet model, or the question of what does or doesn't expand in a rubber universe. 

It will depend on the situation on how you will want to address a model's shortcomings. You certainly do not want to confront your audience with all the potential problems up front -- you do not want the simplicity of the model and its main message(s) to get lost in a sea of caveats, its impact diluted. You might not even want to bring up all the possible pitfalls yourself. After all, if your audience does not choose to go down a certain misleading path by themselves, they might even be better off not learning about that path's existence. Caveats are most difficult to handle in a public talk, or a video, where you do not have other than short-term interactions with your audience. In a classroom setting, where you can follow up and actively try to find out what your students have taken away from your teaching, you have a better chance of discovering, and correcting, misconceptions.

It is certainly advisable to create ``model awareness'' -- to openly talk about the fact that what you will be talking about now is a simplified model, and has certain limitations. (And yes, this is probably advisable even in a situation where you think that it is obvious you will be talking about a limited model.) Otherwise, you run the risk of someone among your audience encountering some difficulty with a model, possibly an apparent contradiction, and concluding that this is their own fault or, worst case, that general relativity, or physics/mathematics in general, are not for them. Awareness that these are limited models will give them an out: when their own thinking within a model leads to unsatisfactory results, it is possibly the model's fault. 

Whether or not you deem a particular model suitable or unsuitable for your teaching will depend on a variety of factors: the model's specific advantages and disadvantages, but also your specific audience, and the context you intend to create. 

My impression is that progress in inventing and understanding suitable models is not being disseminated as widely as it should be.\footnote{As an example, consider the discussion about apparent superluminal recession speeds $z>1$ and the role of parallel transport in defining a relativistic relative speed --- an argument that has been re-discovered, and published, whole or in parts, several times over a few decades \citep{Lanczos1923,Synge1960,Stephani1988,Narlikar1994,Liebscher2007,BunnHogg2009,CookBurns2009,Kaya2011}, while in parallel, texts have been written and published that still describe the cosmological redshift as fundamentally different from a Doppler shift \citep{Kaufmann1988,OdenwaldFienberg1993,BeyversKrusch2009,Higbie2014}.} One of my aims with this review article is to make it easier to find, and then cite, previous work on various models for teaching about general relativity and its applications. An issue that is not completely addressed here is that part of the education literature will not be in English, but in the various local languages --- seeing that the language barrier is a considerably more serious issue for school teachers than it is for scientists. As such, English and German contributions are bound to be overrepresented in the reference section, while what I expect to be valuable materials in other languages are missing. 

In conclusion: Use models responsibly! With great explanatory power comes an obligation to do no cognitive harm, to steer your audience clear of potential pitfalls, and to educate yourself about the models you are about to use.

\section*{Acknowledgements}
\addcontentsline{toc}{section}{Acknowledgements}

I would like to thank Domenico Giulini for giving me the opportunity to talk about the role of models in general relativity education at the 2017 DPG spring meeting in Bremen.

I would also like to thank Thomas M\"uller for helpful comments on different versions of this manuscript, as well as for custom-made visualizations, Benjamin Knispel and Peter R. Saulson for a critical reading of chapter \ref{GWSection}, and Carina Volland for assisting me in setting up some of the photographs.

\raggedright

\end{document}